\documentclass{emulateapj}
\renewcommand{\prc}{PhRevC}
\renewcommand{\prd}{PhRevD}
\renewcommand{\nat}{Natur}
\renewcommand{\apjl}{ApJL}

\shorttitle{Magnetar central engine in short GRBs}
\shortauthors{L\"{u} et al}
\slugcomment{}

\begin{document}

\title{The Millisecond Magnetar Central Engine in short GRBs}
\author{Hou-Jun L\"{u}\altaffilmark{1}, Bing Zhang\altaffilmark{1},
Wei-Hua Lei\altaffilmark{2}, Ye Li\altaffilmark{1}, Paul D
Lasky\altaffilmark{3,4}} \altaffiltext{1}{Department of Physics
and Astronomy, University of Nevada Las Vegas, Las Vegas, NV
89154, USA; lhj@physics.unlv.edu, zhang@physics.unlv.edu}
\altaffiltext{2}{School of Physics, Huazhong University of
Science and Technology, Wuhan, 430074, China}
\altaffiltext{3}{Monash Centre for Astrophysics, School of
Physics and Astronomy, Monash University, VIC 3800, Australia }
\altaffiltext{4}{School of Physics, University of Melbourne,
Parkville, VIC 3010, Australia}

\begin{abstract}
One favored progenitor model for short duration gamma-ray
bursts (GRBs) is the coalescence of two neutron stars
(NS$-$NS). One possible outcome of such a merger would be a
rapidly spinning, strongly magnetized neutron star (known as a
millisecond magnetar). These magnetars may be
``supra-massive,'' implying that would collapse to black holes
after losing centrifugal support due to magnetic dipole spin
down. By systematically analyzing the Burst Alert Telescope
(BAT)-XRT light curves of all short GRBs detected by {\em
Swift}, we test how consistent the data are with this central
engine model of short GRBs. We find that the so-called
``extended emission'' feature observed with BAT in some short
GRBs is fundamentally the same component as the ``internal
X-ray plateau'' as observed in many short GRBs, which is
defined as a plateau in the light curve followed by a very
rapid decay. Based on how likely a short GRB is to host a
magnetar, we characterize the entire {\em Swift} short GRB
sample into three categories: the ``internal plateau'' sample,
the ``external plateau'' sample, and the ``no plateau'' sample.
Based on the dipole spin down model, we derive the physical
parameters of the putative magnetars and check whether these
parameters are consistent with expectations from the magnetar
central engine model. The derived magnetar surface magnetic
field $B_{\rm p}$ and the initial spin period $P_0$ fall into a
reasonable range. No GRBs in the internal plateau sample have a
total energy exceeding the maximum energy budget of a
millisecond magnetar. Assuming that the beginning of the rapid
fall phase at the end of the internal plateau is the collapse
time of a supra-massive magnetar to a black hole, and applying
the measured mass distribution of NS$-$NS systems in our
Galaxy, we constrain the neutron star equation of state (EOS).
The data suggest that the NS EOS is close to the GM1 model,
which has a maximum non-rotating NS mass of $M_{\rm TOV} \sim
2.37 M_\odot$.
\end{abstract}


\section{Introduction}

Gamma-ray bursts (GRBs) are classified into categories of
``long soft'' and ``short hard'' (SGRB) based on the observed
duration ($T_{90}$) and hardness ratio of their prompt
gamma-ray emission (Kouveliotou et al. 1993). Long GRBs are
found to be associated with core-collapse supernovae (SNe; e.g.
Galama et al. 1998; Hjorth et al. 2003; Stanek et al. 2003;
Campana et al. 2006; Xu et al. 2013), and typically occur in
irregular galaxies with intense star formation (Fruchter et al.
2006). They are likely related to the deaths of massive stars,
and the ``collapsar'' model has been widely accepted as the
standard paradigm for long GRBs (Woosley 1993; MacFadyen \&
Woosley 1999). The leading central engine model is a
hyper-accreting black hole (e.g. Popham et al. 1999; Lei et al.
2013). Alternatively, a rapidly spinning, strongly magnetized
neutron star (millisecond magnetar) may be formed during core
collapse. In this scenario, magnetic fields extract the
rotation energy of the magnetar to power the GRB outflow (Usov
1992; Thompson 1994; Dai \& Lu 1998; Wheeler et al. 2000; Zhang
\& M\'esz\'aros 2001; Metzger et al. 2008, 2011; Lyons et al.
2010; Bucciantini et al. 2012; L\"{u} \& Zhang 2014).

In contrast, short GRBs are found to be associated with nearby
early-type galaxies with little star formation (Barthelmy et
al. 2005; Berger et al. 2005; Gehrels et al. 2005; Bloom et al.
2006), to have a large offset from the center of the host
galaxy (e.g. Fox et al. 2005; Fong et al. 2010), and to have no
evidence of an associated SNe (Kann et al. 2011, Berger 2014
and references therein). The evidence points toward an origin
that does not involve a massive star. The leading scenarios
include the merger of two neutron stars (NS$-$NS, Pacz\'ynski
1986; Eichler et al. 1989) or the merger of a neutron star and
a black hole (Pacz\'ynski 1991). For NS$-$NS mergers, the
traditional view is that a BH is formed promptly or following a
short delay of up to hundreds of milliseconds (e.g. Rosswog et
al. 2003; Rezzolla et al. 2011; Liu et al. 2012). Observations
of short GRBs with {\em Swift}, on the other hand, indicated
the existence of extended central engine activity following at
least some short GRBs in the form of extended emission (EE;
Norris \& Bonnel 2006), X-ray flares (Barthelmy et al. 2005;
{Campana et al. 2006), and, more importantly, ``internal
plateaus'' with rapid decay at the end of the plateaus
(Rowlinson et al. 2010, 2013). These observations are difficult
to interpret within the framework of a black hole central
engine, but are consistent with a rapidly spinning millisecond
magnetar as the central engine (e.g. Dai et al. 2006; Gao \&
Fan 2006; Metzger et al. 2008; Rowlinson et al. 2010, 2013;
Gompertz et al. 2013, 2014).

About 20\% of short GRBs detected with {\em Swift} have EE
(Sakamoto et al. 2011) following their initial short, hard
spike. Such EE typically has a lower flux than the initial
spike, but can last for tens of seconds (e.g. Perley et al.
2009). The first short GRB with EE detected with {\em Swift}
was GRB 050724, which had a hard spike of $T_{90}\sim$ 3s
followed by a soft tail with a duration of $\sim$150 s in the
{\em Swift} Burst Alert Telescope (BAT; Barthelmy et al. 2005)
band. The afterglow of this GRB lies at the outskirt of an
early-type galaxy at a redshift of $z$=0.258. It is therefore a
``smoking-gun''  burst of the compact star merger population
(Barthelmy et al. 2005; Berger et al. 2005). GRB 060614 is a
special case with a light curve characterized by a short/hard
spike (with a duration $\sim$ 5s) followed by a series of soft
gamma-ray pulses lasting $\sim$100 s. Observationally, it
belongs to a long GRB without an associated SNe (with very deep
upper limits of the SN light, e.g. Della Valle et al. 2006;
Fynbo et al. 2006; Gal-Yam et al. 2006). Some of its prompt
emission properties, on the other hand, are very similar to a
short GRB (e.g. Gehrels et al. 2006). Through simulations,
Zhang et al. (2007b) showed that if this burst were a factor of
8 less luminous, it would resemble GRB 050724 and appear as a
short GRB with EE. Norris \& Bonnell (2006) found a small
fraction of short GRBs in the BATSE catalog qualitatively
similar to GRB 060614. It is interesting to ask the following
two questions. Are short GRBs with EE different from those
without EE? What is the physical origin of the EE?

{\em Swift} observations of the X-ray afterglow of short GRBs,
on the other hand, provide some interesting clues. A good
fraction of {\em Swift} short GRBs exhibit an X-ray plateau
followed by a very sharp drop with a temporal decay slope of
more than 3. The first case was GRB 090515 (Rowlinson et al.
2010). It showed a nearly flat plateau extending to over 180 s
before rapidly falling off with a decay slope of $\alpha \sim
13$.\footnote{The convention $F_\nu \propto t^{-\alpha}
\nu^{-\beta}$ is adopted throughout the paper.} Such rapid
decay cannot be accommodated by any external shock model, and
so the entire X-ray plateau emission has to be attributed to
the internal dissipation of a central engine wind. Such an
``internal plateau'' has previously been observed in some long
GRBs (e.g. Troja et al. 2007; Lyons et al. 2010), but they are
also commonly observed in short GRBs (Rowlinson et al. 2013).
These plateaus can be interpreted as they internal emission of
a spinning-down magnetar which collapses into a black hole at
the end of the plateau (Troja et al. 2007; Rowlinson et al.
2010; Zhang 2014).

If magnetars are indeed operating in some short GRBs, then
several questions emerge. What fraction of short GRBs have a
millisecond magnetar central engine? What are the differences
between short GRBs with EE and those that have an internal
plateau but no EE? Is the total energy of the magnetar
candidates consistent with the maximum rotation energy of the
magnetars according to the theory? What are the physical
parameters of the magnetar candidates derived from
observational data? How can one use the data to constrain the
equation of state (EOS) of neutron stars?

This paper aims to address these interesting questions through
a systematic analysis of both {\em Swift}/BAT and X-ray
Telescope (XRT) data. The data reduction details and the
criteria for sample selection are presented in \S 2. In \S 3,
the observational properties of short GRBs and their afterglows
are presented. In \S 4, the physical parameters of the putative
magnetars are derived and their statistical properties are
presented. The implications on the NS EOS are discussed. The
conclusions are drawn in \S5 with some discussion. Throughout
the paper, a concordance cosmology with parameters $H_0 = 71$
km s$^{-1}$ Mpc $^{-1}$, $\Omega_M=0.30$, and
$\Omega_{\Lambda}=0.70$ is adopted.

\section{Data reduction and sample selection criteria}

The {\em Swift} BAT and XRT data are downloaded from the {\em
Swift} data
archive.\footnote{http://www.swift.ac.uk/archive/obs.php?burst=1}
We systematically process the BAT and XRT GRB data to extract
light curves and time-resolved spectra. We developed an IDL
script to automatically download and maintain all of the {\em
Swift} BAT data. The HEAsoft package {\em version 6.10},
including {\em bateconvert}, {\em batbinevt} {\em Xspec}, {\em
Xselect}, {\em Ximage}, and the {\em Swift} data analysis tools
are used for data reduction. The details of the data analysis
method can be found in several previous papers (Liang et al.
2007; Zhang et al. 2007c; L\"{u} et al. 2014) in our group, and
Sakamoto et al. (2008).

We analyze 84 short GRBs observed with {\em Swift} between 2005
January and 2014 August. Among these, 44 short GRBs are either
too faint to be detected in the X-ray band, or do not have
enough photons to extract a reasonable X-ray light curve. Our
sample therefore only comprises 40 short GRBs, including 8 with
EE.

We extrapolate the BAT (15-150 keV) data to the XRT band
(0.3-10 KeV) by assuming a single power-law spectrum (see also
O'Brien et al. 2006; Willingale et al. 2007; Evans et al.
2009). We then perform a temporal fit to the light curve with a
smooth broken power law in the rest frame,\footnote{Another
empirical model to fit GRB X-ray afterglow light curves is that
introduced by Willingale et al. (2007, 2010). The function was
found to be a good fit of the external plateaus of long GRBs
(e.g. Dainotti et al. 2010), but cannot fit the internal
plateaus which are likely due to the magnetar origin (e.g.
Lyons et al. 2010). We have tried to use the Willingale
function to fit the data in our sample, but the fits are not
good. This is because our short GRB sample includes a large
fraction of internal plateaus. We therefore do not use the
Willingale function to fit the light curves in this paper.}
\begin{eqnarray}
F = F_{0} \left[\left(\frac{t}{t_b}\right)^{\omega\alpha_1}+
\left(\frac{t}{t_b}\right)^{\omega\alpha_2}\right]^{-1/\omega}
\end{eqnarray}
to identify a possible plateau in the light curve. Here,
$t_{b}$ is the break time, $F_b=F_0 \cdot 2^{-1/\omega}$ is the
flux at the break time $t_b$,  $\alpha_1$ and $\alpha_2$ are
decay indices before and after the break, respectively, and
$\omega$ describes the sharpness of the break. The larger the
$\omega$ parameter, the sharper the break. An IDL routine named
``mpfitfun.pro'' is employed for our fitting (Markwardt 2009).
This routine performs a Levenberg-Marquardt least-square fit to
the data for a given model to optimize the model parameters.

Since the magnetar signature typically invokes a plateau phase
followed by a steeper decay (Zhang \& M\'esz\'aros 2001), we
search for such a signature to decide how likely a GRB is to be
powered by a magnetar. Similar to our earlier work (L\"u \&
Zhang 2014), we introduce three grades to define the likelihood
of a magnetar engine.
\begin{itemize}
\item The internal plateau(Internal) sample: this sample
     is defined by those bursts that exhibit a plateau
     followed by a decay with $t^{-2}$ or steeper than 3.
     The $t^{-2}$ decay is expected by the magnetar dipole
     spin down model (Zhang \& M\'esz\'aros 2001), while a
     slope steeper than three is an indication that the
     emission is powered by internal dissipation of the
     magnetar wind, since essentially no external shock
     model can account for such a steep decay. This sample
     is similar to the ``Gold'' sample defined by L\"u \&
     Zhang (2014)\footnote{L\"u \& Zhang (2014) studied the
     magnetar engine candidates for long GRBs. The grades
     defined in that paper were based on the following
     criteria. Gold sample: those GRBs which dispaly an
     ``internal plateau''; Silver sample: those GRBs which
     display an ``external plateau,'' whose energy
     injection parameter $q$ is consistent with being 0, as
     predicted by the dipole spin down model of GRBs;
     Aluminum sample: those GRBs which display an external
     plateau, but the derived $q$ parameter is not
     consistent with 0; Non-magnetar sample: those GRBs
     which do not show a clear plateau feature.}, but with
     the inclusion of two GRBs with a $t^{-2}$ decay
     following the plateau. These two GRBs (GRB 061201 and
     GRB 070714B) also have a plateau index close to 0 as
     demanded by the magnetar spin down model, and
     therefore are strong candidates for magnetar internal
     emission. For those cases with a post-plateau decay
     index steeper than three, the rapid decay at the end
     of plateau may mark the implosion of the magnetar into
     a black hole (Troja et al. 2007; Zhang 2014).
     Altogether there are 20 short GRBs identified as
     having such a behavior, 13 of which have redshift
     measurements and 7 of which are short GRBs with EE.
     For these latter GRBs, the extrapolated X-ray light
     curves from the BAT band in the EE phase resemble the
     internal plateaus directly detected in the XRT band in
     other GRBs. The light curves of these 22 GRBs are
     presented in Fig.1, along with the smooth broken
     power-law fits. The fitting parameters are summarized
     in Table 1.
\item The external plateau (External) sample: this sample
     includes those GRBs with a plateau phase followed by a
     normal decay segment with a post-decay index close to
     -1. The pre- and post-break temporal and spectral
     properties are consistent with the external forward
     shock model with the plateau phase due to continuous
     energy injection into the blastwave. This sample is
     similar to the Silver and Aluminum samples in L\"u \&
     Zhang (2014). We identified 10 GRBs in this
     group.\footnote{The SN-less long GRB 060614 is
     included in this category. It has EE and an additional
     external plateau at late times.} The XRT light curves
     are presented in Figure 2 along with the smooth broken
     power-law fits. The fitting results are presented in
     Table 1.
\item No plateau (Non) sample: we identify 8 GRBs that do
     not have a significant plateau behavior. They either
     have a single power-law decay, or have erratic flares
     that do not present a clear magnetar signature.
\end{itemize}

Figure 3 collects all of the light curves of the GRBs in our
samples. The Internal sample with or without EE is collected in
Figures 3(a) and 3(b); the External sample (without EE) is
collected in Figure 3(c); and the Non sample are collected in
Figure 3(d).

\section{Derived physical parameters and statistics}

In this section, we derive physical parameters of the short GRBs in various
samples, and perform some statistics to compare among different samples.

\subsection{EE and internal plateau}

Our first task is to investigate whether short GRBs with EE are
fundamentally different from those without EE. The EE has been
interpreted within the magnetar model as the epoch of tapping
the spin energy of the magnetar (Metzger et al. 2008;
Bucciantini et al. 2012). On the other hand, a good fraction of
short GRBs without EE have an internal plateau that lasts for
hundreds of seconds, which can also be interpreted as the
internal emission of a magnetar during the spin-down phase
(Troja et al. 2007; Yu et al. 2010; Rowlinson et al. 2013;
Zhang 2014). It would be interesting to investigate whether or
not there is a connection between the two groups of bursts.

Analyzing the whole sample, we find that the short GRBs with EE
do not show a plateau in the XRT band (except GRB 060614, which
shows an external plateau at a later epoch). Extrapolating the
BAT data to the XRT band, the EE appears as an internal plateau
(Figure.1). Fitting the joint light curve with a broken
power-law model, one finds that there is no significant
difference in the distribution and cumulative distribution of
the plateau durations for the samples with and without EE
(Figure 4(a)). The probability ($p$) that the two samples are
consistent with one another, as calculated using a student's
t-test, is 0.65.\footnote{The hypothesis that the two
distributions are from a same parent sample is statistically
rejected if $p < 0.05$. The two samples are believed to have no
significant difference if $p > 0.05$.}  Figure 4(b) shows the
redshift distributions of those short GRBs in our sample which
have redshift measurements. Separating the sample into EE and
non-EE sub-samples does not reveal a noticeable difference. In
Figure 4(c), we show the flux distribution of the plateau at
the break time. It is shown that the distribution for the EE
sub-sample (mean flux $\log F_{b}=-8.74\pm0.12 ~{\rm
ergs~s^{-1} ~cm^{-2}}$) is systematically higher than that for
the non-EE sub-sample (mean flux $\log F_{b}=-9.84\pm0.07 ~{\rm
ergs~s^{-1} ~cm^{-2}}$). However, the combined sample (Figure
4(d)) shows a single-component log-normal distribution with a
mean flux of $\log F_{b}=-9.34\pm0.07 ~{\rm ergs~s^{-1}
~cm^{-2}}$, with a student's t-test probability of $p=0.76$ of
belonging to the same parent sample. This suggests that the EE
GRBs are simply those with brighter plateaus, and the detection
of EE is an instrumental selection effect. We also calculate
the luminosity of the internal plateau at the break time for
both the GRBs with and without EE. If no redshift is measured,
then we adopt $z=$0.58, the center value for the measured
redshift distribution (Figure 4(b)). We find that the plateau
luminosity of the EE ($\log L_{0}=49.41\pm0.07 ~{\rm
ergs~s^{-1}}$) is systematically higher than the no-EE sample
($\log L_{0}=48.68 \pm0.04 ~{\rm ergs~s^{-1}}$), see Figure
4(e). However, the joint sample is again consistent with a
single component ($\log L_0 = 48.91\pm0.07 ~{\rm ergs~s^{-1}}$,
Figure 4(f)), with a student's t-test probability of $p=0.74$.
For the samples with the measured redshifts only, our results
(shown in the inset of Figure 4(e) and 4(f)), the results are
similar.

The distributions of the plateau duration, flux, and luminosity
suggest that the EE and X-ray internal plateaus are
intrinsically the same phenomenon. The different plateau
luminosity distribution, along with the similar plateau
duration distribution, suggest that the fraction of short GRBs
with EE should increase with softer, more sensitive detectors.
The so-called ``EE" detected in the BAT band is simply the
internal plateau emission when the emission is bright and hard
enough.

\subsection{The host offset and local environment of Internal
and External samples}

One curious question is why most (22) short GRBs have an
internal plateau, whereas some others (10) show an external
plateau. One naive expectation is that the External sample may
have a higher circumburst density than the Internal sample, and
so the external shock emission will be greatly enhanced. It has
been found that short GRBs typically have a large offset from
their host galaxies (Fong et al. 2010; Fong \& Berger 2013;
Berger 2014), so that the local interstellar medium (ISM)
density may be much lower than that of long GRBs (e.g. Fan et
al. 2005; Zhang et al. 2009; Kann et al. 2011). This is likely
due to the asymmetric kicks during the SNe explosions of the
binary systems when the two compact objects (NS or BH) were
born (e.g. Bloom et al. 1999, 2002). If the circumburst density
is the key factor in creating a difference between the Internal
and External samples, then one would expect that the offset
from the host galaxy is systematically smaller for the External
sample than the Internal sample.

With the data collected from the literature (Fong et al 2010,
Leibler \& Berger 2010, Fong \& Berger 2013, Berger 2014), we
examine the environmental effect of short GRBs within the
Internal and External samples. The masses, ages, and specific
star formation rates of the host galaxies do not show
statistical differences between the two samples. The physical
offsets and the normalized offsets\footnote{The normalized
offsets are defined as the physical offsets normalized to
$r_e$, the characteristic size of a galaxy defined by Equation
(1) of Fong et al. (2010).} of these two samples are shown in
the left and right panels of Figure 5. It appears that the
objects in the External sample tend to have smaller offsets
than those in the Internal sample, both for the physical and
normalized offsets. This is consistent with the above
theoretical expectation. Nonetheless, the two samples are not
well separated in the offset distributions. Some GRBs in the
External sample still have a large offset. This may suggest a
large local density in the ISM or intergalactic medium (IGM)
far away from the galactic center, or that some internal
emission of the nascent magnetars may have observational
signatures similar to the external shock emission.

\subsection{Energetics and luminosity}

Similar to L\"{u} \& Zhang (2014), we derive the isotropic
$\gamma$-ray energy ($E_{\rm \gamma,iso}$) and isotropic
afterglow kinetic energy ($E_{\rm K,iso}$) of all of the short
GRBs in our sample. To calculate $E_{\rm \gamma,iso}$, we use
the observed fluence in the detector's energy band and
extrapolate it to the rest-frame $1-10^4$ keV using spectral
parameters with $k$-correction (for details, see L\"{u} \&
Zhang, 2014). If no redshift is measured, then we use $z=$0.58
(see Table 2).

To calculate $E_{\rm K,iso}$, we apply the method described in
Zhang et al. (2007a). Since no stellar wind environment is
expected for short GRBs, we apply a constant density model. One
important step is to identify the external shock component. If
an external plateau is identified, then it is straightforward
to use the afterglow flux to derive $E_{\rm K,iso}$. The
derived $E_{\rm K,iso}$ is constant during the normal decay
phase, but it depends on the time during the shallow decay
phase (Zhang et al. 2007a). We therefore use the flux in the
normal decay phase to calculate $E_{\rm K,iso}$. For the Non
sample, no plateau is derived and we use any epoch during the
normal decay phase to derive $E_{\rm K,iso}$. For GRBs in the
Internal sample, there are two possibilities. (1) In some
cases, a normal decay phase is detected after the internal
plateau, e.g. GRBs 050724, 062006, 070724A, 071227, 101219A,
and 111121A in Figure 1. For these bursts, we use the flux at
the first data point during the normal decay phase to derive
$E_{\rm K,iso}$. (2) For those bursts whose normal decay
segment is not observed after the rapid decay of the internal
plateau at later times (the rest of GRBs in Figure 1), we use
the last data point to place an upper limit to the underlying
afterglow flux. An upper limit of $E_{\rm K,iso}$ is then
derived.

We adopt two typical values of the circumburst density to
calculate the afterglow flux, $n=1 ~{\rm cm^{-3}}$ (a typical
density of the ISM inside a galaxy) and $n=10^{-3} ~{\rm
cm^{-3}}$ (a typical density in the ISM/IGM with a large offset
from the galaxy center). For the late epochs we are discussing,
fast cooling is theoretically disfavored and we stick to the
slow cooling ($\nu_m < \nu_c$) regime. Using the spectral and
temporal information from the X-ray data, we can diagnose the
spectral regime of the afterglow based on the closure relations
(e.g. Zhang \& M\'esz\'aros 2004; see Gao et al. 2013a for a
complete review). Most GRBs belong to the $\nu>{\rm
max}(\nu_m,\nu_c)$ regime, and we use Equations(10) and (11) of
Zhang et al. (2007a) to derive $E_{\rm K,iso}$. In some cases,
the spectral regime $\nu_m < \nu < \nu_c$ is inferred and
Equation (13) of Zhang et al. (2007a) is adopted to derive
$E_{\rm K,iso}$

In order to place an upper limit of $E_{\rm K,iso}$ for the
Internal sample GRBs without a detected external shock
component, one needs to assume the spectral regime and decay
slope of the normal decay. To do so, we perform a statistical
analysis of the decay slope and spectral index in the normal
decay phase using the External and Non samples (Figure 6).
Fitting the distributions with a Gaussian distribution, we
obtain center values of $\alpha_{0,c}=1.21\pm 0.04$ and
$\beta_{X,c}=0.88\pm 0.05$. We adopt these values to perform
the calculations. Since $2\alpha_{0}\approx 3\beta_{X}$ is
roughly satisfied, the spectral regime belongs to $\nu_m < \nu
< \nu_c$, and again Equation (13) of Zhang et al. (2007a) is
again used to derive the upper limit of $E_{\rm K,iso}$.

In our calculations, the microphysics parameters of the shocks
are assigned to standard values derived from the observations
(e.g. Panaitescu \& Kumar 2002; Yost et al. 2003):
$\epsilon_{e}$=0.1 and $\epsilon_{B}=0.01$. The Compton
parameter is assigned to a typical value of $Y=1$. The
calculation results are shown in Table 2.

After obtaining the break time $t_b$ through light curve
fitting, we derive the bolometric luminosity at the break time $t_b$:
\begin{eqnarray}
L_b = 4\pi D_L^2 F_b \cdot k,
\end{eqnarray}
where $F_b$ is the X-ray flux at $t_b$ and $k$ is the
$k$-correction factor. For the Internal sample, we derive the
isotropic internal plateau energy, $E_{X, iso}$, using the
break time and break luminosity (L\"u \& Zhang 2014), i.e.
\begin{eqnarray}
E_{X, iso} \simeq L_b \cdot \frac{t_b}{1+z}
\end{eqnarray}
This energy is also the isotropic emission energy due to internal energy
dissipation.

Comparisons of the statistical properties of various derived
parameters for the Internal and External samples are presented
in Figure 7. Figure 7(a) and (b) show the distributions of the
internal plateau luminosity and duration. For the External
sample, no internal plateau is detected and we place an upper
limit on the internal plateau luminosity using the observed
luminosity of the external plateau. The internal plateau
luminosity of the Internal sample is $L_b \sim 10^{49} ~{\rm
ergs~s^{-1}}$. The distribution of the upper limits of $L_b$ of
the External sample peaks at a smaller value of $L_b \sim
10^{47.5} ~{\rm ergs~s^{-1}}$. This suggests that the
distribution of internal plateau luminosity $L_b$ has an
intrinsically very broad distribution (Figure 7(a)). The
distribution of the duration of the plateaus for the Internal
sample peaks around 100 s, which is systematically smaller than
the duration of the plateaus in the External sample, which
peaks around $10^{3.3}$ s. In Figure 7(a) and (b), we also
compare the plateau luminosity and duration distributions of
our sample with those of long GRBs (Dainotti et al. 2015) and
find that the Internal sample is quite different from long
GRBs, whereas the External sample resembles the distributions
of the long GRBs well. According to our interpretation, the
duration of the internal plateaus is defined by the collapse
time of a supra-massive neutron star (Troja et al. 2007; Zhang
2014). For the external plateaus, the duration of the plateau
is related to the minimum of the spin-down time and the
collapse time of the magnetar. Therefore, by definition, the
External sample should have a higher central value for the
plateau duration than the Internal sample. The observations are
consistent with this hypothesis.

Figure 7(c) and (d) show the distribution of $\gamma-$ray
energy ($E_{\gamma, iso}$) and the internal dissipation energy
($E_{X, iso}$). The $E_{\gamma, iso}$ of the Internal sample is
a little bit less than that of the External sample, but $E_{X,
iso}$ is much larger (for the External sample, only an upper
limit of $E_{X, iso}$ can be derived). This means that internal
dissipation is a dominant energy release channel for the
Internal sample. Figures 7(e) and (f) show the distributions of
the blastwave kinetic energy ($E_{K, iso}$) for different
values of the number density, $n = 1~{\rm cm^{-3}}$ and
$n=10^{-3}~{\rm cm^{-3}}$. In both cases, $E_{K, iso}$ of the
Internal sample is systematically smaller than that for the
External sample. The results are presented in Tables 2 and 3.

In Figures 7(g) and (h) (for $n=1,10^{-3}~{\rm cm^{-3}}$,
respectively), we compare the inferred total energy of GRBs
($E_{\rm total} = E_{\gamma} + E_{\rm X} + E_{\rm K}$) with the
total rotation energy $E_{\rm rot}$ of the millisecond
magnetar:
\begin{eqnarray}
E_{\rm rot} = \frac{1}{2} I \Omega_{0}^{2}
\simeq 3.5 \times 10^{52}~{\rm erg}~
M_{2.46} R_6^2 P_{0,-3}^{-2},
\label{Erot}
\end{eqnarray}
where $I$ is the moment of inertia, $R$, $P_0$, and $\Omega_0$
are the radius, initial period, and initial angular frequency
of the neutron star, and $M$ is normalized to the sum of the
masses of the two NSs ($2.46 M_\odot$) in the observed NS$-$NS
binaries in our Galaxy.\footnote{Strictly speaking, $M$ is
normalized to the mean of the sum of masses of binary NS
systems, taking into account the conservation of rest mass
(Lasky et al. 2014), and ignoring the negligible mass lost
during the merger process (e.g., Hotokezaka et al. 2013).}
Hereafter, the convention $Q = 10^x Q_x$ is adopted in cgs
units for all of the parameters except the mass. It is found
that the total energy of the GRBs are below the $E_{\rm rot}$
line if the medium density is high ($n=1~{\rm cm^{-3}}$). This
energy budget is consistent with the magnetar hypothesis,
namely, all of the emission energy ultimately comes from the
spin energy of the magnetar. For a low-density medium
($n=10^{-3}~{\rm cm^{-3}}$), however, a fraction of GRBs in the
External sample exceed the total energy budget. The main reason
for this is that a larger $E_{\rm K,iso}$ is needed to
compensate a small $n$ in order to achieve a same afterglow
flux. If these GRBs are powered by a magnetar, then the data
demand a relatively high $n$. This is consistent with the
argument that the External sample has a large $n$ and so the
external shock component is more dominant.

Figure 8(a) shows the observed X-ray luminosity at $t=10^{3}$ s
($L_{t=10^3 {\rm s}}$) as a function of the decay slope
$\alpha_{2}$. Figures 8(b) and 8(c) show the respective
distributions of $L_{t=10^3 s}$ and $\alpha_{2}$. The Internal
and External samples are marked in red and black, respectively.
On average, the Internal sample has a relatively smaller
$L_{t=10^3 s}$ than that of the External sample (Figure 8(b)).
The fitting results from the distributions of various
parameters are collected in Table 3.

\section{The millisecond magnetar central engine model and implications}

In this section, we place the short GRB data within the
framework of the millisecond magnetar central engine model and
derive the relevant model parameters of the magnetar, and
discuss the physical implications of these results.

\subsection{The millisecond magnetar central engine model}

We first briefly review the millisecond magnetar central engine
model of short GRBs. After the coalescence of the binary NSs,
the evolutionary path of the central post-merger product
depends on the unknown EOS of the neutron stars and the mass of
the protomagnetar, $M_{\rm p}$. If $M_{\rm p}$ is smaller than
the non-rotating Tolman-Oppenheimer-Volkoff maximum mass
$M_{\rm TOV}$, then the magnetar will be stable in equilibrium
state (Cook et al. 1994; Giacomazzo \& Perna 2013, Ravi \&
Lasky 2014). If $M_{\rm p}$ is only slightly larger than
$M_{\rm TOV}$, then it may survive to form a supra-massive
neutron star (e.g. Duez et al. 2006), which would be supported
by centrifugal force for an extended period of time, until the
star is spun down enough so that centrifugal force can no
longer support the star. At this epoch, the neutron star would
collapse into a black hole.

Before the supra-massive neutron star collapses, it would spin
down due to various torques, the most dominant of which may be
the magnetic dipole spin down (Zhang \& M\'esz\'aros
2001).\footnote{Deviations from the simple dipole spin-down
formula may be expected (e.g. Metzger et al. 2011; Siegel et
al. 2014), but the dipole formula may give a reasonable
first-order approximation of the spin-down law of the nascent
magnetar.} The characteristic spin-down timescale $\tau$ and
characteristic spin-down luminosity $L_0$ depend on $\Omega_0 =
2\pi/P_0$ and the surface magnetic field at the pole $B_p$,
which read (Zhang \& M\'esz\'aros 2001)
\begin{eqnarray}
 \tau &=&\frac{3c^{3}I}{B_{p}^{2}R^{6}\Omega_{0}^{2}}= \frac{3c^{3}IP_{0}^{2}}{4\pi^{2}B_{p}^{2}R^{6}}
 \nonumber\\
 &=&2.05 \times 10^3~{\rm s}~ (I_{45} B_{p,15}^{-2} P_{0,-3}^2 R_6^{-6}),
\label{tau}
\end{eqnarray}
\begin{eqnarray}
 L_0 &=&\frac{I\Omega_{0}^{2}}{2\tau}
 =1.0 \times 10^{49}~{\rm erg~s^{-1}} (B_{p,15}^2 P_{0,-3}^{-4} R_6^6).
\label{L0}
\end{eqnarray}
For a millisecond magnetar, the open field line region opens a
very wide solid angle, and so the magnetar wind can be
approximated as roughly isotropic.

Another relevant timescale is the collapse time of a
supra-massive magnetar, $t_{\rm col}$. For the Internal sample,
the observed break time $t_b$ either corresponds to $t_{\rm
col}$ or $\tau$, depending on the post-break decay slope
$\alpha_2$. If $\alpha_2 \simeq 2$, then the post-break decay
is consistent with a dipole spin-down model, so that $t_b$ is
defined by $\tau$ and one has $t_{\rm col} > \tau$. On the
other hand, if the post-decay slope is steeper than 3, i.e.
$\alpha_2
> 3$, then one needs to invoke an abrupt cessation of the GRB
central engine to interpret the data (Troja et al. 2007;
Rowlinson et al. 2010, 2013; Zhang 2014). The break time is
then defined by the collapse time $t_{\rm col}$, and one has
$t_{\rm col} \leq \tau$. Overall, one can write
\begin{eqnarray}
 \tau \left\{
   \begin{array}{ll}
    =t_{\rm b}/(1+z), & \alpha_{2}=2, \\
    \geq t_{\rm b}/(1+z), & \alpha_{2}>3.
   \end{array}
  \right.
\label{tau1}
\end{eqnarray}
and
\begin{eqnarray}
 t_{col} \left\{
   \begin{array}{ll}
    > t_{\rm b}/(1+z), & \alpha_{2}=2, \\
    =t_{\rm b}/(1+z), & \alpha_{2}>3.
   \end{array}
  \right.
\label{tcol}
\end{eqnarray}
In both cases, the characteristic spin-down luminosity is
essentially the plateau luminosity, which may be estimated as
\begin{eqnarray}
 L_0 \simeq L_{\rm b}
\label{L0b}
\end{eqnarray}

\subsection{Magnetar parameters and correlations}

With the above model, one can derive magnetar parameters and
perform their statistics. Two important magnetar parameters to
define magnetar spin down, i.e. the initial spin period $P_0$
and the surface polar cap magnetic field $B_{\rm p}$, can be
solved from the characteristic plateau luminosity $L_0$
(Equation (\ref{L0})) and the spin-down timescale $\tau$
(Equation (\ref{tau}); (Zhang \& M\'esz\'aros 2001), i.e.
\begin{eqnarray}
 B_{\rm p,15} = 2.05~{\rm G} (I_{45} R_6^{-3} L_{0,49}^{-1/2} \tau_{3}^{-1}),
\label{Bp}
\end{eqnarray}
\begin{eqnarray}
 P_{0,-3} = 1.42~{\rm s}~ (I_{45}^{1/2} L_{0,49}^{-1/2} \tau_{3}^{-1/2}).
\label{P0}
\end{eqnarray}
Since the magnetar wind is likely isotropic for short GRBs (in
contrast to long GRBs; L\"u \& Zhang 2014), the measured $L_0$
and $\tau$ can be directly used to derive these two parameters.
For the Internal sample, both $P_0$ and $B_{\rm p}$ can be
derived if $\alpha_{2}=2$. If $\alpha_{2}>3$, we can derive the
upper limit for $P_0$ and $B_{\rm p}$. The results are
presented in Table 2 and Figure 9(a).\footnote{The derived
magnetar parameters of most GRBs are slightly different from
those derived by Rowlinson et al. (2013). One main discrepancy
is that they used $M_p = 1.4 M_\odot$ to calculate the
protomagnetar's moment of inertia $I$, wheareas we used $M_p
=2.46 M_\odot$, which is more relevant for post-merger
products. The different data selection criteria and fitting
methods also contribute to the discrepancies between the two
works. }

Figure 9(b) shows the distribution of the collapse times for
our Internal sample. For GRB 061201 and GRB 070714B, the decay
slope following the plateau is $\alpha_2\sim 2$, which means
that we never see the collapsing feature. A lower limit of the
collapse time can be set by the last observational time, so
that the stars should be stable long-lived magnetars. For the
collapsing sample, the center value of the $t_{\rm col}$
distribution is $\sim 100$ s, but the half width spans about
one order of magnitude.

Figure 10(a) presents an anti-correlation between $L_{0}$ and
$t_{\rm col}$,  i.e.
\begin{eqnarray}
\log L_{0,49}=(-2.79 \pm 0.39) \log t_{col,2}-(0.45 \pm 0.28) \nonumber \\
\end{eqnarray}
with $r=0.87$ and $p<0.0001$. This suggests that a longer
collapse times tends to have a lower plateau luminosity. This
is consistent with the expectation of the magnetar central
engine model. The total spin energy of the millisecond
magnetars may be roughly standard. A stronger dipole magnetic
field tends to power a brighter plateau, making the magnetar
spin down more quickly, and therefore giving rise to a shorter
collapse time (see also Rowlinson et al. 2014).

Figure 10(b) presents an anti-correlation between $E_{\rm
total,iso}$ and $t_{\rm col}$.
\begin{eqnarray}
\log E_{\rm total,iso,52}=(-1.08 \pm 0.27) \log t_{\rm col,2}+(0.11 \pm 0.18)
\nonumber \\
\end{eqnarray}
with $r=0.71$ and $p=0.0009$. This may be understood as the follows. A
higher plateau luminosity corresponds to a shorter spin-down
timescale. It is possible that in this case, the collapse time
is closer to the spin-down timescale, and so most energy is
already released before the magnetar collapses to form a black
hole. A lower plateau luminosity corresponds to a longer
spin-down timescale, and it is possible that the collapse time
can be much shorter than the spin-down timescale, so that only
a fraction of the total energy is released before the collapse.

Empirically, (Dainotti et al. 2008, 2010, 2013) discovered an
anti-correlation between $L_b$ and $t_b$ for long GRBs. In
Figure 10(c) we plot our short GRB Internal + External sample
and derive an empirical correlation of
\begin{equation}
\log L_{b,49}=(-1.41 \pm 0.14) \log t_{b,3}-(0.46 \pm 0.37),
\end{equation}
with $r=0.88$ and $p<0.001$. The slope of the correlation is
slightly steeper than that of the ``Dainotti relation'' (e.g.
Dainotti et al. 2008, data see gray dots in Fig.10(c)). This is
probably related to different progenitor systems for long and
short GRBs, in particular, the dominance of Internal plateaus
in our sample. Rowlinson et al. (2014) performed a joint
analysis of both long and short GRBs taking into account for
the intrinsic slope of the luminosity-time correlation
(Dainotti et al. 2013). We focus on short GRBs only but studied
the Internal and External sub-samples separately.

\subsection{Constraining the neutron star EOS}

The inferred collapsing time can be used to constrain the
neutron star EOS (Lasky et al. 2014; Ravi \& Lasky 2014). The
basic formalism is as follows.

The standard dipole spin-down formula gives (Shapiro \& Teukolsky 1983)
\begin{eqnarray}
P(t) &=& P_{0} (1+\frac{4\pi ^{2}}{3c^{3}}\frac{B_{p}^{2}R^{6}}{I P_{0}^{2}}t)^{1/2}\nonumber \\
&=&P_{0} (1+\frac{t}{\tau})^{1/2}.
\label{Pt}
\end{eqnarray}
For a given EOS, a maximum NS mass for a non-rotating NS, i.e.
$M_{\rm TOV}$, can be derived. When an NS is supra-massive but
rapidly rotating, a higher mass can be sustained. The maximum
gravitational mass ($M_{\rm max}$) depends on spin period,
which can be approximated as (Lyford et al. 2003)
\begin{eqnarray}
M_{\rm max} = M_{\rm TOV}(1+\hat{\alpha} P^{\hat{\beta}})
\label{Mt1}
\end{eqnarray}
where $\hat{\alpha}$ and $\hat{\beta}$ depend on the EOS. The
numerical values of $\hat{\alpha}$ and $\hat{\beta}$ for
various EOSs have been worked out by Lasky et al. (2014), and
are presented in Table 4 along with $M_{\rm TOV}$, $R$, and
$I$.

As the neutron star spins down, the maximum mass $M_{\rm max}$
gradually decreases. When $M_{\rm max}$ becomes equal to the
total gravitational mass of the protomagnetar, $M_p$, the
centrifugal force can no longer sustain the star, and so the NS
will collapse into a black hole. Using equation
Equation(\ref{Pt}) and Equation (\ref{Mt1}), one can derive the
collapse time:
\begin{eqnarray}
t_{\rm col} &=& \frac{3c^{3}I}{4\pi^{2}B_{\rm p}^{2}R^{6}}[(\frac{M_{\rm p}-M_{\rm TOV}}{\hat{\alpha} M_{\rm
TOV}})^{2/\hat{\beta}}-P_{0}^{2}]\nonumber \\
&=&\frac{\tau}{P_{\rm 0}^{2}}[(\frac{M_{\rm p}-M_{\rm TOV}}{\hat{\alpha} M_{\rm
TOV}})^{2/\hat{\beta}}-P_{0}^{2}].
\label{tcol}
\end{eqnarray}
As noted, one can infer $B_p$, $P_0$, and $t_{\rm col}$ from
the observations.  Moreover, as the Galactic binary NS
population has a tight mass distribution (e.g., Valentim et al.
2011; Kiziltan et al. 2013), one can infer the expected
distribution of protomagnetar masses, which is found to be
$M_p=2.46^{0.13}_{-0.15}M_\odot$ (for details see Lasky et al.
2014). The only remaining variables in Equation (16) are
related to the EOS, implying that the observations can be used
to derive constraints on the EOS of nuclear matter. For most
GRBs in our Internal sample, only the lower limit of $\tau$ is
derived from $t_b$ (Equation (\ref{tau1})). One can also infer
the maximum $\tau$ by limiting $P_0$ to the break-up limit.
Considering the uncertainties related to gravitational wave
radiation, we adopt a rough limit of 1 millisecond. By doing
so, one can then derive a range of $\tau$, and hence a range of
$M_p$ based on the data and a given EOS.

Figure 11 presents the collapse time ($t_{\rm col}$) as a
function of protomagnetar mass ($M_p$) for each short GRB in
the Internal sample with redshift measurements. Five NS
equations of state, i.e. SLy (black, Douchin \& Haensel. 2001),
APR (red, Akmal et al. 1998), GM1 (green, Glendenning \&
Moszkowski. 1991), AB-N, and AB-L (blue and cyan, Arnett \&
Bowers. 1997) are shown in different vertical color bands. The
gray shaded region is the protomagnetar mass distribution,
$M_p$, discussed above. The horizontal dashed line is the
observed collapse time for each short GRB. Our results show
that the GM1 model gives an $M_p$ band fall in the 2$\sigma$
region of the protomagnetar mass distribution, so that the
correct EOS should be close to this model. The maximum mass for
non-rotating NS in this model is $M_{\rm TOV} = 2.37 M_\odot$.

Lasky et al. (2014) applied the observational collapse time of
short GRBs to constrain the NS EOS (see also a rough treatment
by Fan et al. 2013a). Our results are consistent with those of
Lasky et al. (2014), but using a larger sample. Another
improvement is that we introduce a range of $\tau$ rather than
one single $\tau$ to derive the range of plausible $M_p$, since
the observed collapse time only gives the lower limit of
$\tau$. This gives a range of the allowed $M_p$ (rather than a
fine-tuned value for the single $\tau$ scenario) for each GRB
for a given observed $t_b$.

\section{Conclusions and Discussion}
In this paper, by systematically analyzing the BAT-XRT light
curves of short GRBs detected by {\em Swift} before 2014
August, we systematically examine the millisecond magnetar
central engine model of short GRBs. About 40 GRBs have bright
X-ray afterglows detected with {\em Swift}/XRT, 8 0f which have
EE detected with {\em Swift}/BAT. Based on the existence of
plateaus, their observation properties, and how likely a GRB is
powered by a millisecond magnetar central engine, we
characterized short GRBs into three samples: Internal
(plateau), External (plateau), and Non (plateau). We compared
the statistical properties of our samples and derived or placed
limits on the magnetar parameters $P_0$ and $B_p$ from the
data. Using the collapse time $t_{\rm col}$ of the
protomagnetar inferred from the plateau break time $t_b$ in the
Internal sample, we went on to constrain the NS EOS. The
following interesting results are obtained.
\begin{itemize}
 \item At least for the Internal sample, the data seem to
     be consistent with the expectations of the magnetar
     central engine model. Assuming isotropic emission, the
     derived magnetar parameters $B_p$ and $P_0$ fall into
     a reasonable range. The total energy (sum of $E_{\rm
     \gamma}$, $E_{\rm X}$, and $E_{\rm K}$) is within the
     budget provided by the spin energy of the millisecond
     magnetar ($E_{\rm rot}\sim 3.5\times 10^{52} ~{\rm
     erg}$). The $L_0 - t_{col}$ anti-correlation is
     generally consistent with the hypothesis that the
     total spin energy of the magnetar may be standard, and
     a higher dipolar magnetic field powers a brighter but
     shorter plateau.
 \item The so-called EE following some short GRBs is
     essentially the brightest internal plateau commonly
     observed in short GRBs. A more sensitive and softer
     detector would detect more EE from short GRBs.
 \item The External sample may also be consistent with
     having a magnetar central engine, even though the
     evidence is not as strong. If both the Internal and
     External samples are powered by a millisecond magnetar
     central engine, then the difference between the two
     samples may be related to the circumburst medium
     density. The physical and host-normalized offsets of
     the afterglow locations for the Internal sample is
     somewhat larger than those of the External sample,
     even though the separation between the two samples is
     not clear cut. In any case, it is consistent with this
     expectation. The total energy budget of the GRB is
     within the magnetar energy budget for the External
     sample only if the ambient density is relatively
     large, and hence powers a strong external shock
     emission component. There is no significant difference
     between those two groups for the star formation rate,
     metallicity, and age of the host galaxy.
 \item Using the collapse time of supra-massive
     protomagnetar to form a black hole and the
     distribution of the total mass of NS$-$NS binaries in
     the Galaxy, one can constrain the NS EOS. The data
     point toward an EOS model close to GM1, which has a
     non-spinning maximum NS mass $M_{\rm TOV} \sim 2.37
     M_\odot$.
\end{itemize}

The short GRB data are consistent with the hypothesis that the
post-merger product of an NS$-$NS merger is a supra-massive
neutron star. The existence of such a long-lived post-merger
product opens some interesting prospects in the multi-messenger
era. In particular, the dipole spin-down power of the
supra-massive NS can power bright electromagnetic radiation
even if the short GRB jet does not beam toward earth, and so
some interesting observational signatures are expected to be
associated with gravitational wave signals in the Advanced
LIGO/Virgo era (Fan et al. 2013b; Gao et al. 2013; Yu et al.
2013; Zhang 2013; Metzger \& Piro 2014). Another interesting
possibility is that a fast radio burst (e.g. Lorimer et al.
2007; Thornton et al. 2013) may be released when the
supra-massive magnetar collapses into a black hole (Falcke \&
Rezzolla 2014; Zhang 2014). The discovery of an FRB following a
GRB at the end of the internal plateau (see Bannister et al.
2012) would nail down the origin of FRBs, although such
observations require fast telescope response times given the
expected distribution of collapse times following SGRBs (see
figure 9(b) and Ravi \& Lasky 2014). The GRB-FRB associations,
if proven true, would be invaluable for cosmology studies (Deng
\& Zhang 2014; Gao et al. 2014; Zheng et al. 2014 Zhou et al.
2014).

Recently, Rezzolla \& Kumar (2014) and Ciolfi \& Siegel (2014)
proposed a different model to interpret the short GRB
phenomenology. In their model, the post-merger product is also
a supra-massive NS, but the collapse time is allocated as the
epoch of the short GRB itself, rather than the end of the
Internal plateau. Our conclusions drawn in this paper do not
apply to that model. A crucial observational test to
differentiate between our model and theirs is whether or not
there exists strong X-ray emission before the short GRB itself.
This may be tested in the future with a sensitive wide-field
XRT.

\acknowledgments We thank an anonymous referee for helpful
suggestions, Hui Sun, and Luciano Rezzolla for useful comments.
We acknowledge the use of the public data from the Swift data
archive and the UK Swift Science Data Center. This work is
supported by the NASA ADAP program under grant NNX14AF85G [BZ],
the National Natural Science Foundation of China under grants
U1431124, 11361140349(China-Israel jointed program) [WHL], and
an Australian Research Council Discovery Project DP140102578
[PDL].

%*******************************************************************************************

\begin{center}
\begin{deluxetable}{lllllllllllll}
%\rotate
\tablewidth{500pt} \tabletypesize{\footnotesize}
\tabletypesize{\tiny} \tablecaption{Observed properties of
short GRBs in our samples.} \tablenum{1}
\tablehead{\colhead{GRB}& \colhead{$z$\tablenotemark{a}}&
\colhead{$T_{90}$/EE\tablenotemark{b}}&\colhead{$\Gamma_{\gamma}$\tablenotemark{c}}&
\colhead{$\beta_{X}$\tablenotemark{d}}& \colhead{Host
Offset\tablenotemark{e}} & \colhead{Host
Offset\tablenotemark{e}}& \colhead{$t_{b}$\tablenotemark{f}}&
\colhead{$\alpha_1$\tablenotemark{f}}&
\colhead{$\alpha_2$\tablenotemark{f}}& \colhead{$\chi^2$/dof}&
\colhead{Ref}
\\\colhead{name}&\colhead{}&
\colhead{(s)}&\colhead{}&
\colhead{}&\colhead{(kpc)}&\colhead{$(r_{e})$}&
\colhead{(s)}&\colhead{}&\colhead{}&\colhead{}&\colhead{}}

\startdata
Internal\\
\hline
050724	&	0.2576	&	3/154	&	1.89$\pm$0.22	&	0.58$\pm$0.19	&	2.76$\pm$0.024	&	---      	  &	139$\pm$9	&

0.20$\pm$0.1	&	4.16$\pm$0.05   & 980/835	&(1, 2)\\
051210	&	(0.58)	&	1.27/40	&	1.06$\pm$0.28	&	1.1$\pm$0.18	&	24.9$\pm$24.6	&	4.65$\pm$4.6  &	67$\pm$4	&

0.15$\pm$0.04	&	2.96$\pm$0.09   &118/132	&(1, 2)\\
051227	&	(0.58)	&	3.5/110	&	1.45$\pm$0.24	&	1.1$\pm$0.4  	&	---	            &	---      	  &	89$\pm$5	&

0.10$\pm$0.05	&	3.19$\pm$0.13	&681/522    &(3, 4, 5)\\
060801	&	1.13	&	0.49/N	&	1.27$\pm$0.16	&	0.43$\pm$0.12	&	---	            &	---           &	212$\pm$11	&

0.10$\pm$0.11	&	4.35$\pm$0.26   &81/75	    &(1)\\
061006	&	0.4377	&	0.5/120	&	1.72$\pm$0.17	&	0.76$\pm$0.28	&	1.3$\pm$0.24	&	0.35$\pm$0.07 &	99$\pm$7	&

0.17$\pm$0.03	&	9.45$\pm$1.14   &111/138	&(1, 2)\\
061201	&	0.111	&	0.76/N	&	0.81$\pm$0.15	&	1.2$\pm$0.22	&	32.47$\pm$0.06	& 14.91$\pm$0.03  &	2223$\pm$43	&

0.54$\pm$0.06	&	1.84$\pm$0.08	&20/24      &(1, 6)\\
070714B	&	0.9224	&	3/100	&	1.36$\pm$0.19	&	1.01$\pm$0.16	&	12.21$\pm$0.53	&	5.55$\pm$0.24 &	82$\pm$2	&

0.10$\pm$0.07	&	1.91$\pm$0.03   &672/581	&(1, 6)\\
070724A	&	0.46	&	0.4/N	&	1.81$\pm$0.33	&	0.5$\pm$0.3 	&	5.46$\pm$0.14	&	1.5$\pm$0.04  &	77$\pm$6	&

0.01$\pm$0.1	&	6.45$\pm$0.46   &256/222	&(1, 6)\\
071227	&	0.381	&	1.8/100	&	0.99$\pm$0.22	&	0.8$\pm$0.3	    &	15.5$\pm$0.24	&	3.28$\pm$0.05 &	69$\pm$8	&

0.27$\pm$0.08	&	2.92$\pm$0.06   &244/212	&(1, 6)\\
080702A	&	(0.58)	&	0.5/N	&	1.34$\pm$0.42	&	1.03$\pm$0.35	&	---	            &	---      	  &	586$\pm$14	&

0.51$\pm$0.22	&	3.56$\pm$0.31   &3/5	    &(7)\\
080905A	&	0.122	&	1/N 	&	0.85$\pm$0.24	&	0.45$\pm$0.14	&	17.96$\pm$0.19	&	10.36$\pm$0.1 &	13$\pm$3	&

0.19$\pm$0.09	&	2.37$\pm$0.07   &43/52   	&(6, 7)\\
080919	&	(0.58)	&	0.6/N	&	1.11$\pm$0.26	&	1.09$\pm$0.36	&	---	            &	---           &	340$\pm$26	&

0.40$\pm$0.14	&	5.20$\pm$0.55   &7/5    	&(7)\\
081024A	&	(0.58)	&	1.8/N	&	1.23$\pm$0.21	&	0.85$\pm$0.3	&	---      	    &	---      	  &	102$\pm$5	&

0.27$\pm$0.02	&	5.89$\pm$0.3    &50/42  	&(7)\\
090510	&	0.903	&	0.3/N	&	0.98$\pm$0.21	&	0.75$\pm$0.12	&	10.37$\pm$2.89	&	1.99$\pm$0.39 &	1494$\pm$87	&

0.69$\pm$0.04	&	2.33$\pm$0.11   &112/132	&(6, 7)\\
090515	&	(0.58)	&	0.036/N	&	1.61$\pm$0.22	&	0.75$\pm$0.12	&	75.03$\pm$0.15	&	15.53$\pm$0.03&	178$\pm$3	&

0.10$\pm$0.08	&	12.62$\pm$0.5   &42/38  	&(6, 7)\\
100117A	&	0.92	&	0.3/N	&	0.88$\pm$0.22	&	1.1$\pm$0.26	&	1.32$\pm$0.33	&	0.57$\pm$0.13 &	252$\pm$9	&

0.55$\pm$0.03	&	4.59$\pm$0.13   &84/92  	&(6, 7, 8)\\
100625A	&	0.425	&	0.33/N	&	0.91$\pm$0.11	&	1.3$\pm$0.3	    &	---      	    &	---      	  &	200$\pm$41	&

0.26$\pm$0.44	&	2.47$\pm$0.18   &3/6    	&(7, 9)\\
100702A	&	(0.58)	&	0.16/N	&	1.54$\pm$0.15	&	0.88$\pm$0.11	&	---      	    &	---      	  &	201$\pm$6	&

0.62$\pm$0.13	&	5.28$\pm$0.23   &82/69  	&(7)\\
101219A	&	0.718	&	0.6/N	&	0.63$\pm$0.09	&	0.53$\pm$0.26	&	---      	    &	---      	  &	197$\pm$10	&

0.13$\pm$0.19	&	20.52$\pm$8.01  &3/5    	&(7)\\
111121A	&	(0.58)	&	0.47/119&	1.66$\pm$0.12	&	0.75$\pm$0.2	&	---      	    &	---      	  &	56$\pm$9	&

0.10$\pm$0.13	&	2.26$\pm$0.04   &274/289	&(7)\\
120305A	&	(0.58)	&	0.1/N	&	1.05$\pm$0.09	&	1.4$\pm$0.3 	&	---      	    &	---      	  &	188$\pm$14	&

0.73$\pm$0.14	&	6.49$\pm$0.63   &14/18  	&(7)\\
120521A	&	(0.58)	&	0.45/N	&	0.98$\pm$0.22	&
0.73$\pm$0.19	&	---      	&	---      	&	270$\pm$55
& 0.30$\pm$0.27	&	10.74$\pm$4.76  &3/7	&(7)\\                    		
\hline 																					
External\\
\hline 	
051221A	&    0.55	&	1.4/N	&	1.39$\pm$0.06	&	1.07$\pm$0.13	&	1.92$\pm$0.18	&	0.88$\pm$0.08	& 25166$\pm$870

&  0.12$\pm$0.13	&	1.43$\pm$0.04   &52/63	&(1, 2, 7)\\
060313	&	(0.58)	&	0.71/N	&	0.71$\pm$0.07	&	1.06$\pm$0.15	&	2.28$\pm$0.5	&	1.23$\pm$0.23	&	2294$\pm$65

&  0.3$\pm$0.15 	&	1.52$\pm$0.04   &54/45	&(1, 2, 7)\\
060614	&	0.1254	&	5/106	&	2.02$\pm$0.04	&	1.18$\pm$0.09	&	---             &	---             &	
49840$\pm$3620	&  0.18$\pm$0.06	&	1.9$\pm$0.07    &70/54	&(1,10)\\
070714A	&	(0.58)	&	2/N 	&	2.6$\pm$0.2 	&	1.1$\pm$0.3	    &	---             &	---             &	892$\pm$34

&  0.11$\pm$0.09	&	0.95$\pm$0.06   &15/18	&(7)\\
070809	&	0.219	&	1.3/N	&	1.69$\pm$0.22	&	0.37$\pm$0.21	&	33.22$\pm$2.71	&	9.25$\pm$0.75	&	8272$\pm$221

&  0.18$\pm$0.06	&	1.31$\pm$0.17   &17/22	&(6, 7)\\
080426	&	(0.58)	&	1.7/N	&	1.98$\pm$0.13	&	0.92$\pm$0.24	&	---             &	---             &	566$\pm$97

&  0.11$\pm$0.16	&	1.29$\pm$0.05   &28/21	&(7)\\
090426	&	2.6 	&	1.2/N	&	1.93$\pm$0.22	&	1.04$\pm$0.15	&	0.45$\pm$0.25	&	0.29$\pm$0.14	&	208$\pm$53

&  0.12$\pm$0.07	&	1.04$\pm$0.04   &15/11	&(6, 7)\\
100724A	&	1.288	&	1.4/N	&	1.92$\pm$0.21	&	0.94$\pm$0.23	&	---             &	---             &	5377$\pm$331

&  0.72$\pm$0.08	&	1.61$\pm$0.12   &16/19	&(11, 12)\\
130603B	&	0.356	&	0.18/N	&	1.83$\pm$0.12	&	1.18$\pm$0.18	&	5.21$\pm$0.17	&	1.05$\pm$0.04	&	3108$\pm$356

&  0.4$\pm$0.02 	&	1.69$\pm$0.04   &126/109&(6, 13, 14)\\
130912A	&	(0.58)	&	0.28/N	&	1.21$\pm$0.2	&	0.56$\pm$0.11	&	---             &	---             &	231$\pm$54

&  0.04$\pm$0.39	&	1.34$\pm$0.04   &28/21	&(15)\\
\enddata

\tablenotetext{a}{The measured redshift are from the published
papers and GNCs. When the redshift is not known, 0.58 is used.}
\tablenotetext{b}{The duration of the GRB without and with
extended emission (if EE exists). ``N'' denotes no EE.}
\tablenotetext{c}{The photon index in the BAT band (15-150keV)
fitted using a power law.} \tablenotetext{d}{The spectral index
of the absorbed power-law model for the normal segments.}
\tablenotetext{e}{Physical and host-normalized offsets for the
short GRBs with \emph {Hubble Space Telescope (HST)}
observations.}\tablenotetext{f}{The break time of the light
curves from our fitting, $\alpha_1$ and $\alpha_2$ are the
decay slopes before and after the break time.}

\tablerefs{(1)Zhang et al.(2009),(2)Fong et
al.(2010),(3)Hullinger et al.(2005),(4)Butler et
al.(2007),(5)Gompertz et al.(2014),(6)Fong \&
Berger(2013),(7)Rowlinson et al.(2013),(8)Fong et
al.(2011),(9)Fong et al.(2013),(10)L\"{u} \&
Zhang.(2014),(11)Thoene et al.(2010),(12)Markwardt et
al.(2010),(13)Cucchiara et al.(2013),(14)Barthelmy et
al.(2013),(15)Krimm et al.(2013).}
\end{deluxetable}
\end{center}

%*******************************************************************************************
\begin{center}
\begin{deluxetable}{lllllllllllll}
%\rotate
\tablewidth{0pt} \tabletypesize{\footnotesize}
\tabletypesize{\tiny} \tablecaption{The derived properties of the short
GRBs in our samples.}\tablenum{2}

\tablehead{ \colhead{GRB}&
\colhead{$E_{\rm\gamma,iso,51}$\tablenotemark{a}}&
\colhead{$L_{b,49}$\tablenotemark{b}}&
\colhead{$\tau_{3}$\tablenotemark{b}}&
\colhead{$B_{p,15}$\tablenotemark{c}}&
\colhead{$P_{0,-3}$\tablenotemark{c}}&
\colhead{$L_{47}(10^{3}s)$\tablenotemark{d}}&
\colhead{$E_{K,iso,51}$\tablenotemark{e}}&
\colhead{$E_{K,iso,51}$\tablenotemark{e}}&
\colhead{$E_{X,iso,51}$\tablenotemark{f}}\\\colhead{Name}&\colhead{($\rm
erg$)}& \colhead{($\rm erg~s^{-1}$)}&\colhead{(s)}&
\colhead{(G)}&\colhead{($s^{-1}$)}&\colhead{($\rm
erg~s^{-1}$)}& \colhead{($n=1,\rm
erg$)}&\colhead{($n=10^{-3},\rm erg$)}& \colhead{($\rm erg$)}}

\startdata
Internal\\
\hline 050724	&       $0.09^{+0.11}_{-0.02}$	&	
1.1$\pm$0.16	&	0.11$\uparrow$	&	17.15$\downarrow$ 	&	
4.04$\downarrow$ 	&	
0.05$\pm$0.006 	&	0.97$\pm$0.13 	&2.37$\pm$0.26&	1.25$\pm$0.14 	\\
051210	&	$0.22^{+0.036}_{-0.036}$&	2.23$\pm$0.26	&	
0.04$\uparrow$	&	32.69$\downarrow$ 	&	4.68$\downarrow$ 	
&	0.32$\downarrow$ 	&	0.34$\downarrow$ 	&1.89$\downarrow$&	0.94$\pm$0.10 	\\
051227	&	$1.20^{+1.6}	_{-0.5}$&	1.89$\pm$0.02	&	
0.05$\uparrow$	&	28.68$\downarrow$ 	&	4.57$\downarrow$ 	
&	0.66$\pm$0.086 	&	2.69$\pm$0.35 	&5.65$\pm$0.26&	0.98$\pm$0.11 	\\
060801	&	$1.70^{+0.2}	_{-0.2}$&	0.73$\pm$0.07	&	
0.14$\uparrow$	&	17.81$\downarrow$ 	&	4.57$\downarrow$ 	
&	0.46$\downarrow$ 	&	3.84$\downarrow$ 	&2.03$\downarrow$&	0.98$\pm$0.16 	\\
061006	&	$2.20^{+1.2}	_{-1.2}$&	3.37$\pm$0.32	&	
0.04$\uparrow$	&	18.31$\downarrow$ 	&	3.16$\downarrow$ 	
&	0.17$\pm$0.022 	&	6.37$\pm$0.83 	&6.37$\pm$0.83&	2.06$\pm$0.23 	\\
061201	&	$0.18^{+0.02}_{-0.01}$	&	(1$\pm$0.11)E-3	&
2$\pm$0.043	&	31.17$\pm$2.36 	&	30.80$\pm$1.97 	&	
0.15$\pm$0.019 	&	0.74$\pm$0.10 	&1.84$\pm$0.21&	0.02$\pm$0.01 	\\
070714B	&	$11.60^{+4.1}	_{-2.2}$&	6.22$\pm$0.09	&	
0.04$\pm$0.002	&	19.12$\pm$1.08 	&	2.77$\pm$0.09 	&	
1.40$\pm$0.182 	&	4.40$\pm$0.57 	&9.47$\pm$0.41&	2.67$\pm$0.29 	\\
070724A	&	$0.03^{+0.01}_{-0.01}$	&	13.1$\pm$7.2	&	
0.05$\uparrow$	&	10.89$\downarrow$ 	&	1.73$\downarrow$ 	
&	0.05$\pm$0.007 	&	7.99$\pm$1.04 	&7.99$\pm$1.04&	6.81$\pm$4.56 	\\
071227	&	$2.20^{+0.8}	_{-0.8}$&	0.77$\pm$0.01	&	
0.05$\uparrow$	&	44.93$\downarrow$ 	&	7.16$\downarrow$ 	
&	0.05$\pm$0.007 	&	0.91$\pm$0.12 	&2.07$\pm$0.23&	0.40$\pm$0.04 	\\
080702A	&	$0.13^{+0.208}_{-0.0556}$&	(7$\pm$0.25)E-3	&
0.37$\uparrow$	&	64.41$\downarrow$ 	&	27.40$\downarrow$ 	
&	0.02$\pm$0.002 	&	0.26$\downarrow$ 	&0.84$\downarrow$&	0.03$\pm$0.01 	\\
080905A	&	$7^{+11}_{-4}$    	&	2.76$\pm$0.9	&	
0.01$\uparrow$	&	102.83$\downarrow$ &	7.87$\downarrow$ 	
&	0.01$\downarrow$ 	&	0.37$\downarrow$ 	&0.72$\downarrow$&	0.33$\pm$0.18 	\\
080919	&	$0.42^{+0.41}_{-0.278}$	&	0.05$\pm$0.01	&	
0.22$\uparrow$	&	41.98$\downarrow$ 	&	13.63$\downarrow$ 	
&	0.11$\pm$0.014 	&	0.20$\downarrow$ 	&1.01$\downarrow$&	0.11$\pm$0.04 	\\
081024A	&	$0.56^{+0.69}_{-0.278}$	&	0.78$\pm$0.14	&	
0.06$\uparrow$	&	36.34$\downarrow$ 	&	6.42$\downarrow$ 	
&	0.41$\downarrow$ 	&	0.50$\downarrow$ 	&0.95$\downarrow$&	0.50$\pm$0.12 	\\
090510	&	$3^{+5}_{-2}$      	&	0.18$\pm$0.03	&	
0.75$\uparrow$	&	6.36$\downarrow$ 	&	3.85$\downarrow$ 	
&	7.90$\pm$1.027 	&	3.71$\pm$0.48 	&7.79$\pm$0.85&	1.38$\pm$0.37 	\\
090515	&	$0.08^{+0.16}_{-0.042}$	&	1.24$\pm$0.05	&	
0.11$\uparrow$	&	16.29$\downarrow$ 	&	3.82$\downarrow$ 	
&	0.28$\downarrow$ 	&	0.83$\downarrow$ 	&0.93$\downarrow$&	1.40$\pm$0.10 	\\
100117A	&	$2.50^{+0.3}_{-0.3}$ 	&	0.45$\pm$0.04	&	
0.16$\uparrow$	&	19.06$\downarrow$ 	&	5.32$\downarrow$ 	
&	0.02$\downarrow$ 	&	0.12$\downarrow$ 	&0.92$\downarrow$&	0.72$\pm$0.09 	\\
100625A	&	$0.64^{+0.031}_{-0.031}$&	0.042$\pm$0.03	&	
0.13$\uparrow$	&	79.67$\downarrow$ &	19.75$\downarrow$ &	
0.02$\pm$0.003 	&	0.07$\downarrow$ 	&0.11$\downarrow$&	0.05$\pm$0.05 	\\
100702A	&	$0.47^{+0.045}_{-0.045}$&	0.97$\pm$0.14	&	
0.13$\uparrow$	&	16.36$\downarrow$ 	&	4.07$\downarrow$ 	
&	1.20$\downarrow$ 	&	1.82$\downarrow$ 	&4.04$\downarrow$&	1.24$\pm$0.22 	\\
101219A	&	$4.80^{+0.3}_{-0.3}$ 	&	0.56$\pm$0.05	&	
0.12$\uparrow$	&	23.84$\downarrow$ 	&	5.65$\downarrow$ 	
&	0.23$\downarrow$ 	&	4.14$\pm$0.54 	&10.03$\pm$1.11&	0.64$\pm$0.10 	\\
111121A	&	$2.80^{+0.25}_{-0.25}$	&	14$\pm$0.8	&	
0.04$\uparrow$	&	15.22$\downarrow$ 	&	2.02$\downarrow$ 	
&	1.57$\pm$0.204 	&	9.80$\pm$1.27 	&22.64$\pm$1.49&	5.04$\pm$0.55 	\\
120305A	&	$0.29^{+0.0112}_{-0.0112}$&	0.48$\pm$0.09	&	
0.13$\uparrow$	&	23.30$\downarrow$ 	&	5.80$\downarrow$ 	
&	0.11$\pm$0.014 	&	0.45$\pm$0.06 	&0.89$\pm$0.10&	0.61$\pm$0.17 	\\
120521A	&	$0.23^{+0.0115}_{-0.0356}$&	0.07$\pm$0.003	&	
0.17$\uparrow$	&	44.68$\downarrow$ &	12.90$\downarrow$ 	&	
1.01$\downarrow$ 	&	2.46$\downarrow$ 	&4.42$\downarrow$&	0.12$\pm$0.03 	\\
\hline
External\\
\hline 	

051221A	&	$2.80^{+2.1}_{-1.1}$	&(1.78$\pm$0.09)E-5	&	
--- 	&	--- 	&	--- 	&	0.63$\pm$0.08 	&	
16.29$\pm$2.12 	&35.56$\pm$3.91&	0.31$\pm$0.032 	\\	060313	
&	$12.90^{+0.889}_{-7.56}$&	(2.74$\pm$0.21)E-2	&	
--- 	&	--- 	&	--- 	&	3.00$\pm$0.39 	&	
8.11$\pm$1.05 	&17.21$\pm$1.89&	0.45$\pm$0.054 	\\	060614	
&	$2.40^{+0.4}_{-0.4}$	&	(2.55$\pm$0.12)E-4	&	
--- 	&	--- 	&	--- 	&	0.04$\pm$0.01 	&	
7.06$\pm$0.92 	&14.56$\pm$1.61&	0.11$\pm$0.012 	\\	070714A	
&	$0.42^{+1.25}_{-0.069}$	&	(1.3$\pm$0.15)E-2	&	
--- 	&	--- 	&	--- 	&	0.70$\pm$0.09 	&	
13.94$\pm$1.81 	&13.94$\pm$1.81&	0.07$\pm$0.013 	\\	070809	
&	$0.01^{+0.01}_{-0.01}$	&	(3.2$\pm$0.31)E-5	&	
--- 	&	--- 	&	--- 	&	0.05$\pm$0.01 	&	
2.25$\pm$0.29 	&5.61$\pm$0.62&	0.02$\pm$0.003 	\\	080426	&	
$0.82^{+1.25}_{-0.0556}$&	(3.53$\pm$1.01)E-2	&	
--- 	&	--- &	--- 	&	0.83$\pm$0.11 	&	
4.71$\pm$0.61 	&10.35$\pm$1.14&	0.13$\pm$0.064 	\\	090426	
&	$4.20^{+5}_{-0.4}$	&	2.46$\pm$0.48   	&	
--- 	&	--- &	--- 	&	12.50$\pm$1.63 	&	
52.84$\pm$6.87 	&128.09$\pm$14.09&	1.43$\pm$0.690 	\\	100724A	
&	$0.7^{+0.1}_{-0.1}$	&	(2.85$\pm$0.32)E-2	&	
--- 	&	--- 	&	--- 	&	5.20$\pm$0.68 	&	
13.83$\pm$1.80 	&30.42$\pm$3.34&	0.67$\pm$0.180 	
\\	130603B	&	$2.20^{+0.2}_{-0.2}$	&	(1.2$\pm$0.05)E-2	
&	--- 	&	--- 	&	--- 	&	1.60$\pm$0.21 	&	
6.12$\pm$0.80 	&13.46$\pm$1.26&	0.27$\pm$0.055 	\\	130912A	
&	$0.73^{+0.08}_{-0.08}$	&	0.21$\pm$0.09   	&	
--- 	&	--- &	--- 	&	1.50$\pm$0.20 	&	
20.15$\pm$2.62 	&49.53$\pm$5.45&	0.30$\pm$0.237 	\\	
																		
\enddata

\tablenotetext{a}{$E_{\rm \gamma, iso}$ is calculated using
fluence and redshift extrapolated into 1-10,000keV (rest frame)
with a spectral model and a $k$-correction, in units of
$10^{51}$ erg.} \tablenotetext{b}{Isotropic luminosity at the
break time (in units of $10^{49}~{\rm erg~s^{-1}}$) and the
spin-down time (in units of $10^3$s).} \tablenotetext{c}{The
dipolar magnetic field strength at the polar cap in units of
$10^{15} G$, and the initial spin period of the magnetar in
units of milliseconds, with an assumption of an isotropic
wind.} \tablenotetext{d}{The luminosity of the afterglow at
$t=1000$ s. The arrow sign indicates the upper limit.}
\tablenotetext{e}{The isotropic kinetic energy measured from
the afterglow flux during the normal decay phase, in units of
$10^{51}$ erg.}\tablenotetext{f}{The isotropic internal
dissipation energy in the X-ray band (also internal plateau),
in units of $10^{51}$ erg.}

\end{deluxetable}
\end{center}

%*******************************************************************************************
\begin{center}
\begin{deluxetable}{lllllllllllll}
%\rotate
\tablewidth{0pt} \tabletypesize{\footnotesize}
%\tabletypesize{\tiny}
\tablecaption{The center values and standard deviations of the
Gaussian fits of various distributions.}\tablenum{3}

\tablehead{ \colhead{Name}& \colhead{Internal}&
\colhead{External} }

\startdata
${\rm log}(L_b)~{\rm erg~s^{-1}}$                              &($49.06\pm0.15$)$~{\rm erg~s^{-1}}$   &($47.55\pm0.16$) $~{\rm
erg~s^{-1}}$ \\
$log (t_{\rm b})~{\rm s}$                                      &($2.01\pm0.06$) $~{\rm s}$            &($3.41\pm0.04$) $~{\rm
s}$ \\
$log (E_{\rm \gamma,iso})~{\rm erg}$                           &($50.78\pm0.16$)$~{\rm erg}$          &($51.25\pm0.08$) $~{\rm
erg}$ \\
$log (E_{\rm X,iso}, n=1 ~{\rm cm^{-3}})~{\rm erg}$            &($50.86\pm0.11$)$~{\rm erg}$          &($51.35\pm0.04$) $~{\rm
erg}$ \\
$log (E_{\rm X,iso}, n=10^{-3} ~{\rm cm^{-3}})~{\rm erg}$      &($51.74\pm0.18$)$~{\rm erg}$          &($52.32\pm0.06$)$~{\rm
erg}$ \\
$log (E_{\rm total,iso}, n=1 ~{\rm cm^{-3}})~{\rm erg}$        &($51.36\pm0.06$)$~{\rm erg}$          &($51.82\pm0.04$)$~{\rm
erg}$ \\
$log (E_{\rm total,iso}, n=10^{-3} ~{\rm cm^{-3}})~{\rm erg}$  &($51.61\pm0.07$)$~{\rm erg}$          &($52.39\pm0.03$)$~{\rm
erg}$ \\
$log (t_{\rm col})~{\rm s}$                                    &($1.96\pm0.02$) $~{\rm s}$            &--- \\
$log (L_{\rm t=10^{3}s})~{\rm erg~s^{-1}}$                     &($46.09\pm0.07$)$~{\rm erg~s^{-1}}$   &($47.08\pm0.09$)$~{\rm
erg~s^{-1}}$ \\
\enddata

\end{deluxetable}
\end{center}

%*******************************************************************************************

\begin{center}
\begin{deluxetable}{lllllllllllll}
%\rotate
\tablewidth{0pt} \tabletypesize{\footnotesize}
%\tabletypesize{\tiny}
\tablecaption{The parameters of various NS EOS models}
\tablenum{4}

\tablehead{ & \colhead{SLy}& \colhead{APR} & \colhead{GM1} &
\colhead{AB-N} & \colhead{AB-L}}

\startdata
$M_{TOV}(M_{\odot}$)                             &2.05    &2.20    &2.37   &2.67    &2.71  \\
R (km)                                            &9.99    &10.0    &12.05  &12.9    &13.7  \\
$I(10^{45}~{\rm g~cm^{2}})$                      &1.91    &2.13    &3.33   &4.30    &4.70  \\
$\hat{\alpha}(10^{-10}~{s^{-\hat{\beta}}})$      &1.60    &0.303   &1.58   &0.112   &2.92  \\
$\hat{\beta}$                                    &-2.75   &-2.95   &-2.84  &-3.22   &-2.82  \\
\enddata

\tablerefs{The neutron star EOS parameters are derived in Lasky
et al. (2014) and Ravi \& Lasky (2014).}
\end{deluxetable}
\end{center}

%*******************************************************************************************

\begin{figure}
\includegraphics[angle=0,scale=0.45]{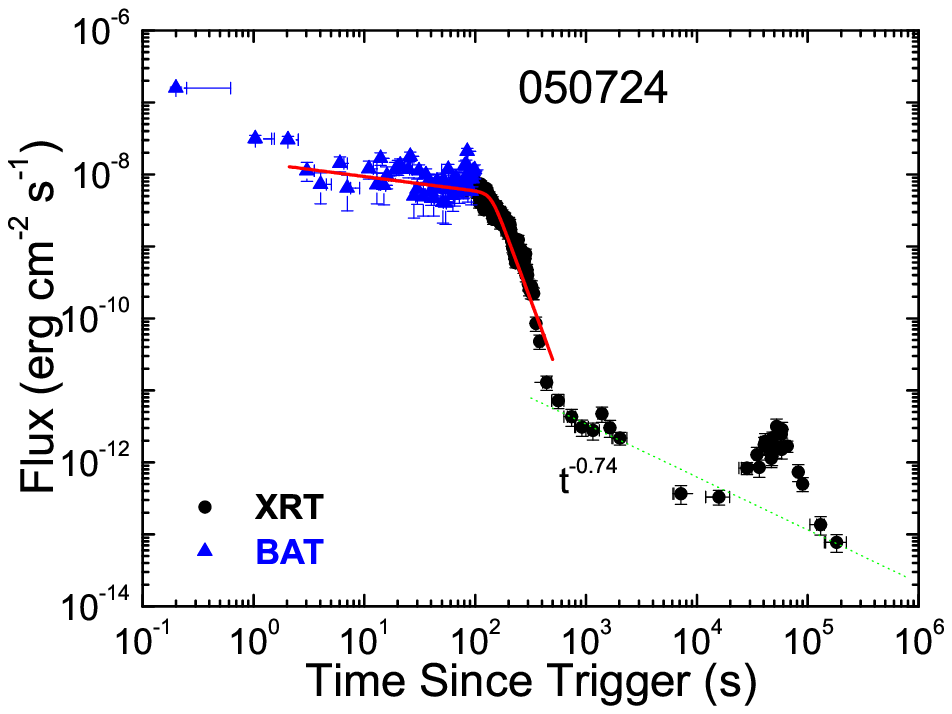}
\includegraphics[angle=0,scale=0.45]{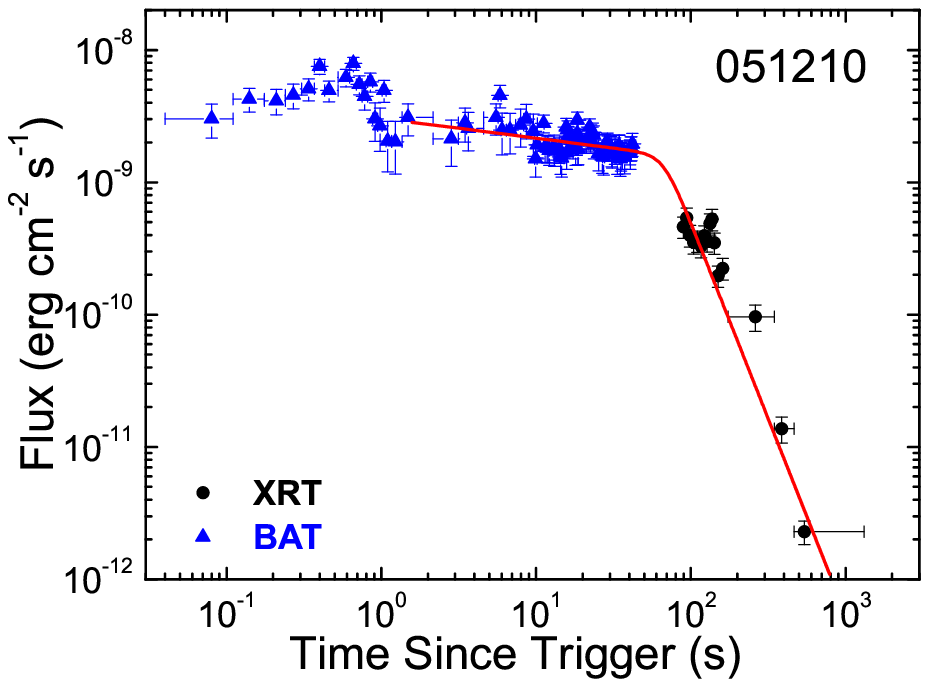}
\includegraphics[angle=0,scale=0.45]{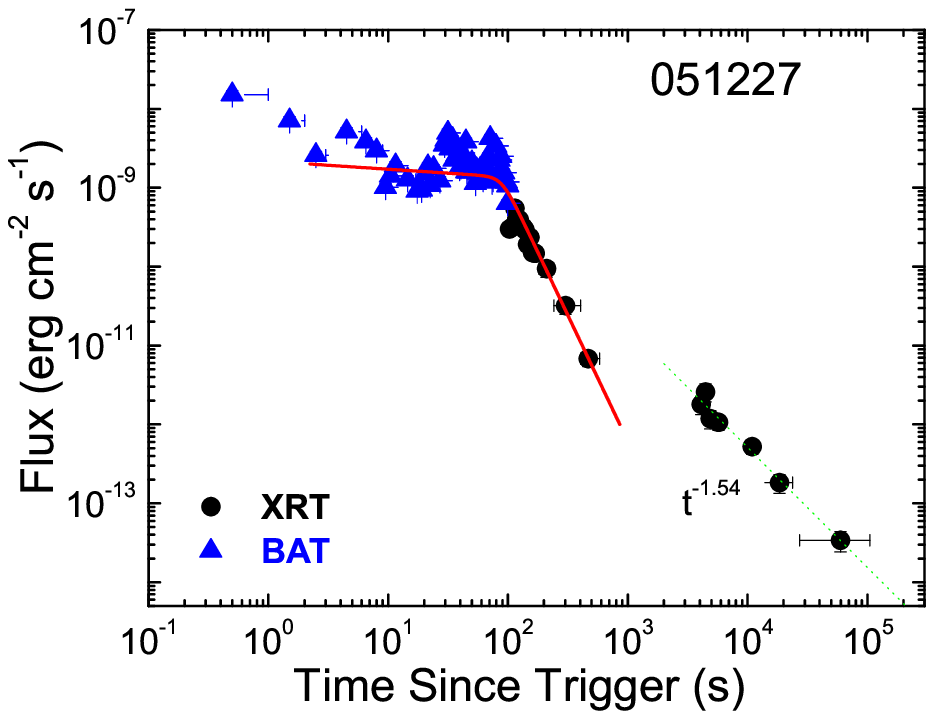}
\includegraphics[angle=0,scale=0.45]{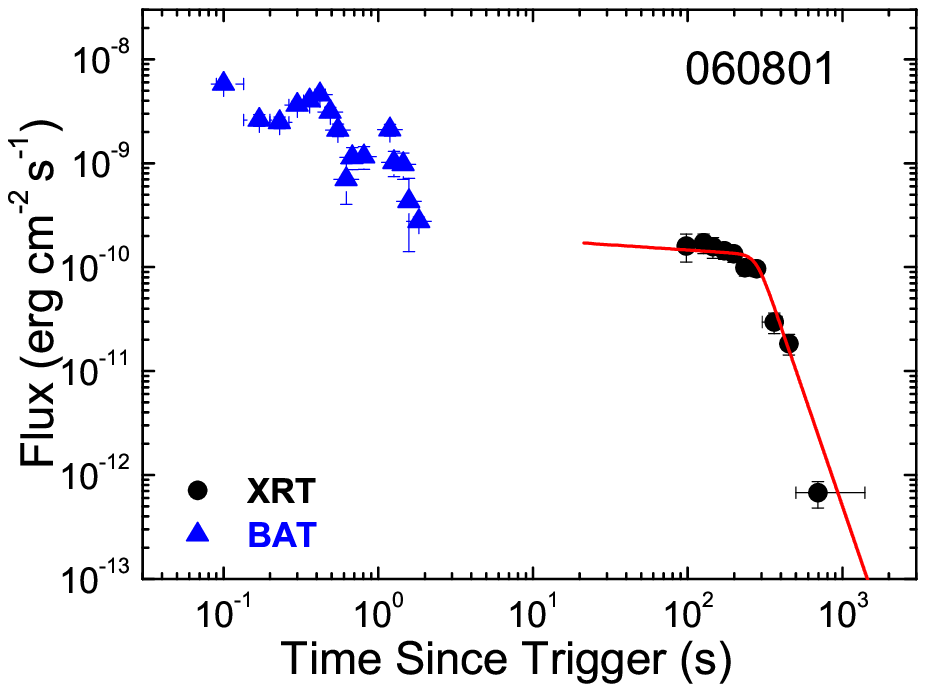}
\includegraphics[angle=0,scale=0.45]{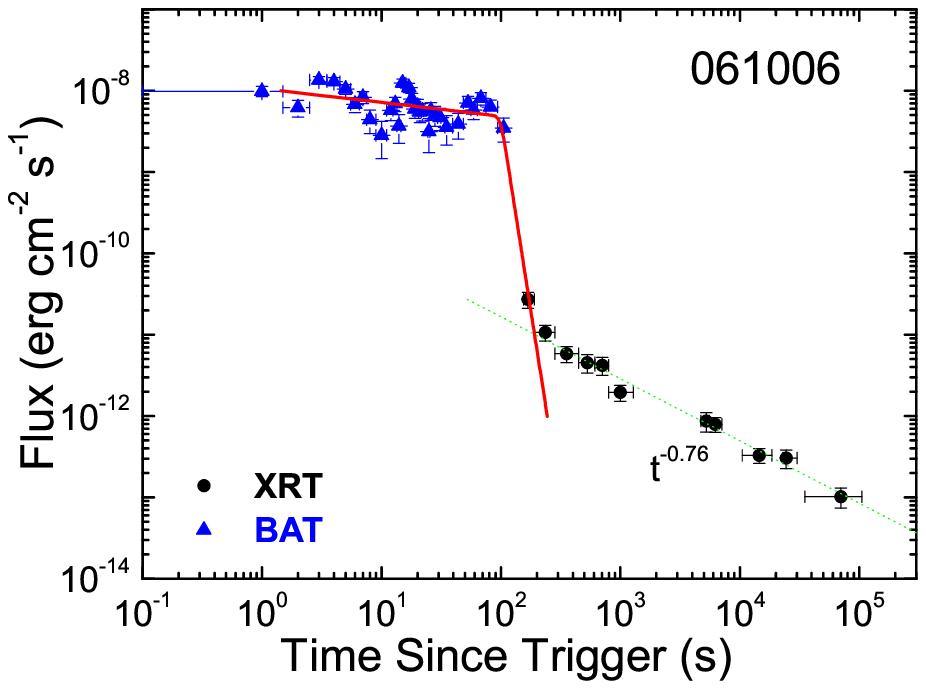}
\includegraphics[angle=0,scale=0.45]{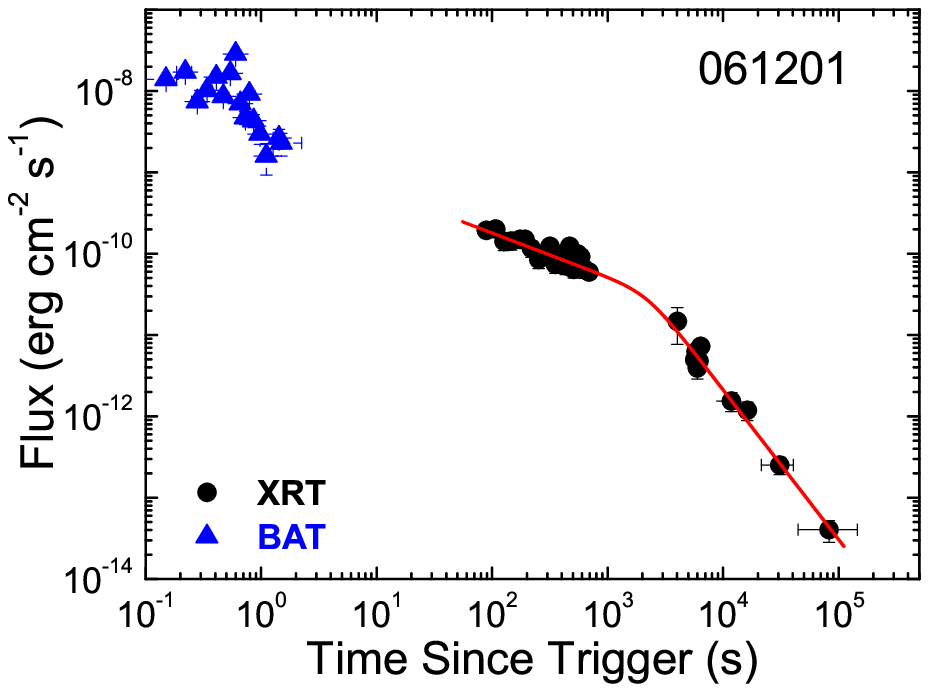}
\includegraphics[angle=0,scale=0.45]{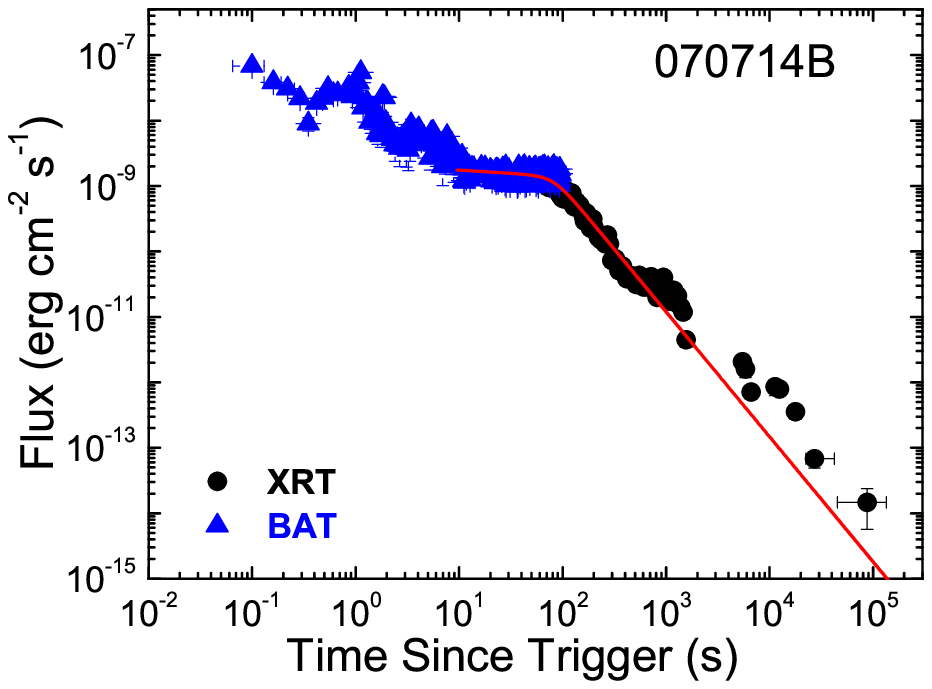}
\includegraphics[angle=0,scale=0.455]{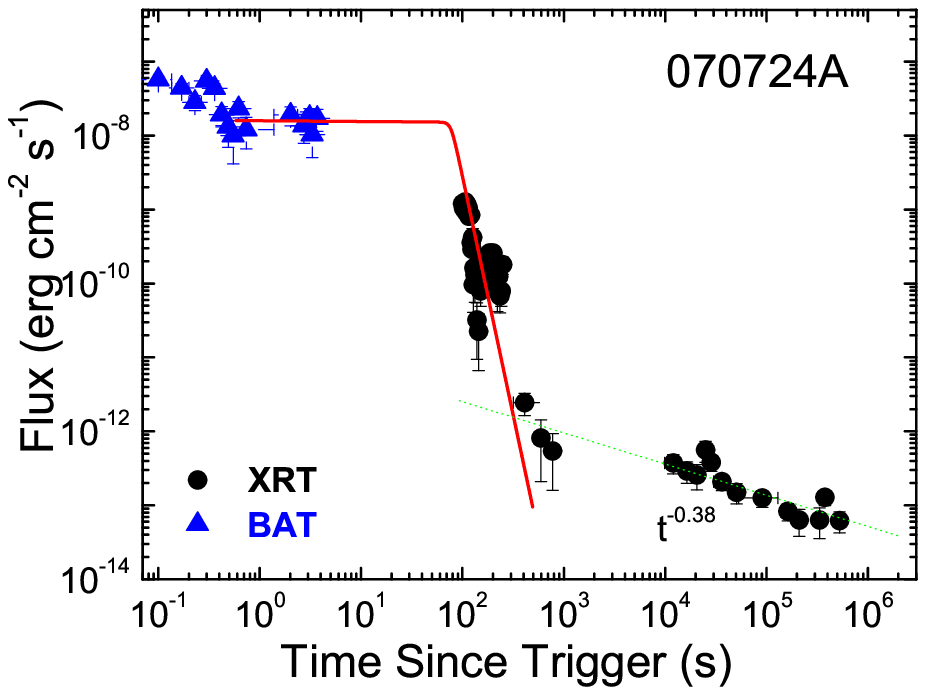}
\includegraphics[angle=0,scale=0.45]{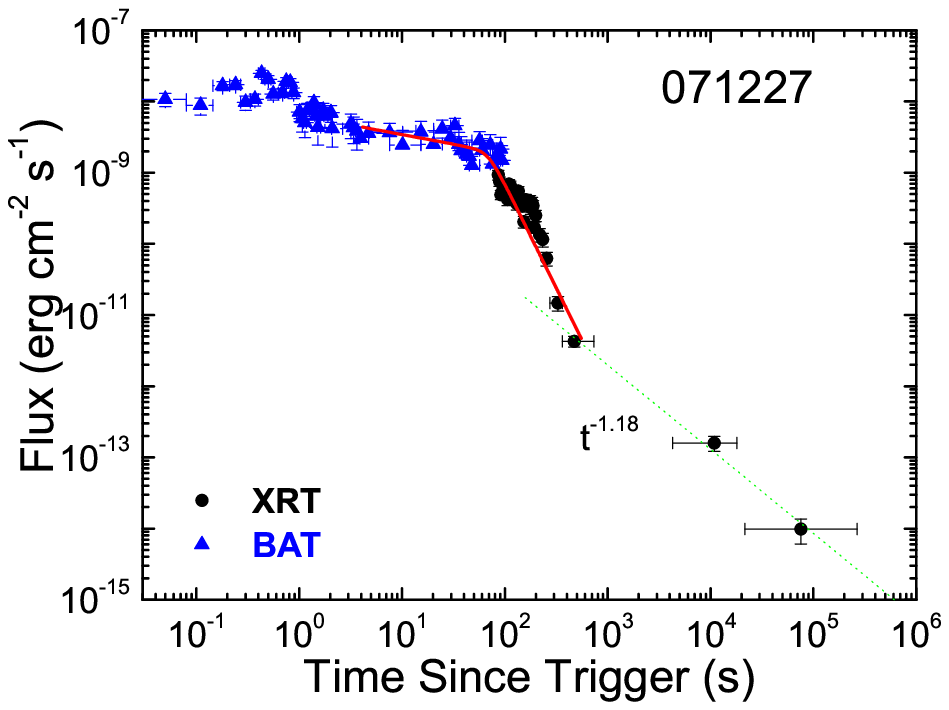}
\includegraphics[angle=0,scale=0.45]{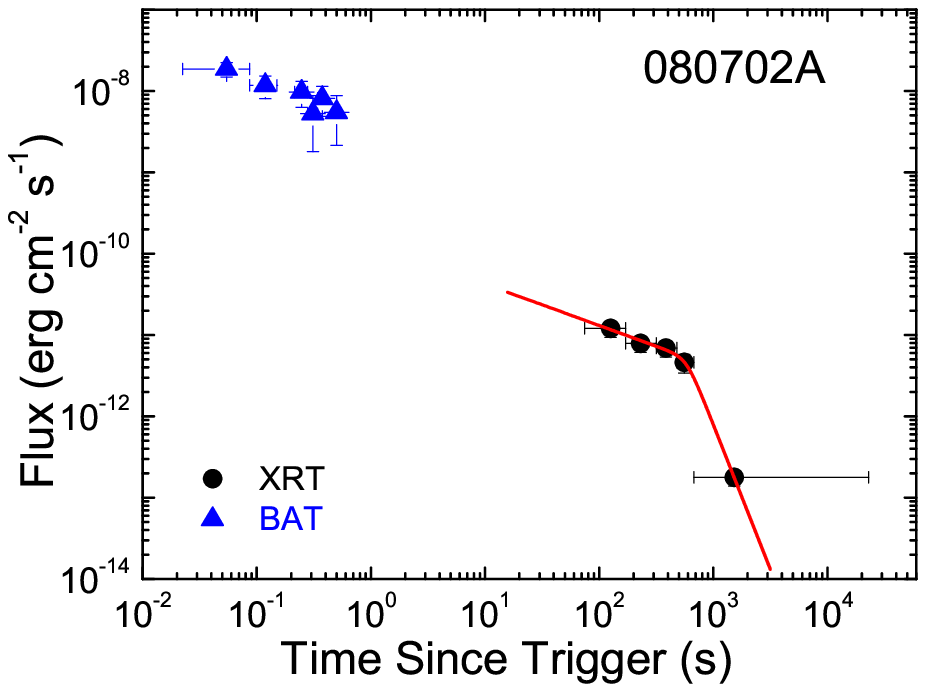}
\includegraphics[angle=0,scale=0.45]{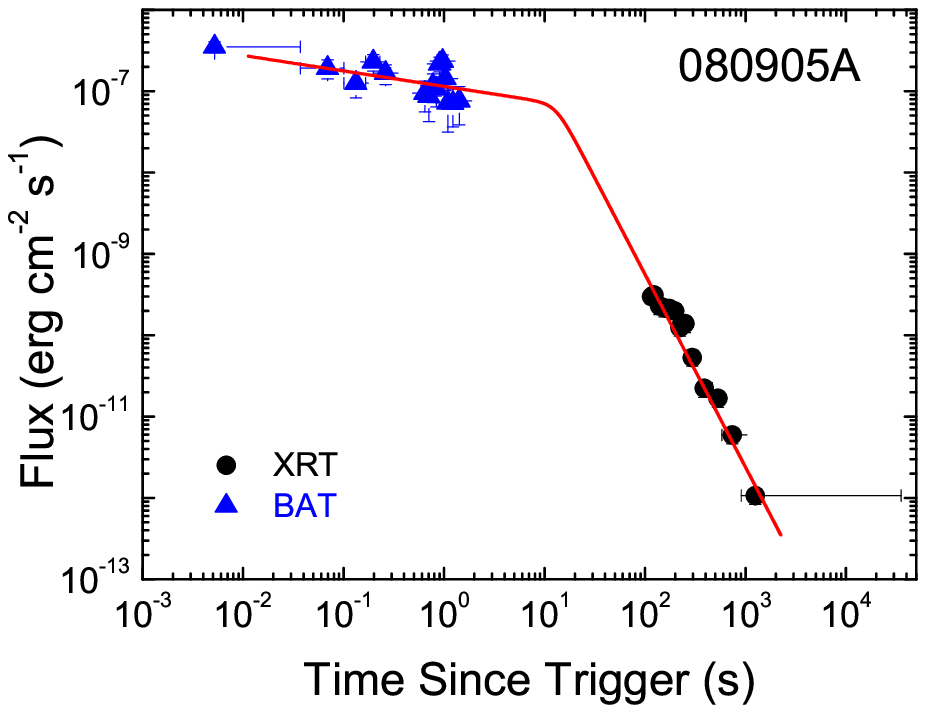}
\includegraphics[angle=0,scale=0.45]{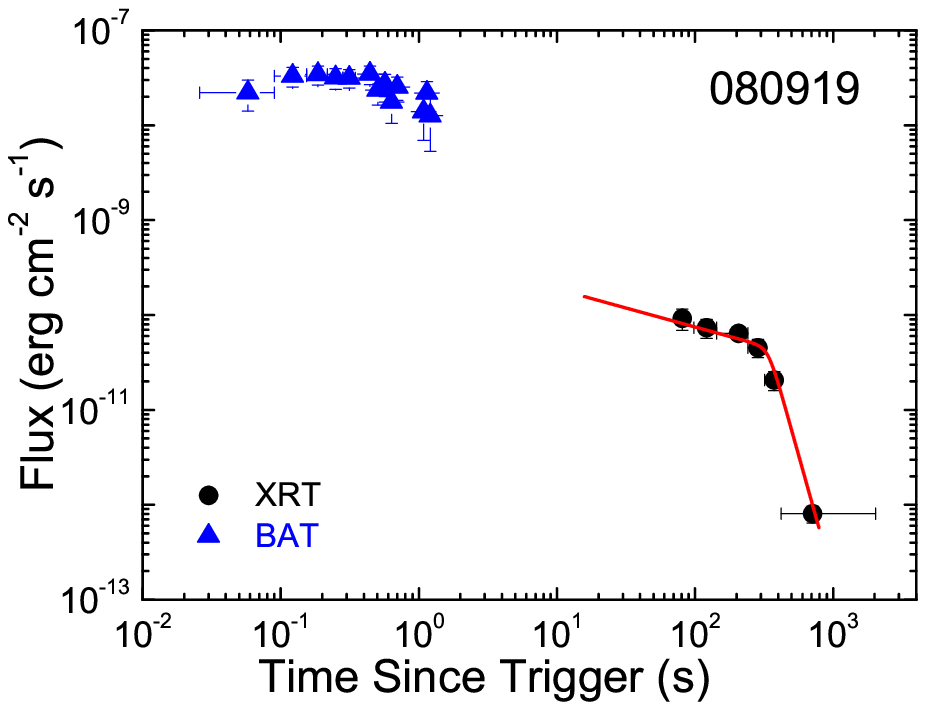}
\includegraphics[angle=0,scale=0.45]{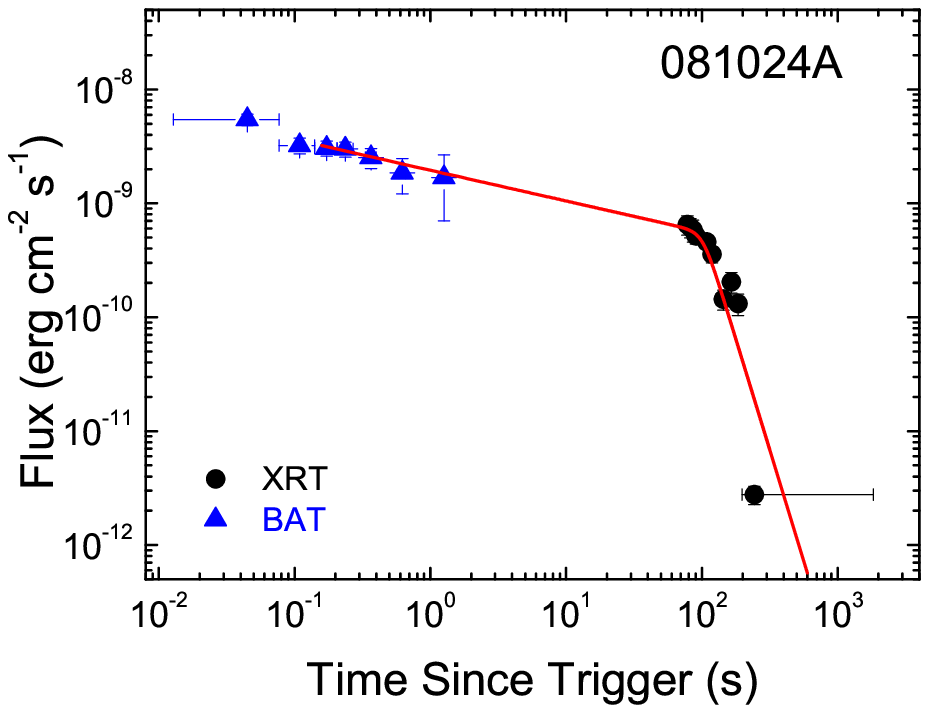}
\includegraphics[angle=0,scale=0.45]{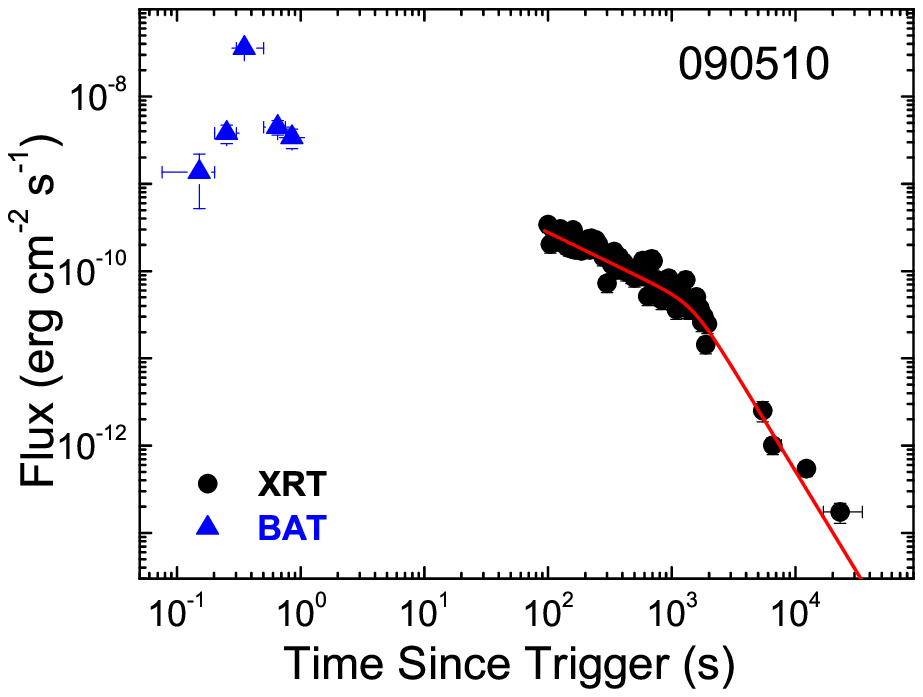}
\hfill
\includegraphics[angle=0,scale=0.45]{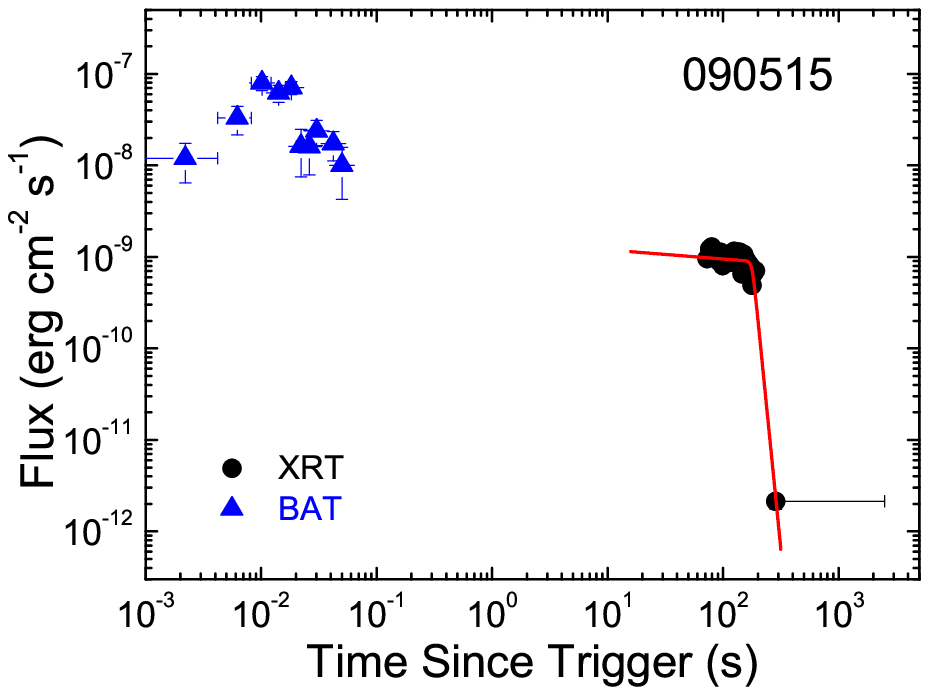}
\center\caption{The BAT-XRT rest-frame light curves of the GRBs
in our Internal sample. Blue triangles are BAT data
extrapolated to the XRT band, and black points (with error
bars) are the XRT data. The red solid curves are the best fits
with a smooth power-law model to the data. The green dotted
lines are the best fits with power-law model after the steeper
decay. }\label{X-ray}
\end{figure}

%*******************************************************************************************
\begin{figure}
\includegraphics[angle=0,scale=0.45]{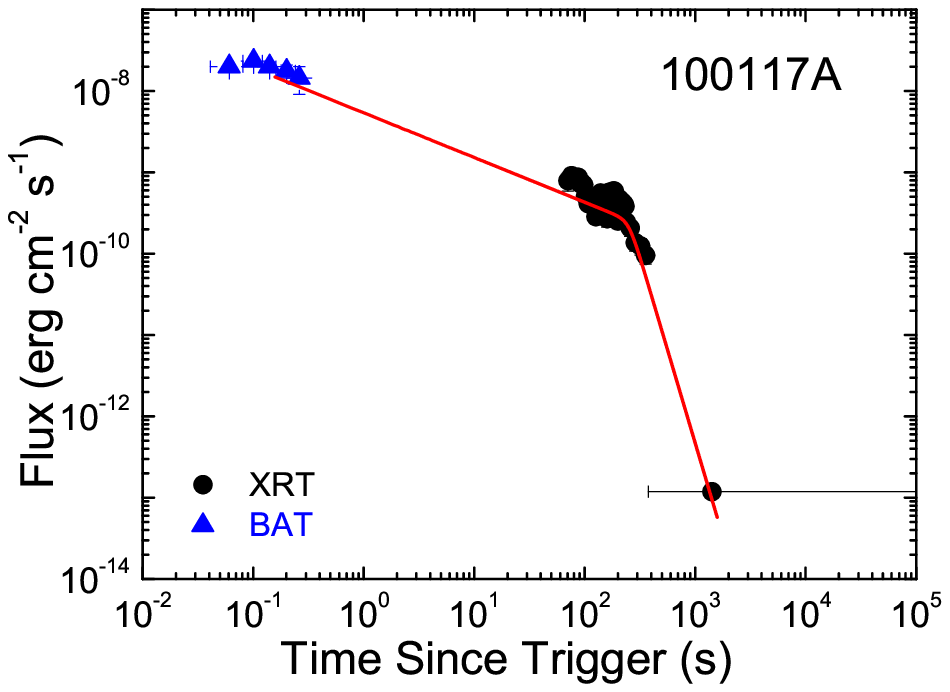}
\includegraphics[angle=0,scale=0.45]{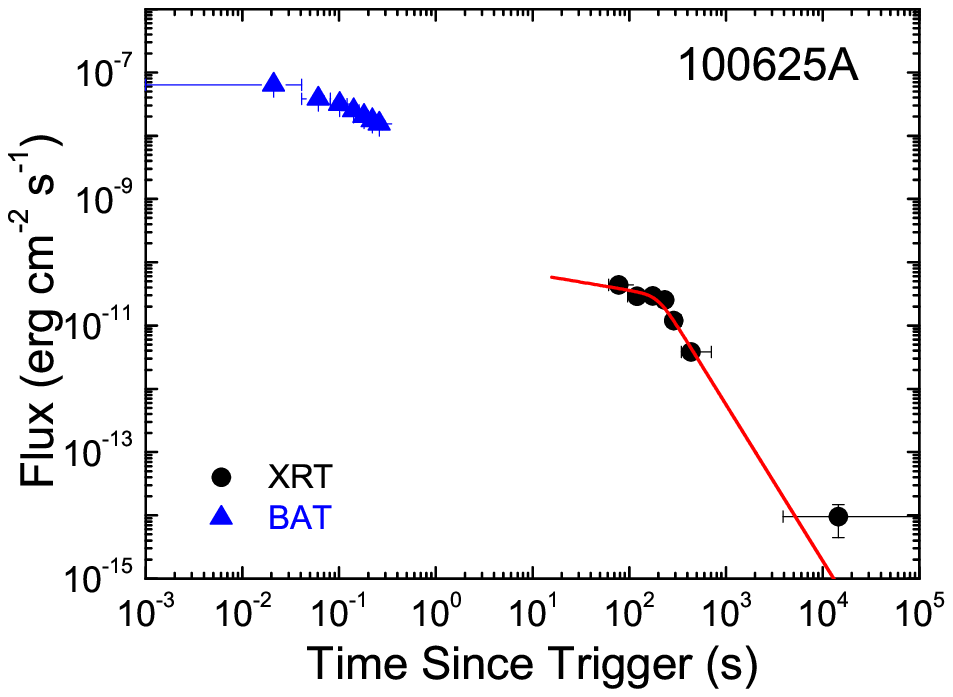}
\includegraphics[angle=0,scale=0.45]{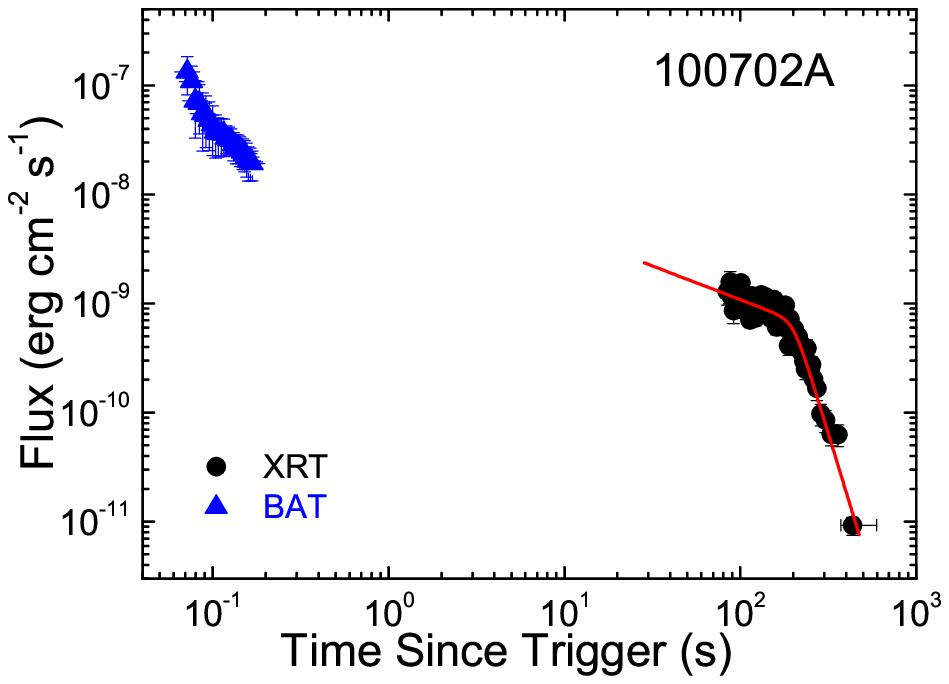}
\includegraphics[angle=0,scale=0.45]{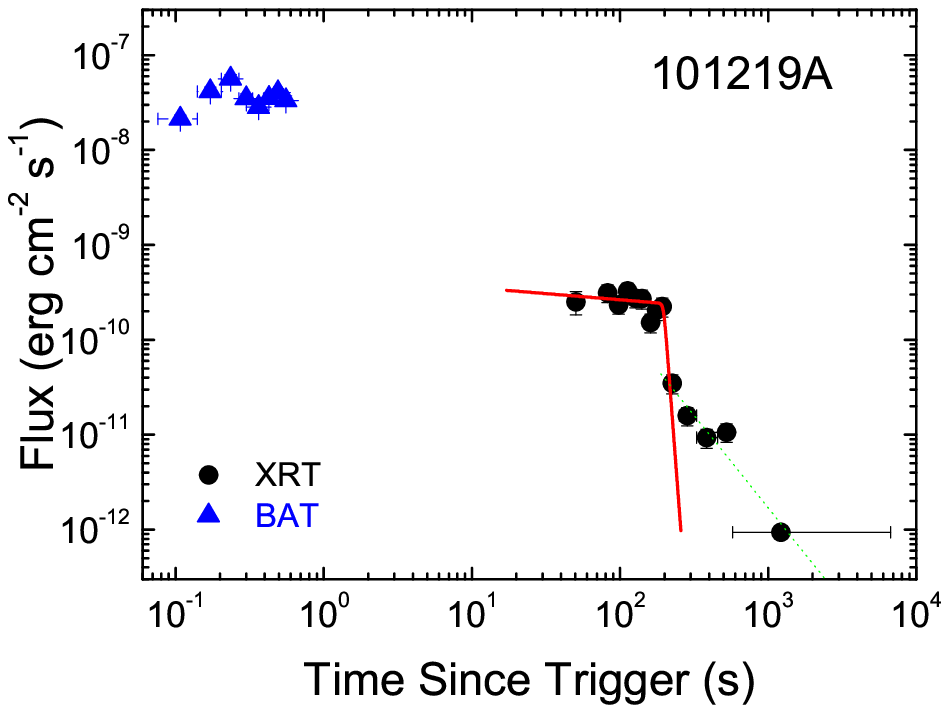}
\includegraphics[angle=0,scale=0.45]{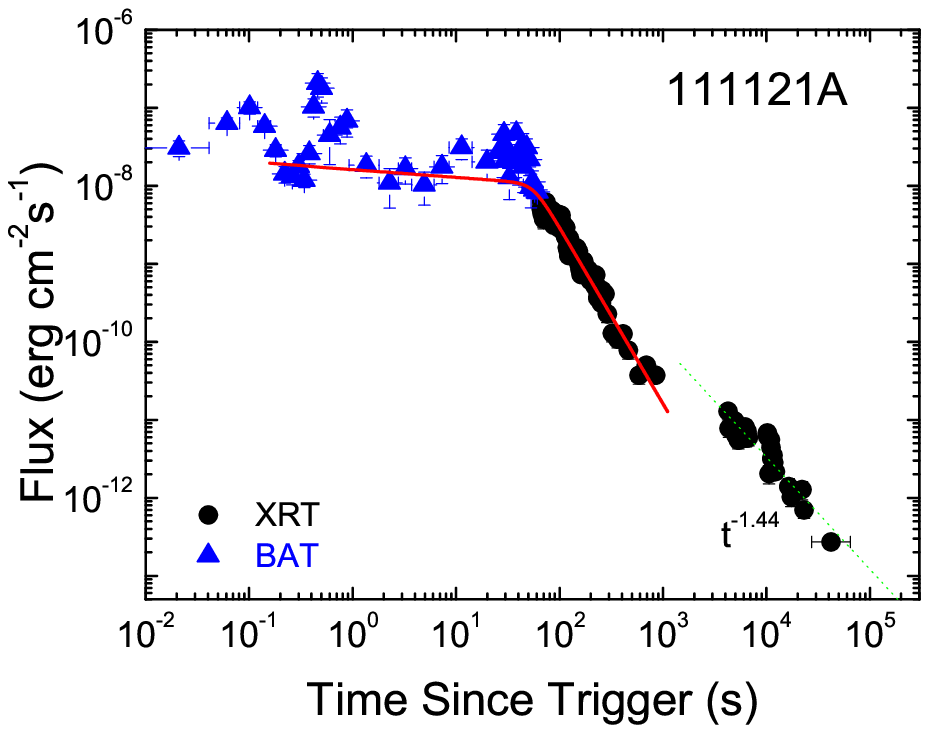}
\includegraphics[angle=0,scale=0.45]{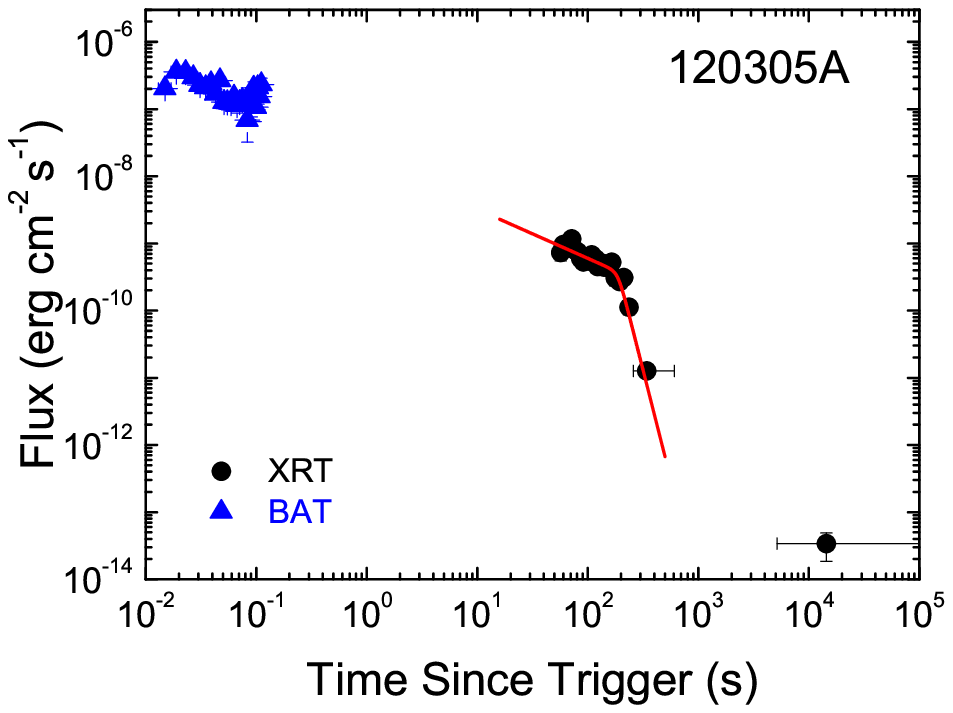}
\hfill
\includegraphics[angle=0,scale=0.45]{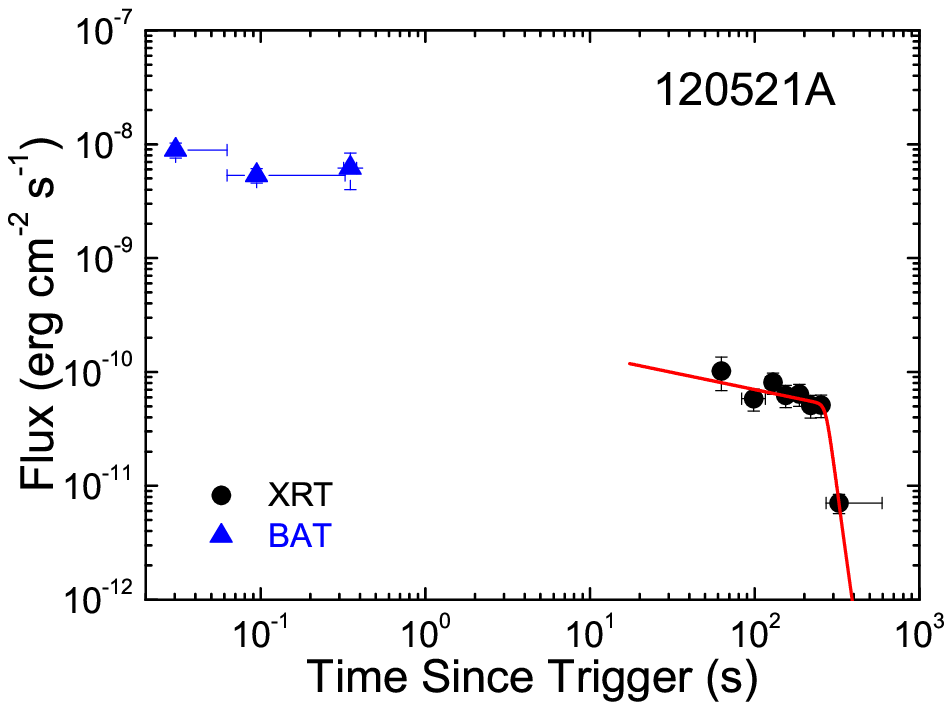}
\center{Fig.1--- continued.}\label{X-ray}
\end{figure}

%*******************************************************************************************
\begin{figure}
\includegraphics[angle=0,scale=0.45]{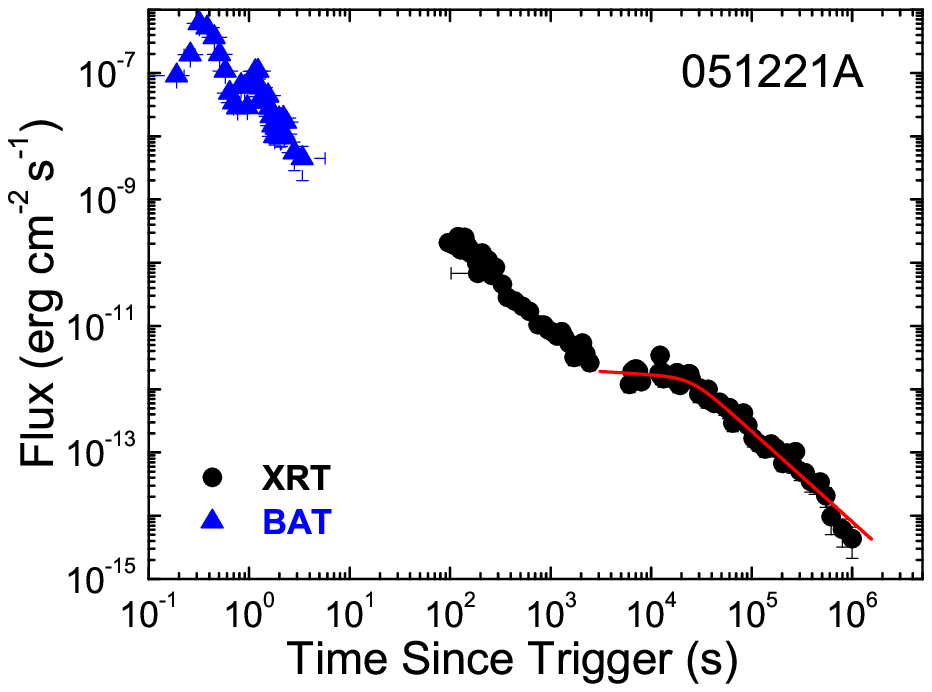}
\includegraphics[angle=0,scale=0.45]{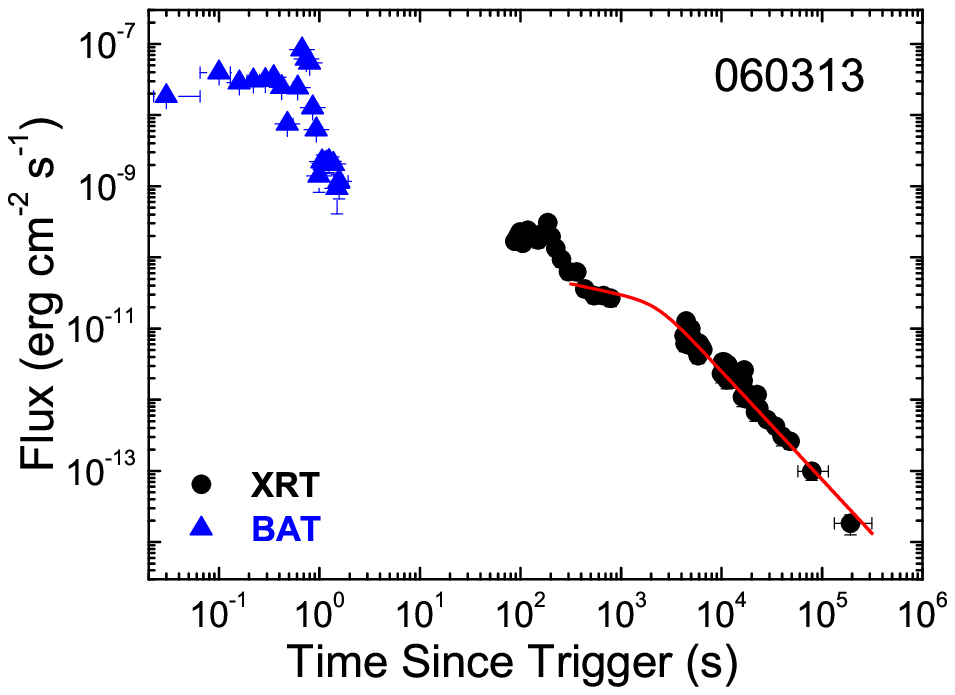}
\includegraphics[angle=0,scale=0.45]{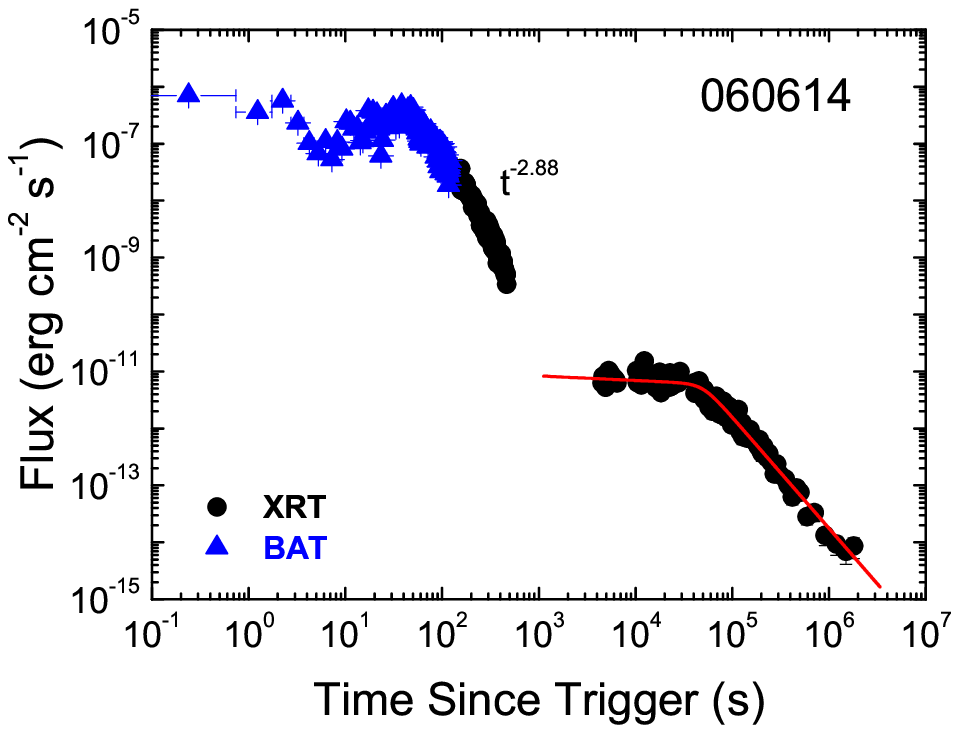}
\includegraphics[angle=0,scale=0.45]{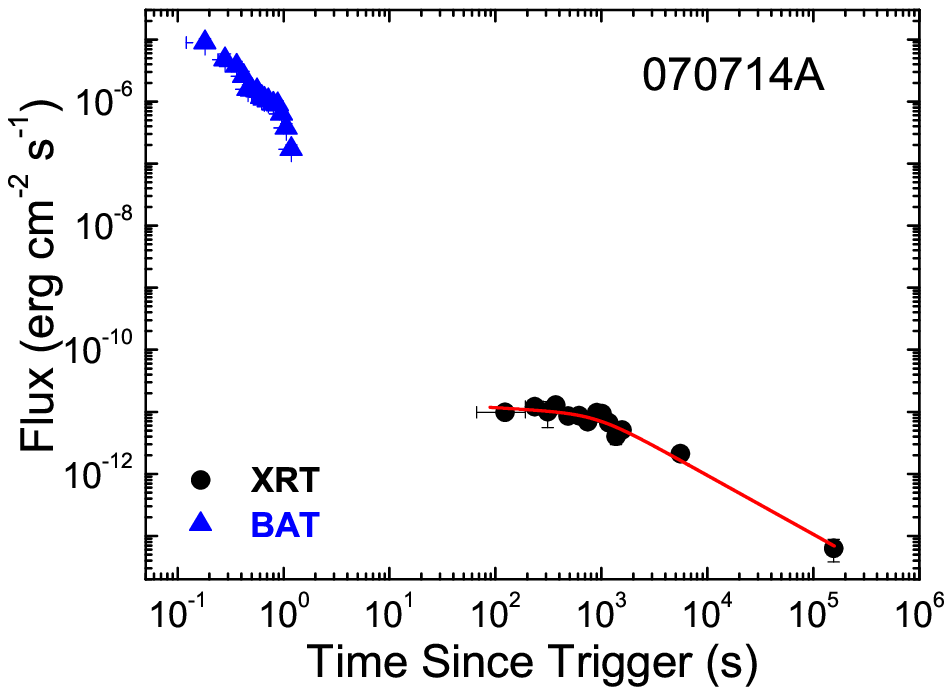}
\includegraphics[angle=0,scale=0.45]{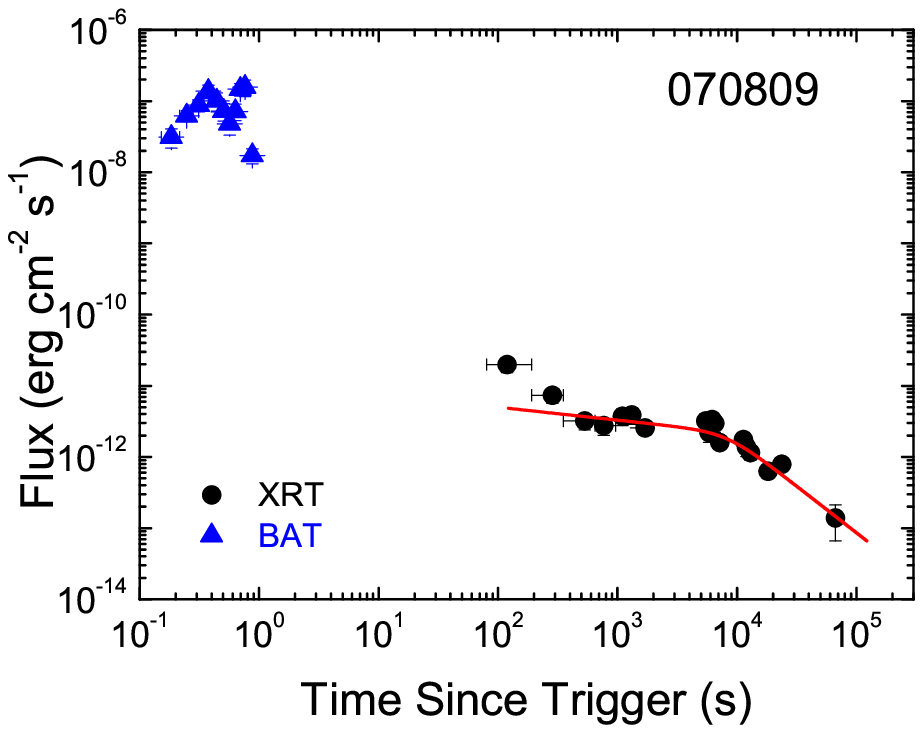}
\includegraphics[angle=0,scale=0.45]{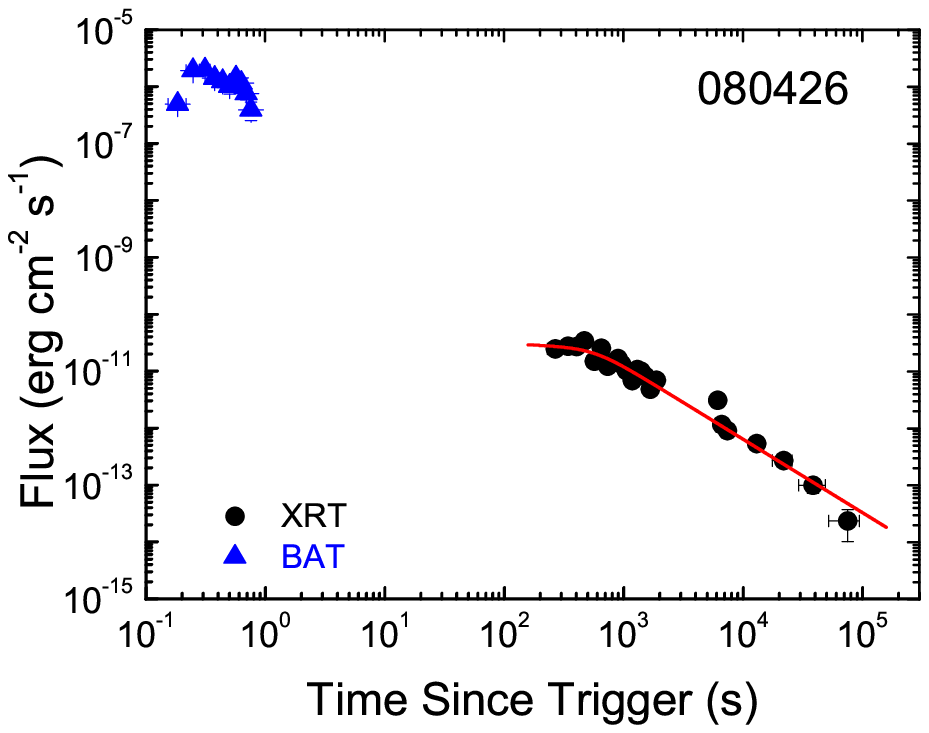}
\includegraphics[angle=0,scale=0.45]{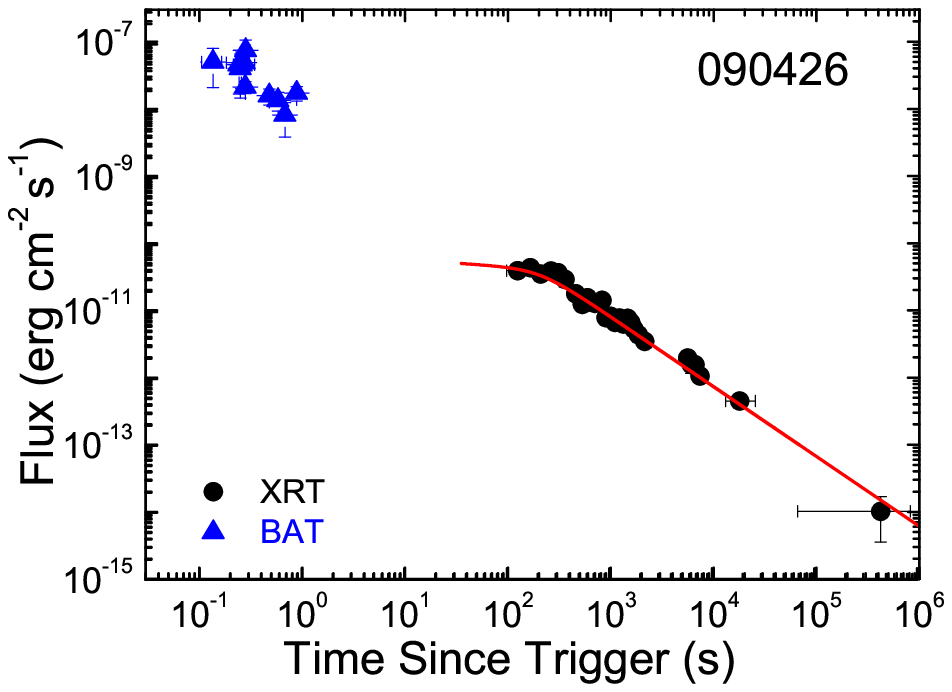}
\includegraphics[angle=0,scale=0.45]{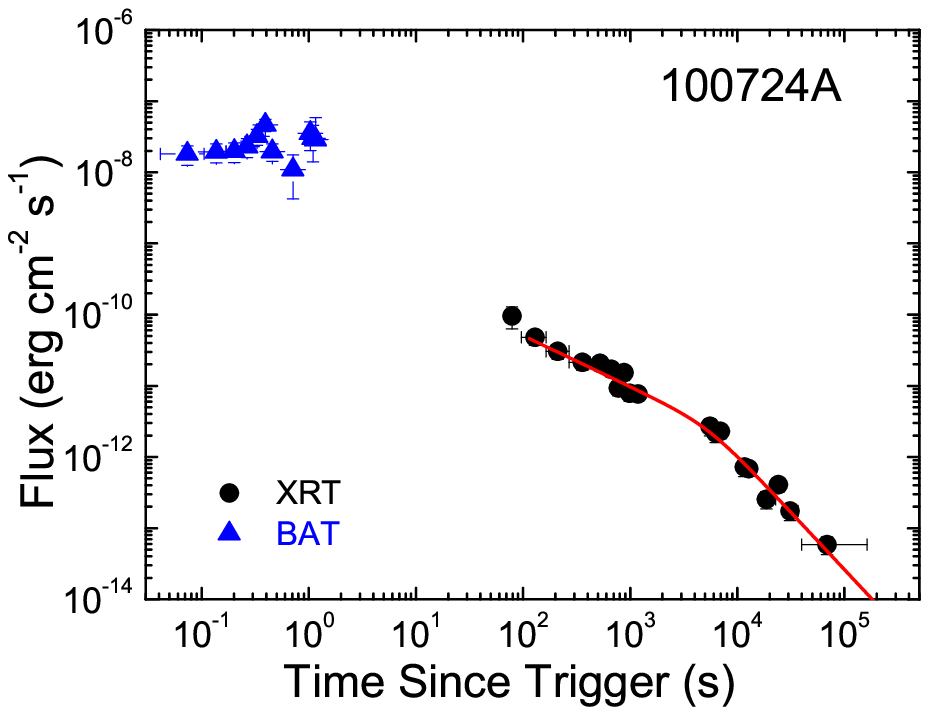}
\includegraphics[angle=0,scale=0.45]{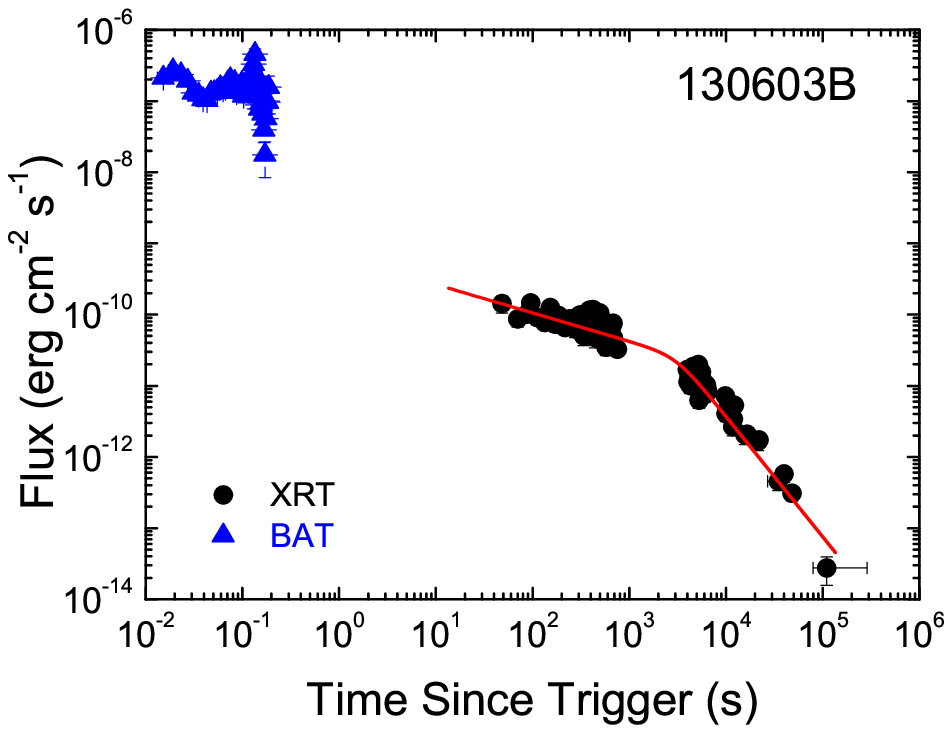}
\hfill
\includegraphics[angle=0,scale=0.45]{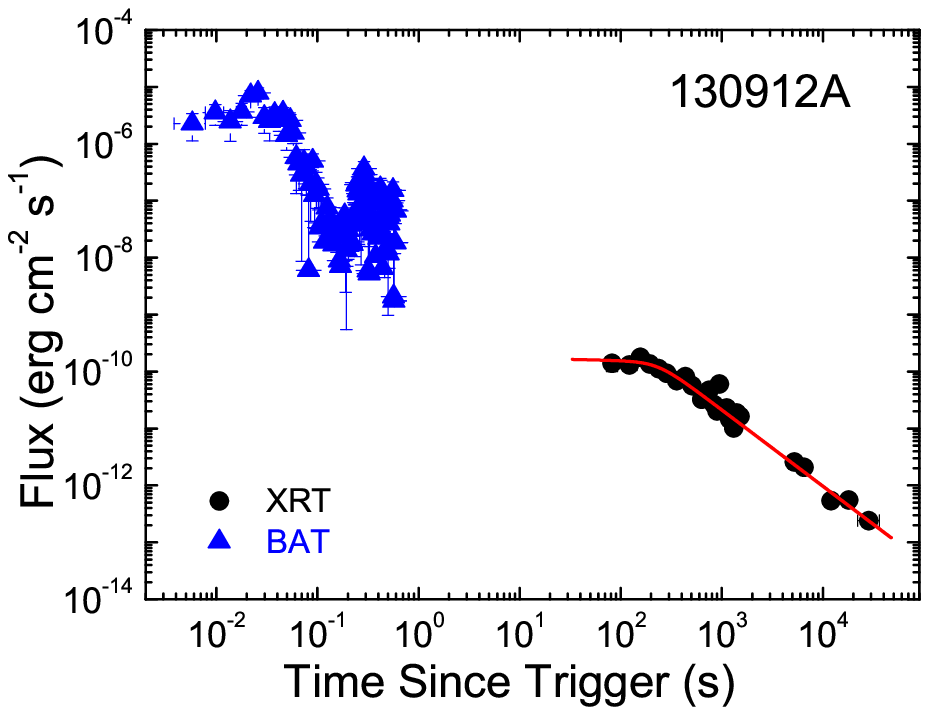}
\center\caption{Similar to Figure 1, but for the External
sample.}
\end{figure}

%*******************************************************************************************
\begin{figure}
\includegraphics[angle=0,scale=0.8]{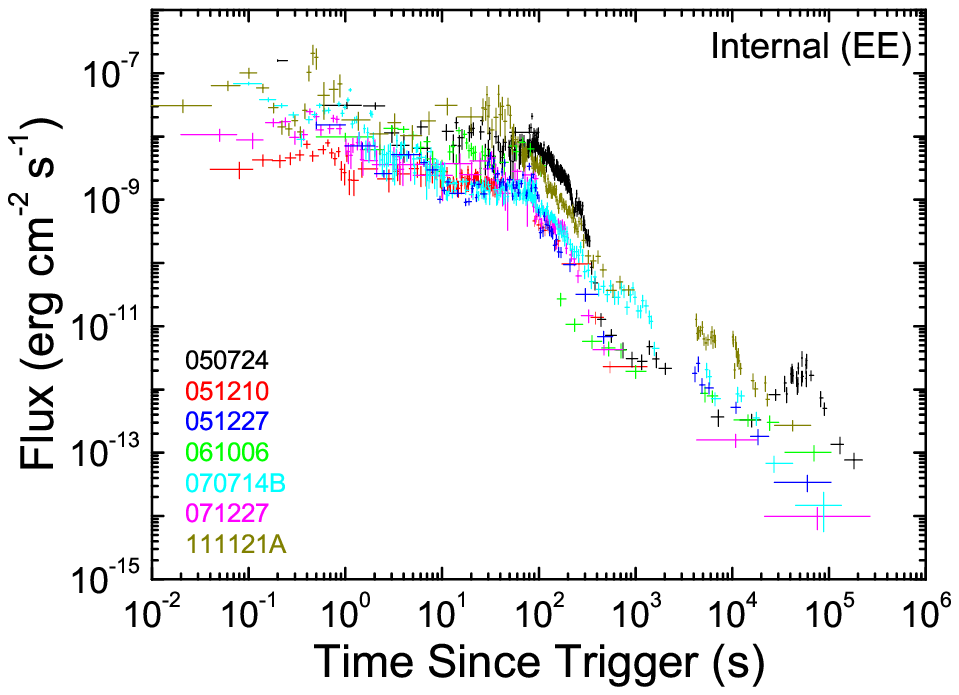}
\includegraphics[angle=0,scale=0.8]{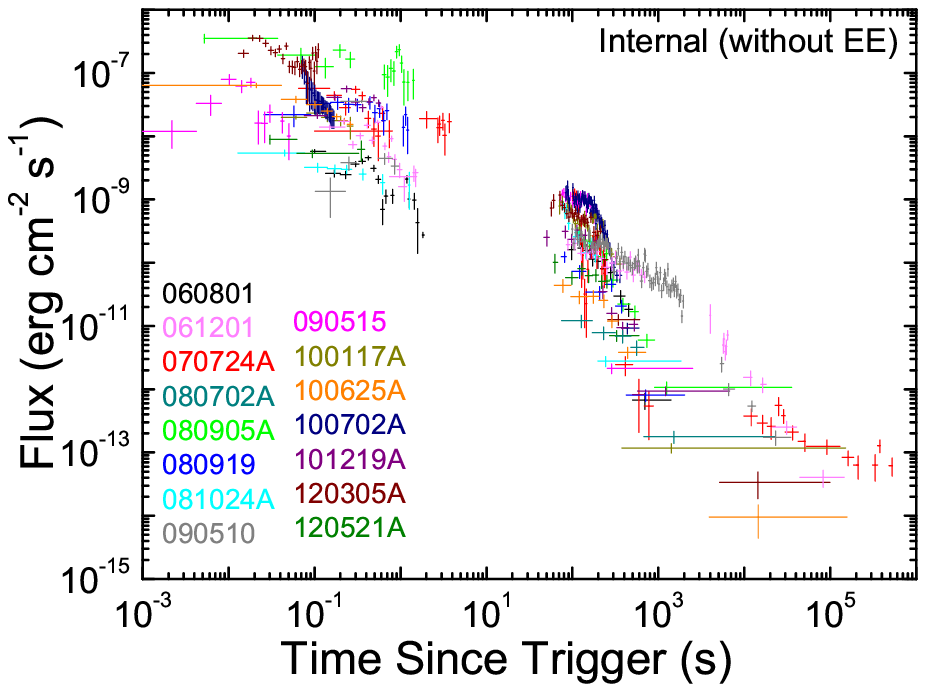}
\includegraphics[angle=0,scale=0.8]{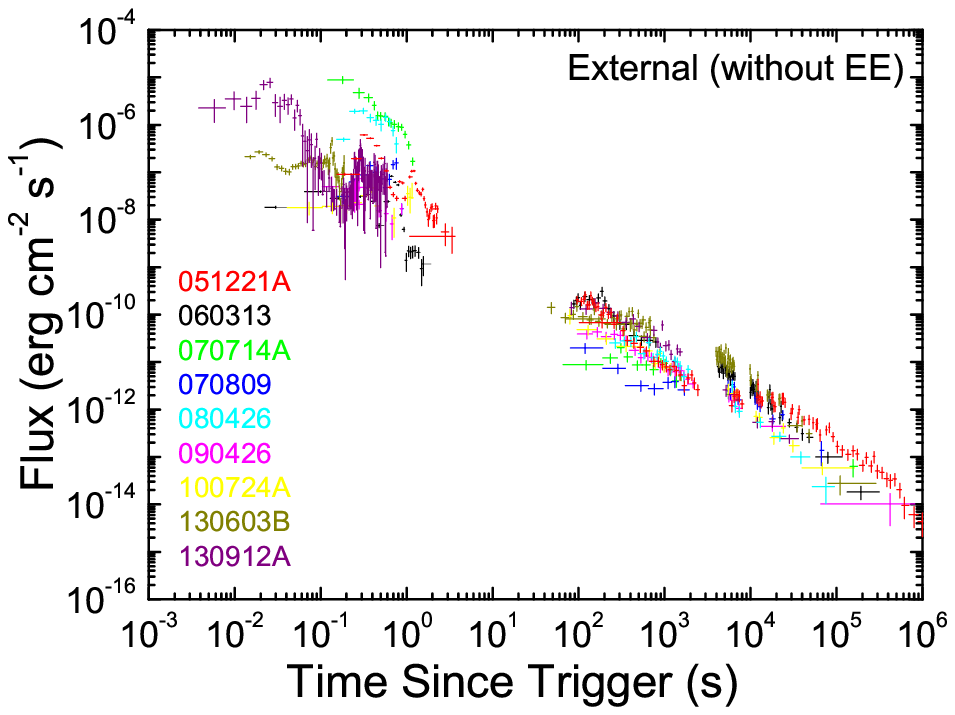}
\hfill
\includegraphics[angle=0,scale=0.8]{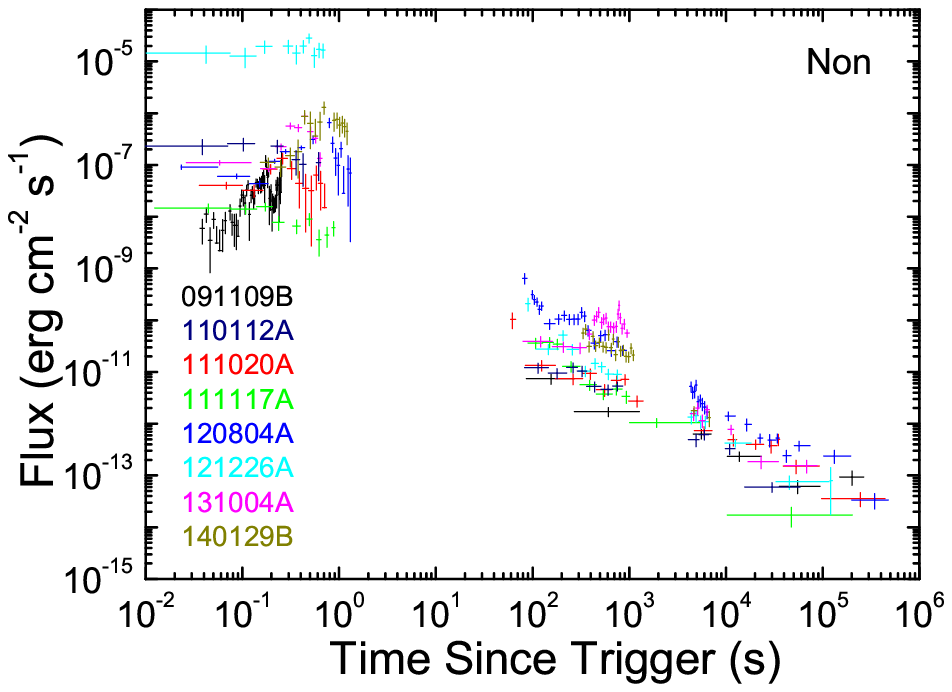}
\caption{Ensemble of X-ray light curves (0.3-10 keV) of the
GRBs in our Internal sample with EE, Internal sample without
EE, External sample, and Non sample.}
\end{figure}

%*******************************************************************************************
\begin{figure}
\includegraphics[angle=0,scale=0.5]{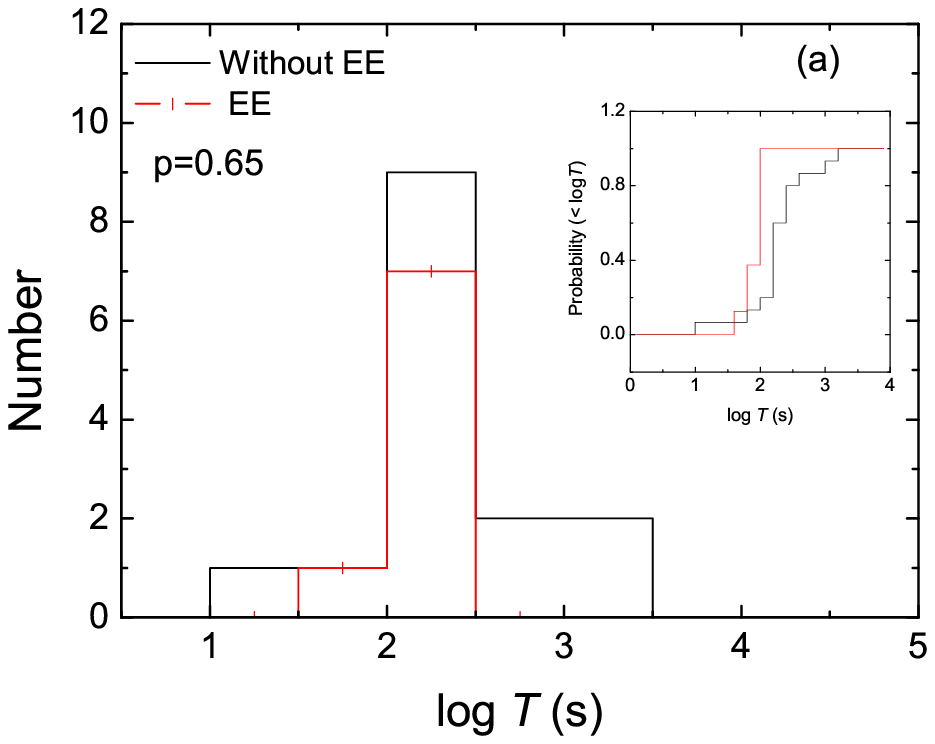}
\includegraphics[angle=0,scale=0.55]{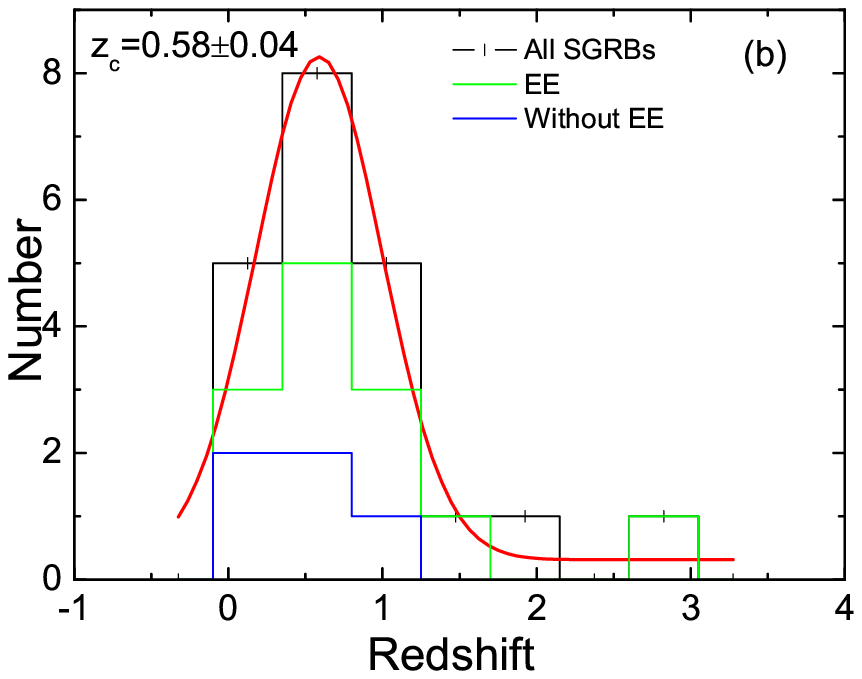}
\includegraphics[angle=0,scale=0.5]{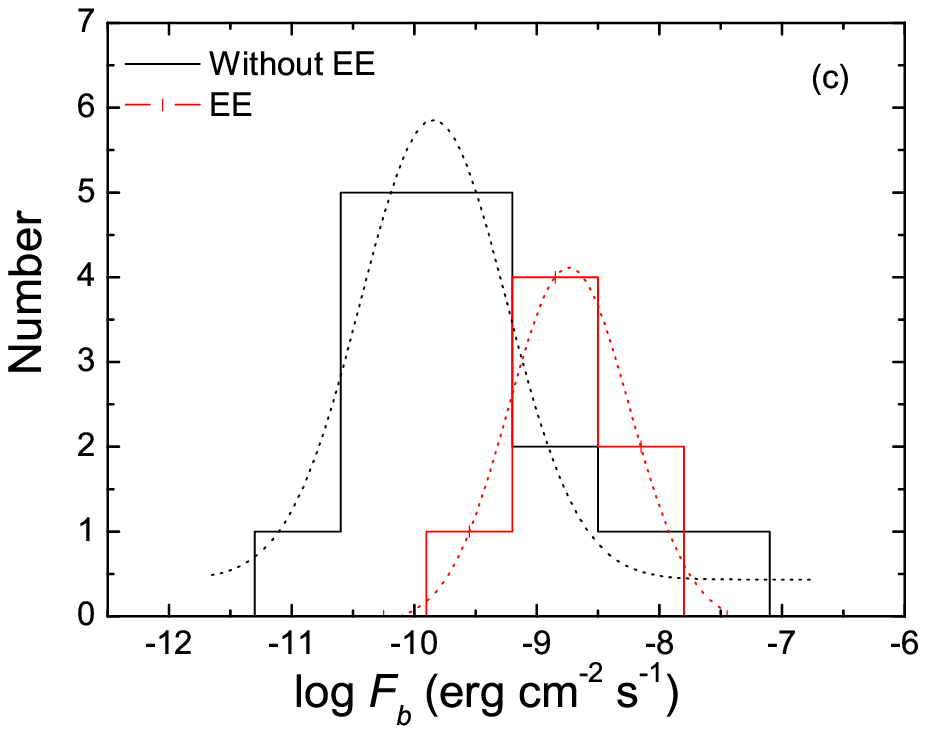}
\hfill
\includegraphics[angle=0,scale=0.5]{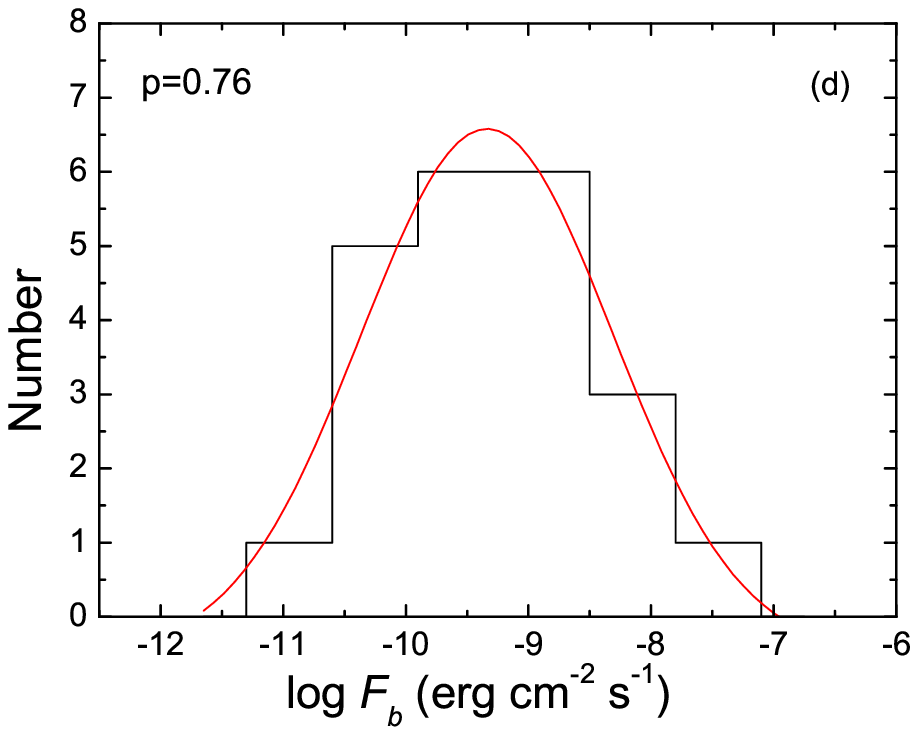}
\includegraphics[angle=0,scale=0.5]{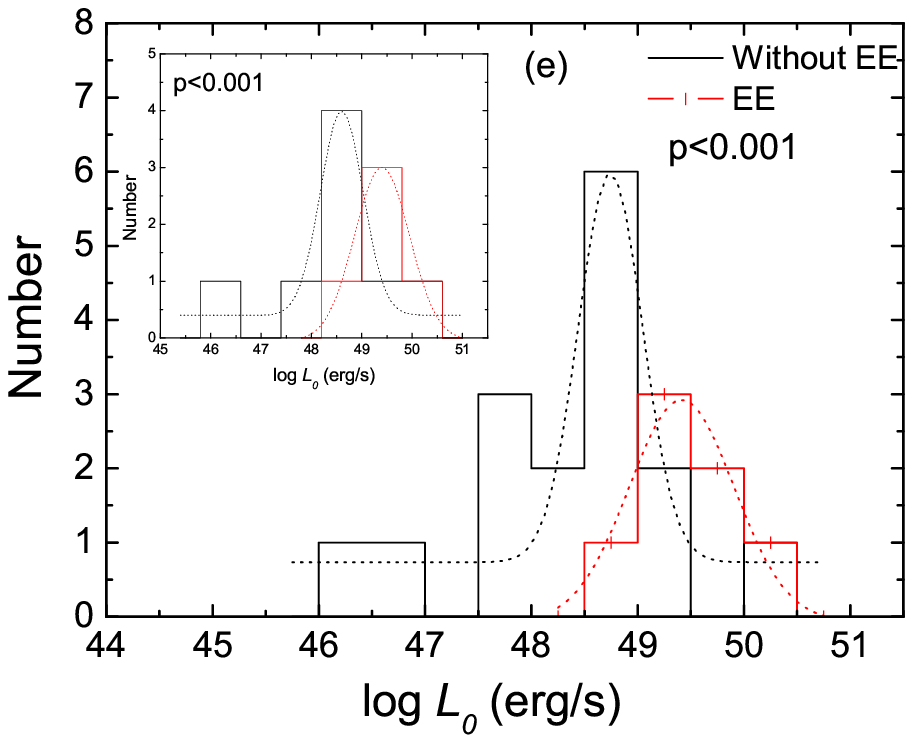}
\hfill
\includegraphics[angle=0,scale=0.5]{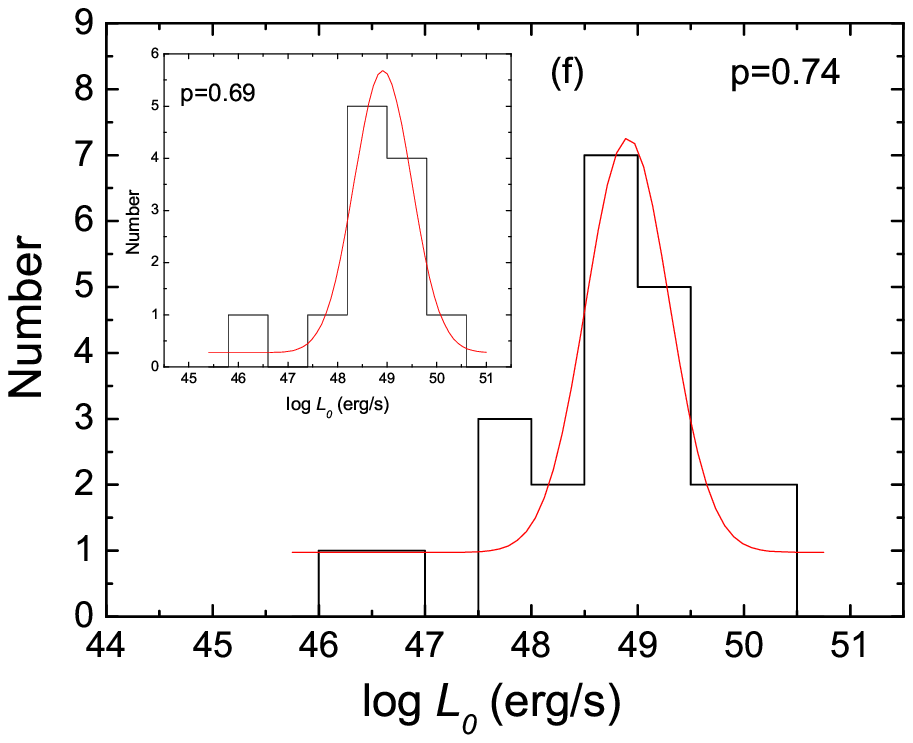}
\caption{(a) Duration distributions of the
extended emission for EE sample, and the internal plateau emission
for the no-EE sample. Inset: the cumulative duration
distributions for the EE and no-EE sub-samples. (b) The redshift distribution
of all short GRBs with $z$ measurements. The red solid line is
the best Gaussian fit with a center value $z_{c}=0.58$. The
green and blue histograms are the redshift distributions for the EE and no-EE
sub-samples, respectively. (c)
The plateau flux distributions of both EE (red, solid line +
bar) and no-EE (black, solid line) GRBs in our Internal sample.
The dotted lines are the best Gaussian fits to the
distributions. (d) A joint fit to the flux distribution of all
the GRBs in the Internal plateau (both EE and no-EE included).
(e) The plateau luminosity distributions of both EE (red,
solid line + bar) and no-EE (black, solid line) GRBs in our
Internal sample. The dotted lines are the best Gaussian fits to
the distributions. (f) A joint fit to the luminosity
distribution of all the GRBs in the Internal plateau (both EE
and no-EE included). The insets in (e) and (f) are for the
GRBs with measured redshifts only.}
\end{figure}

%*******************************************************************************************

\begin{figure}
\includegraphics[angle=0,scale=0.8]{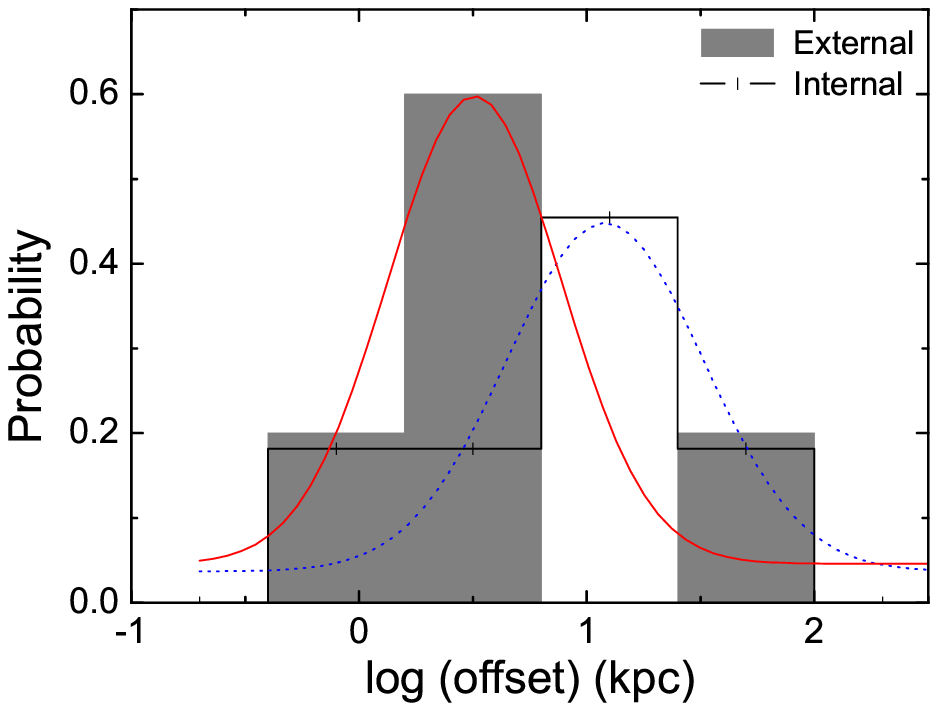}
\includegraphics[angle=0,scale=0.8]{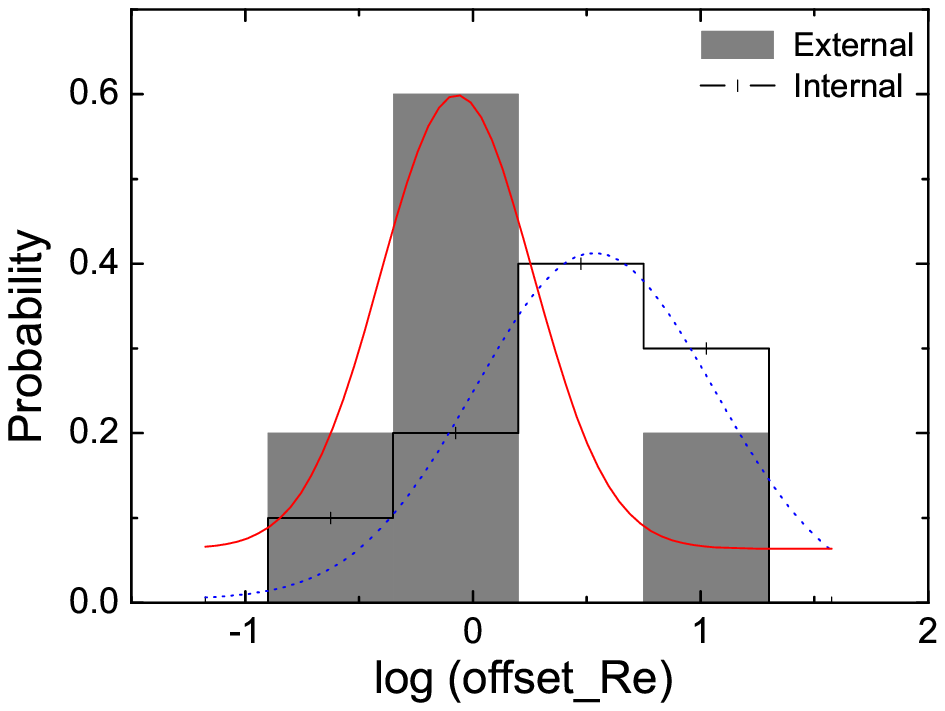}
\hfill\caption{Distributions of the physical offsets and host-normalized
offsets of the Internal and External samples.
The solid and dash lines are the best
Gaussian fitting for the Internal and External samples, respectively.}
\end{figure}

%*******************************************************************************************
\begin{figure}
\includegraphics[angle=0,scale=0.8]{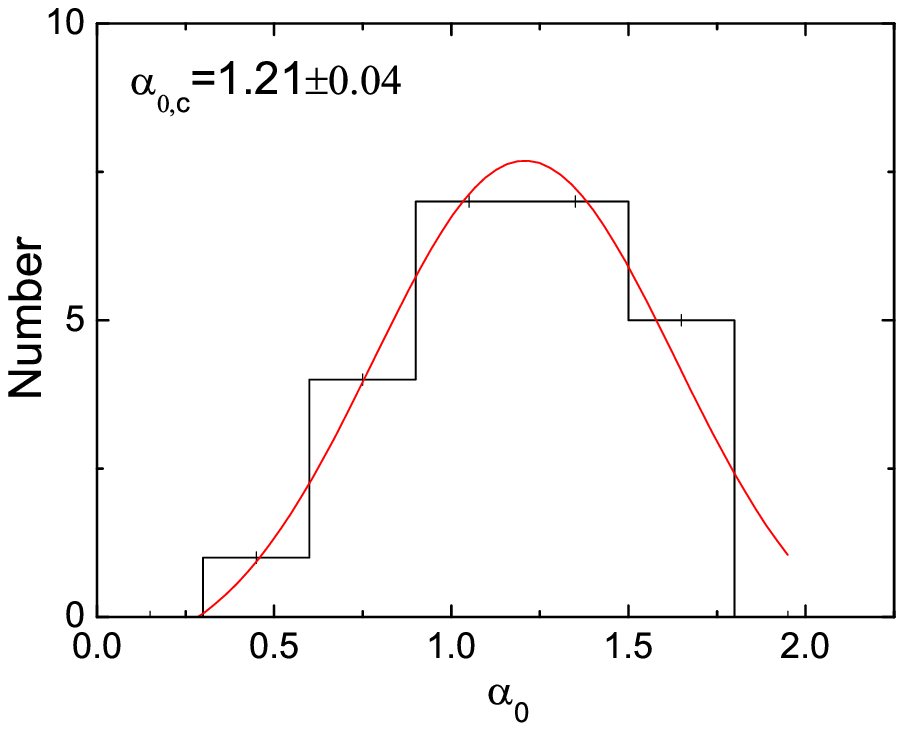}
\includegraphics[angle=0,scale=0.8]{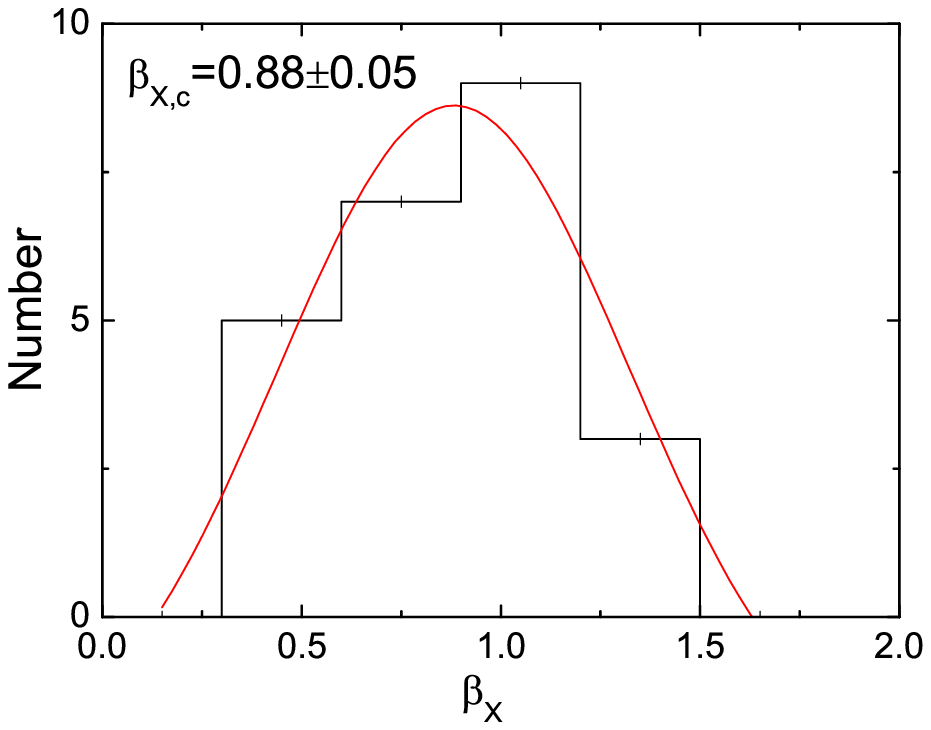}
\hfill\caption{Distributions of the decay slope $\alpha_{0}$ and
spectral index $\beta_{X}$ in the normal decay phase in our
External and Non samples. The solid lines are the best Gaussian
fits to the distributions.}
\end{figure}

%*******************************************************************************************

\begin{figure}
\includegraphics[angle=0,scale=0.45]{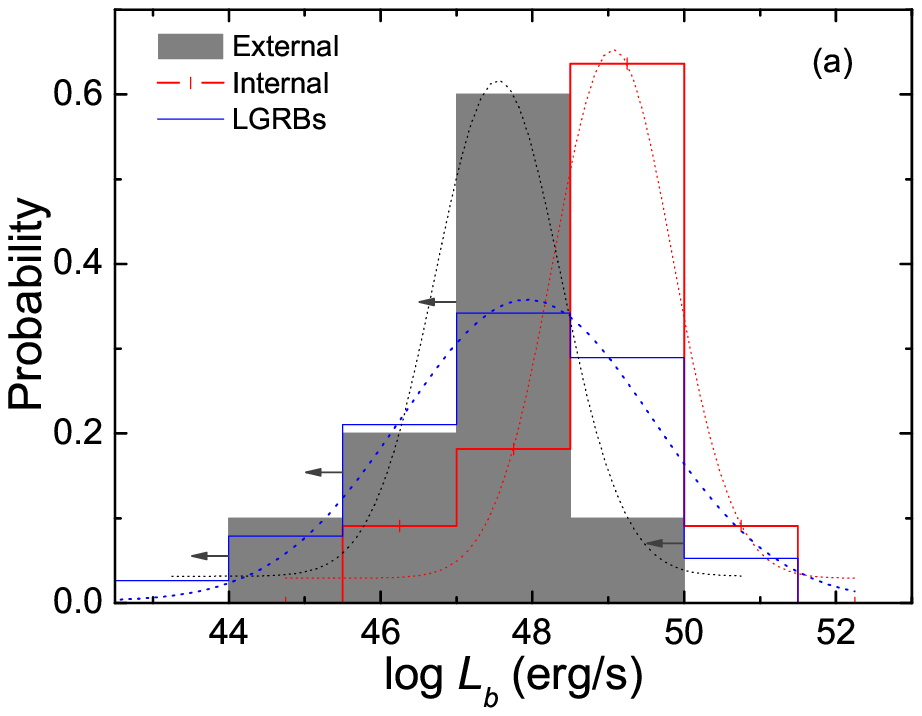}
\includegraphics[angle=0,scale=0.45]{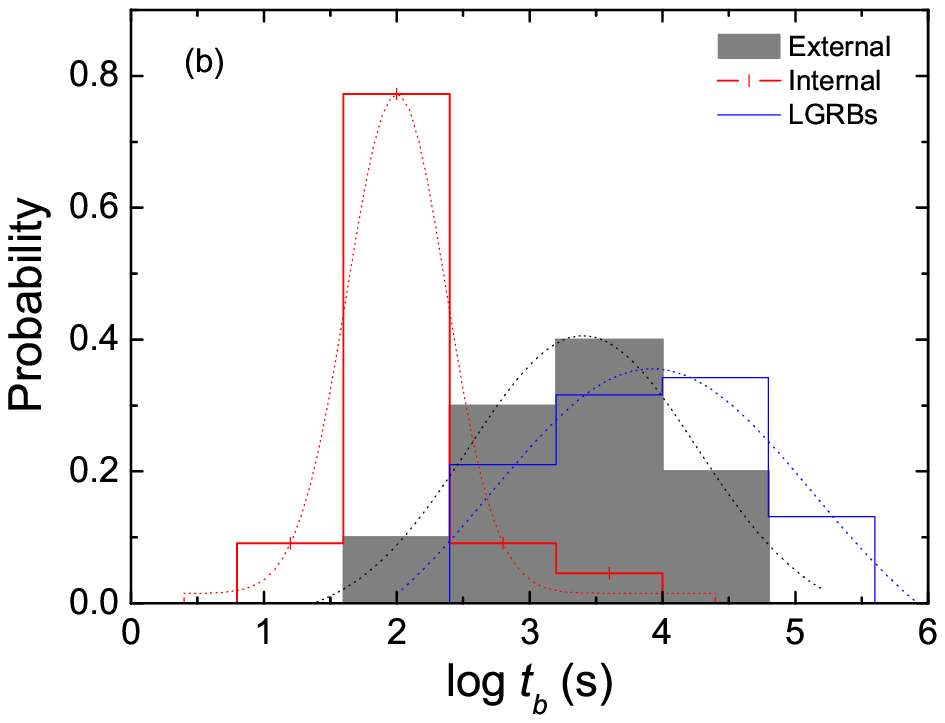}
\includegraphics[angle=0,scale=0.45]{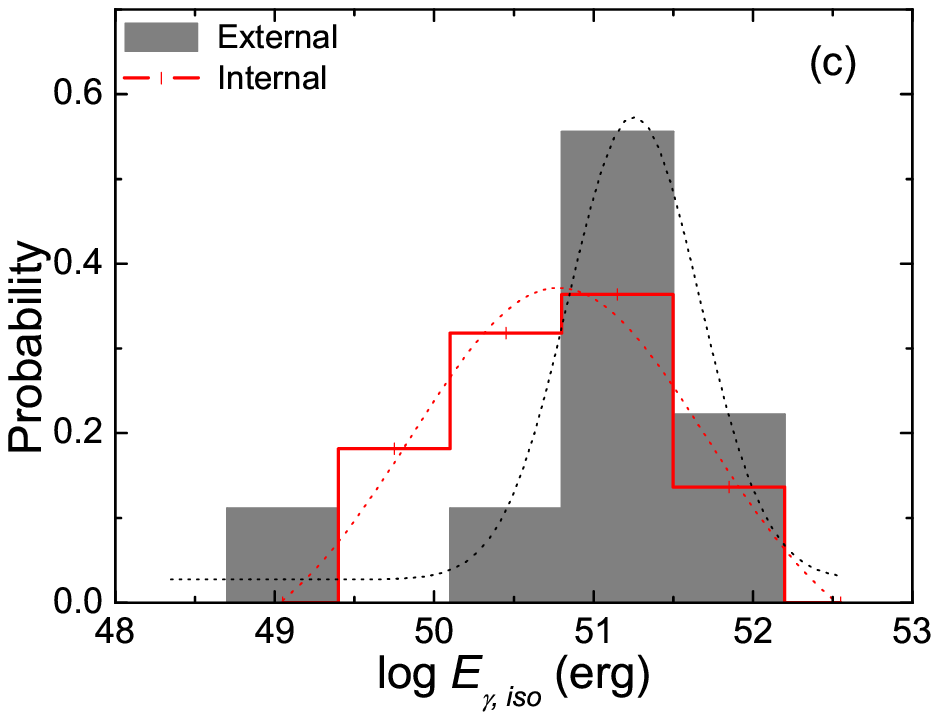}
\includegraphics[angle=0,scale=0.45]{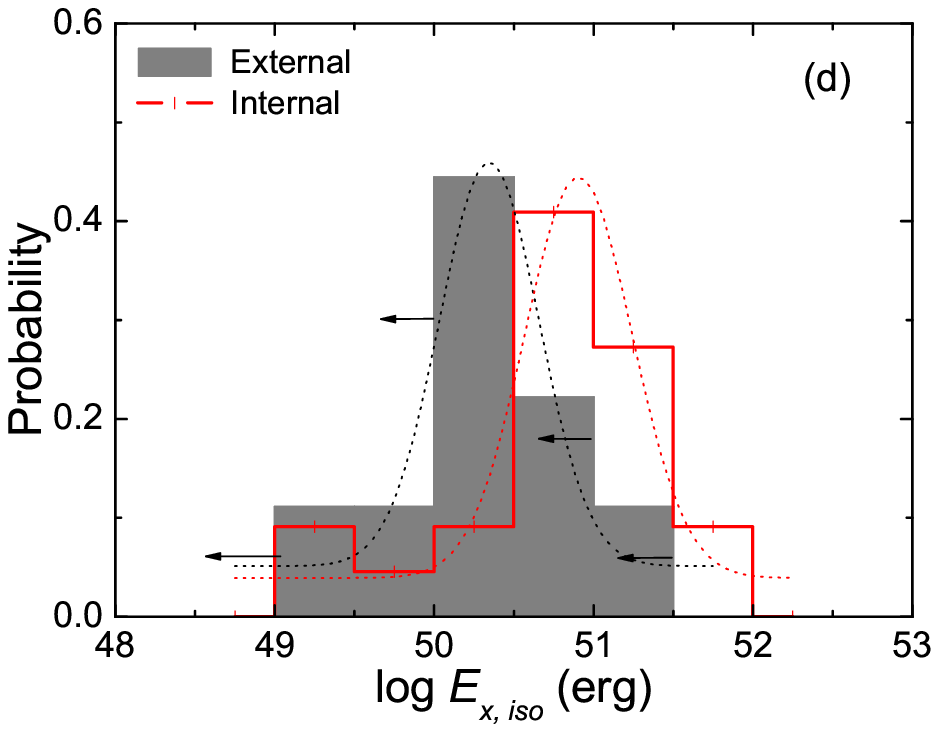}
\includegraphics[angle=0,scale=0.45]{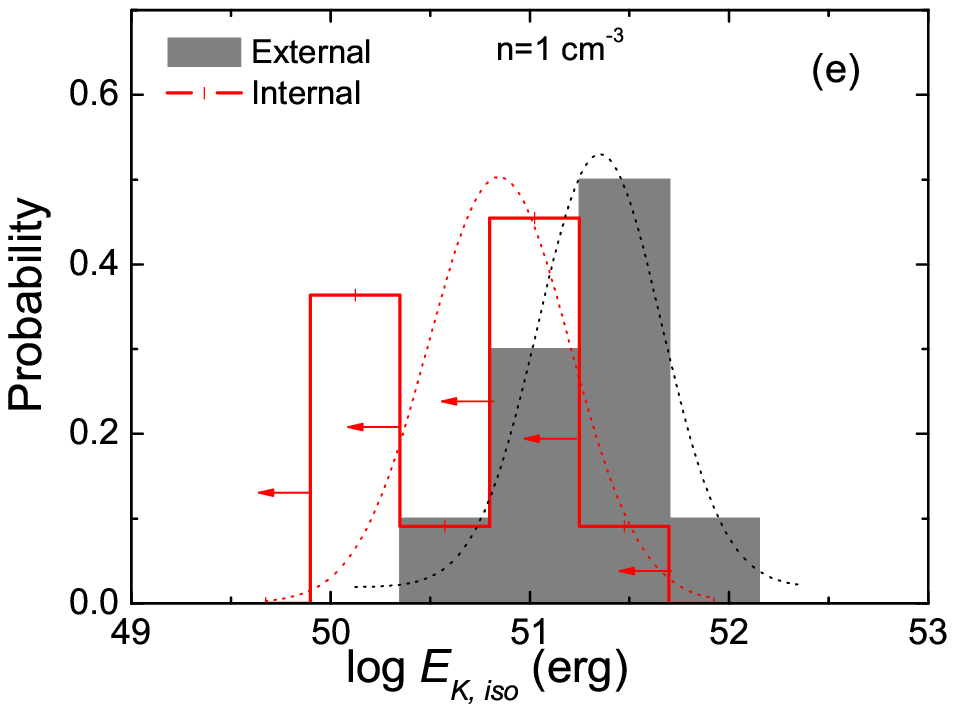}
\includegraphics[angle=0,scale=0.45]{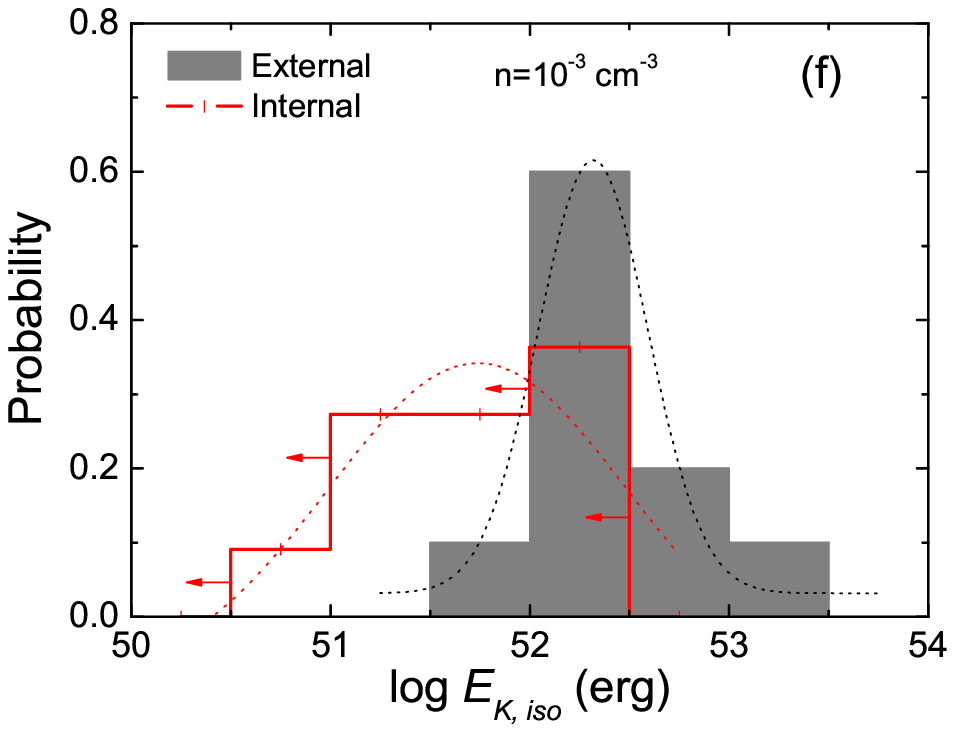}
\includegraphics[angle=0,scale=0.45]{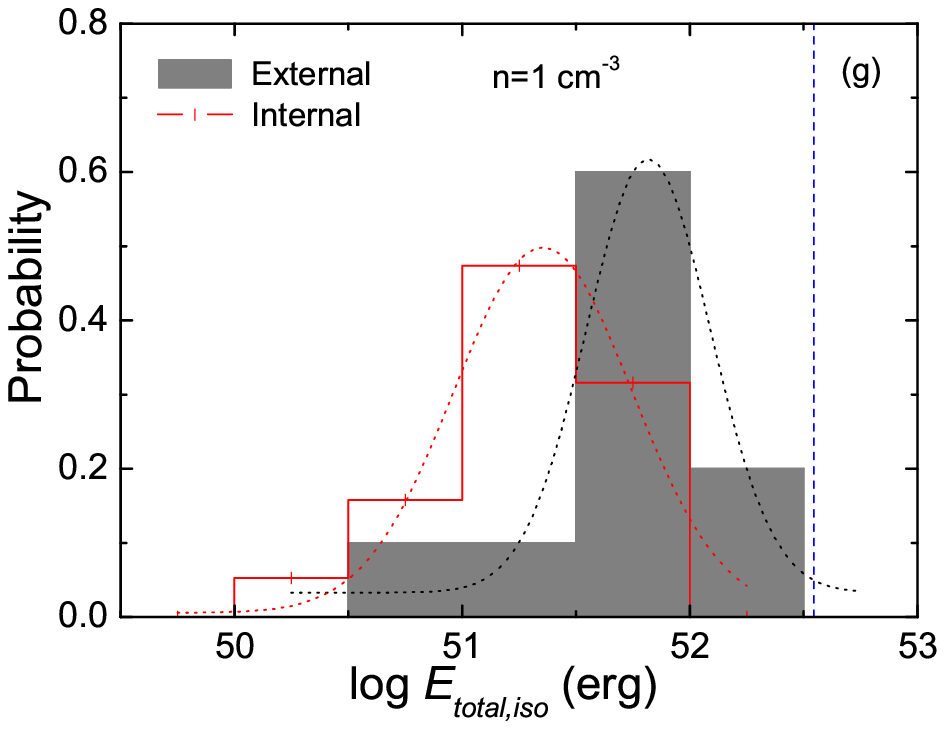}
\hfill
\includegraphics[angle=0,scale=0.45]{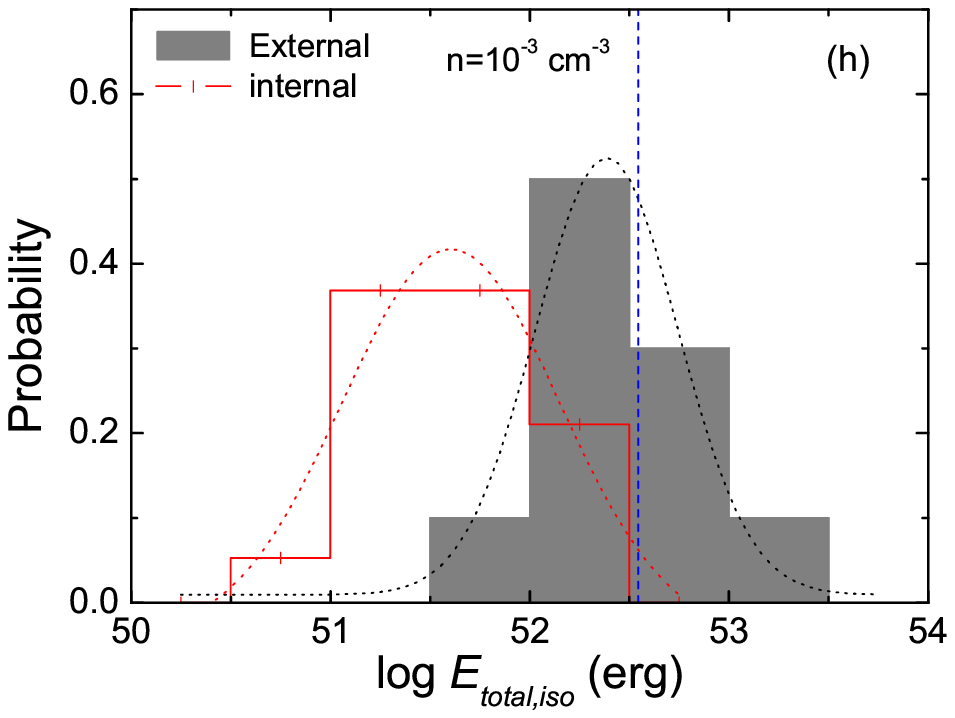}
\caption{Comparisons of various properties between the Internal
(red, open histogram) sample and the External (black, gray
histogram) samples. The best-fit Gaussian profiles are
over-plotted with the respective colors. The eight panels
denote histograms of $L_b$, $t_b$, $E_{\rm \gamma, iso}$,
$E_{\rm X, iso}$, $E_{\rm K, iso}$ and $E_{\rm total, iso}$,
respetively, with the last two parameters plotted twice for two
different medium densities, $n=1, 10^{-3}~{\rm cm^{-3}}$. The
vertical dotted line in panels (g) and (h) denotes the total
rotation energy budget of a millisecond magnetar. If no
redshift is measured, then $z=0.58$ is adopted in the calculations.}
\end{figure}

%*******************************************************************************************
\begin{figure}
\includegraphics[angle=0,scale=0.8]{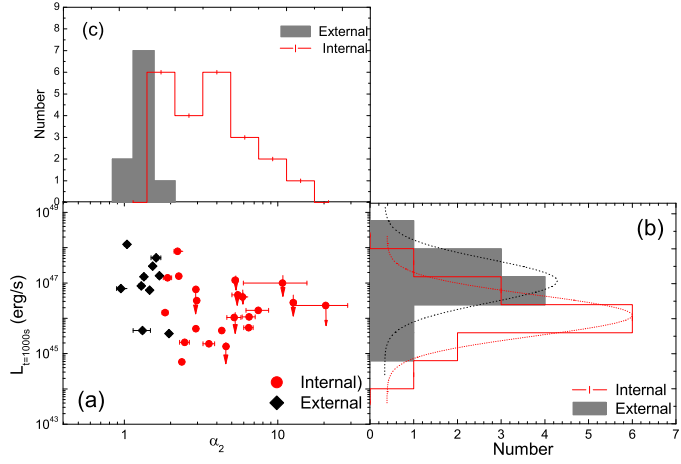}
\caption{One-dimensional (panels (b) and (c)) and two-dimensional (panel (a))
$L(t=1000~{\rm s})-\alpha_{2}$ distributions of the GRBs in our samples.
The red diamonds and black dots denote the Internal
and External samples, respectively, and the arrows indicate the
upper limits.}
\end{figure}

%*******************************************************************************************

\begin{figure}
\includegraphics[angle=0,scale=0.8]{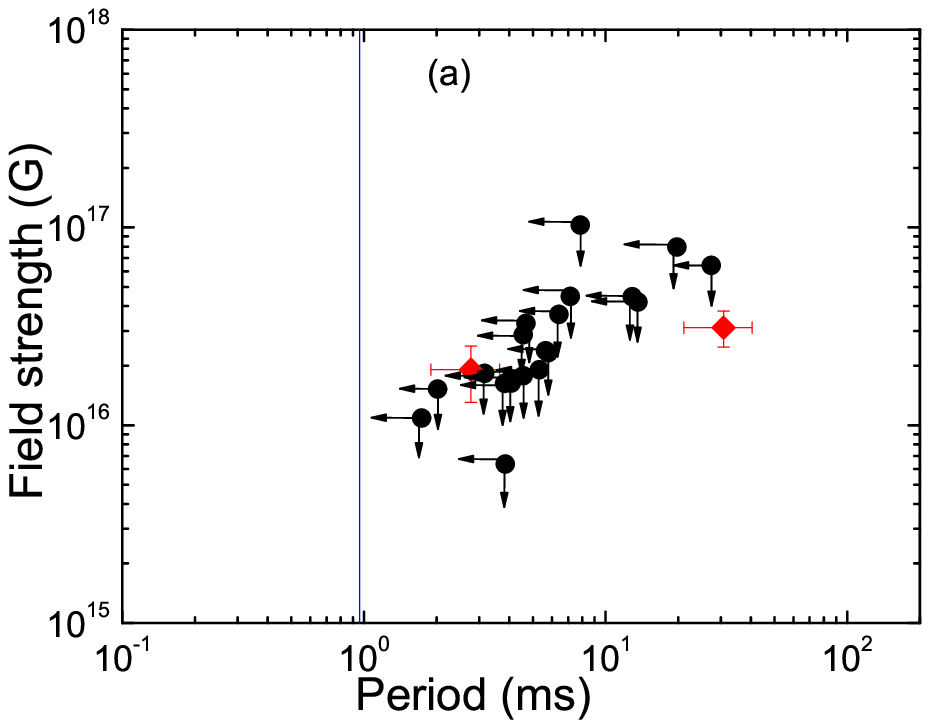}
\includegraphics[angle=0,scale=0.8]{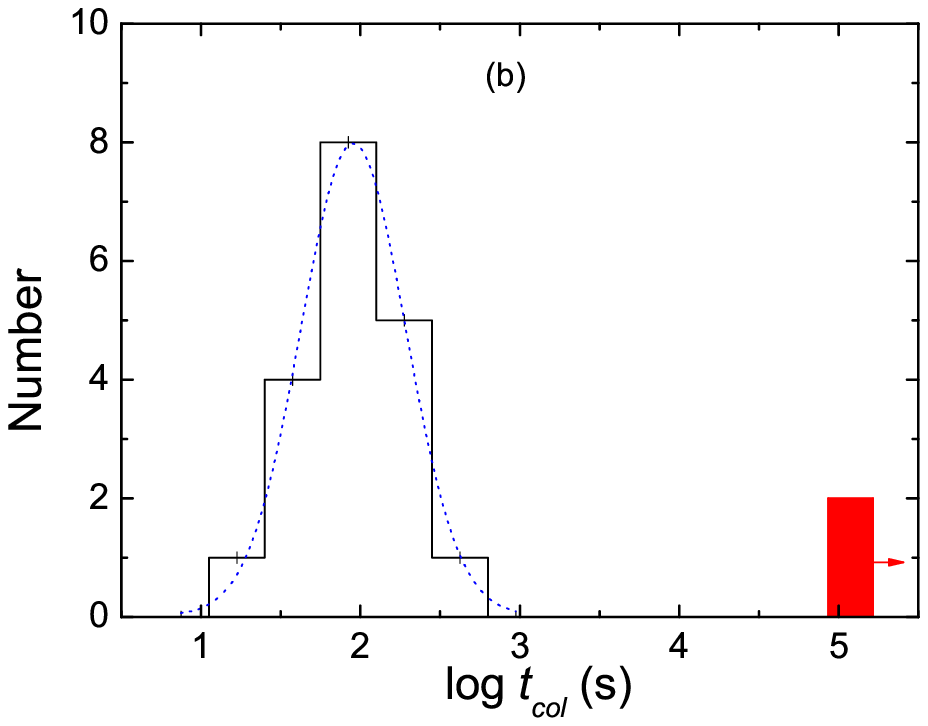}
\caption{(a) Inferred magnetar parameters, initial spin
period $P_0$ vs. surface polar cap magnetic field strength
$B_p$ derived for our Internal sample. The red diamonds
indicate GRB 061201 and GRB 070714B, which have $\tau$ measured
from $t_b$. All the other GRBs only have the lower limit of $\tau$.
The arrows denote upper limits. The vertical solid line is the break-up
spin period limit for a neutron star (Lattimer \& Prakash 2004). (b)
The distribution of the collapse time for our Internal sample. The
dotted line is the best Gaussian profile fit.}
\end{figure}

%*******************************************************************************************

\begin{figure}
\includegraphics[angle=0,scale=0.8]{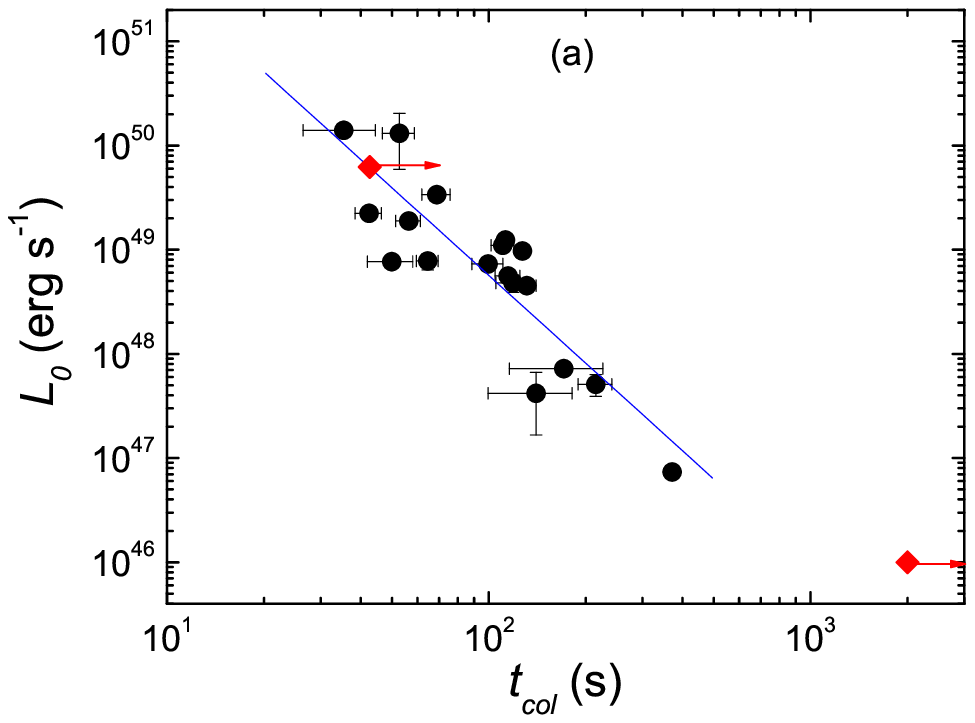}
\includegraphics[angle=0,scale=0.8]{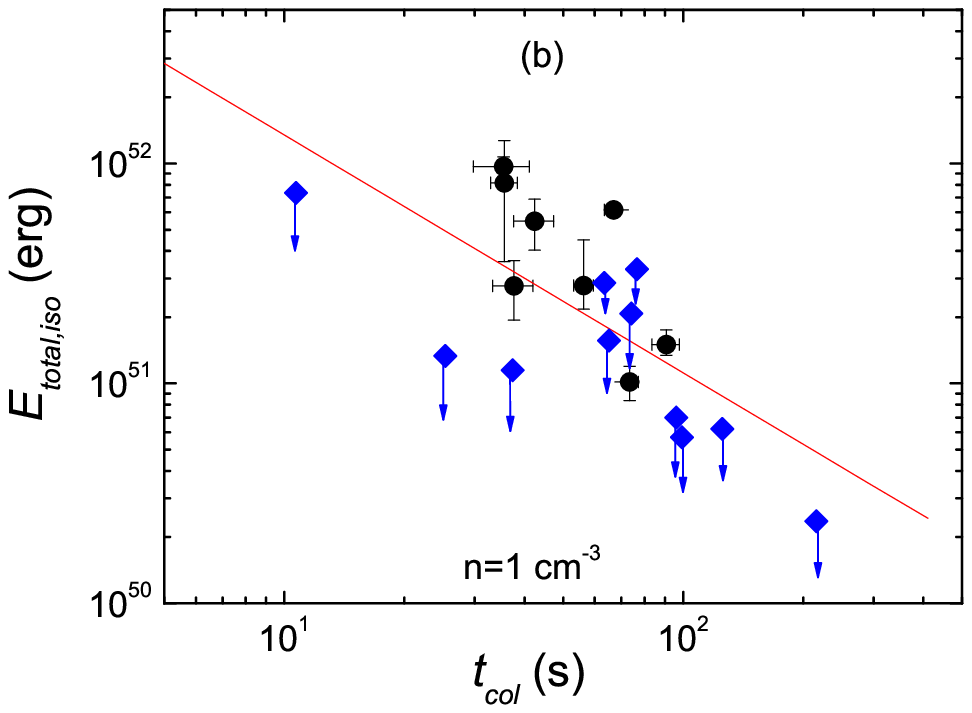}
\includegraphics[angle=0,scale=0.8]{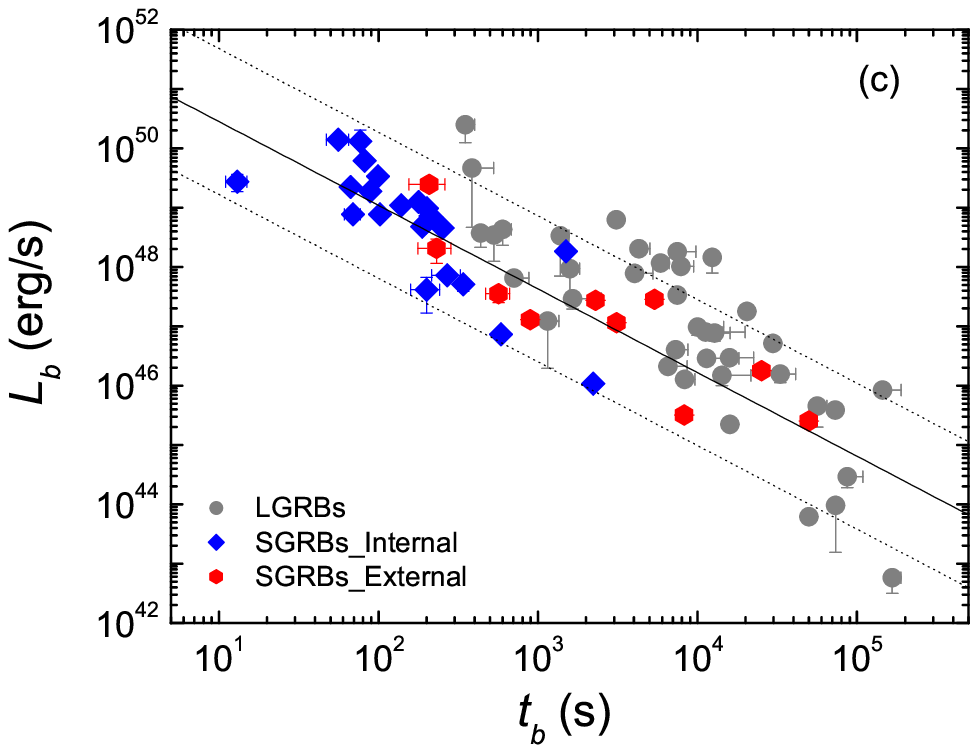}
\caption{(a) $L_{0}-t_{col}$ anti-correlation for our Internal
samples. The red diamonds are GRB 061201 and GRB 070714B, and
the arrows denote the lower limits of the collapse time.
(b) The $E_{\rm total, iso}-t_{col}$ anti-correlation for our
Internal sample using $n=1 ~{\rm cm^{-3}}$
to calculate $E_{\rm K, iso}$. The blue diamonds indicate
the upper limits to calculate $E_{\rm K, iso}$, and the red
solid line is the best-fitting line. (c) The empirical $L_b-t_b$ correlation
derived from the short GRBs in our sample (red for External and blue for Internal
samples) compared with the Dainotti relation for long GRBs (gray).
The solid line is the best power-law fit to the SGRBs sample, and the two dotted
lines denote the 2$\sigma$ region of the fit.}
\end{figure}

%*******************************************************************************************

\begin{figure}
\includegraphics[angle=0,scale=0.45]{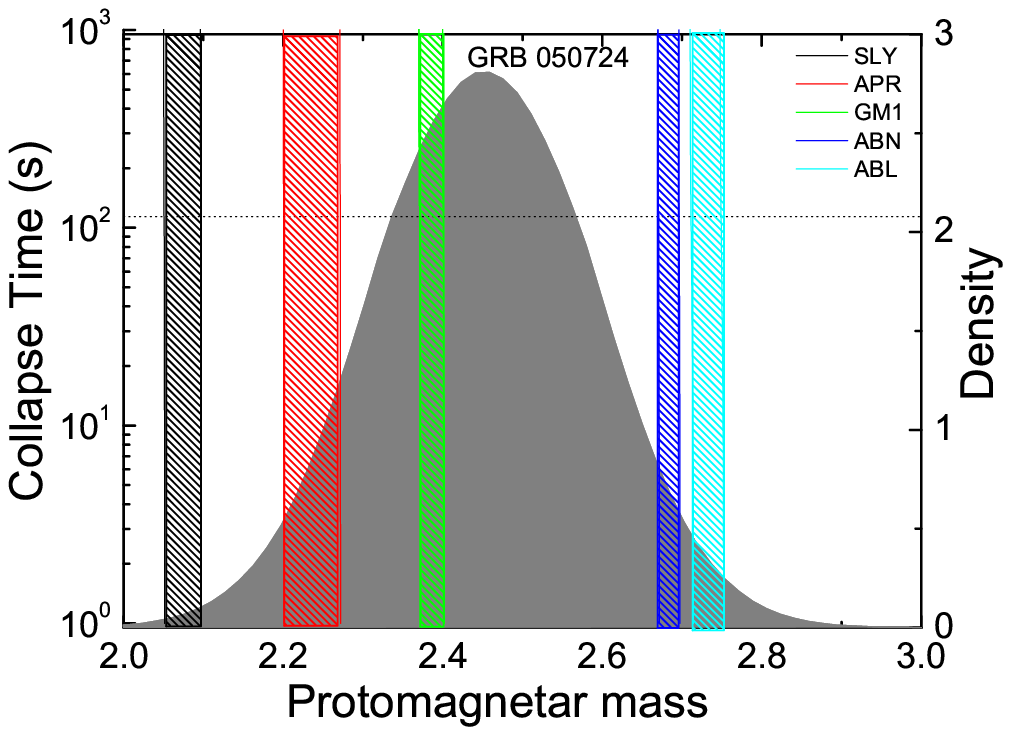}
\includegraphics[angle=0,scale=0.45]{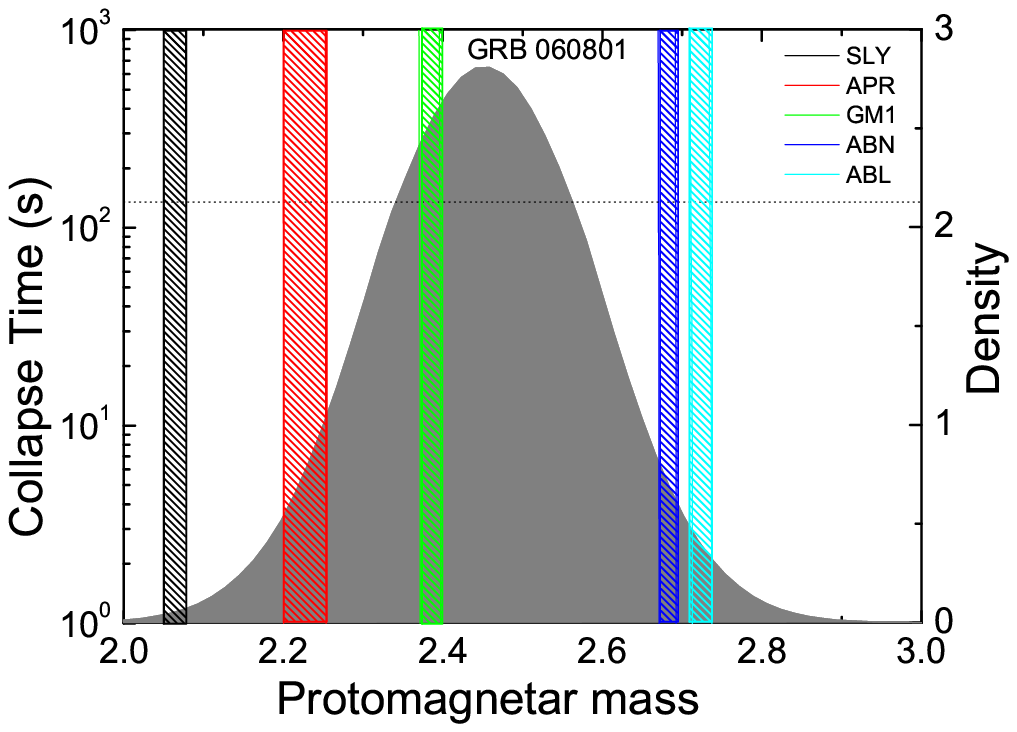}
\includegraphics[angle=0,scale=0.45]{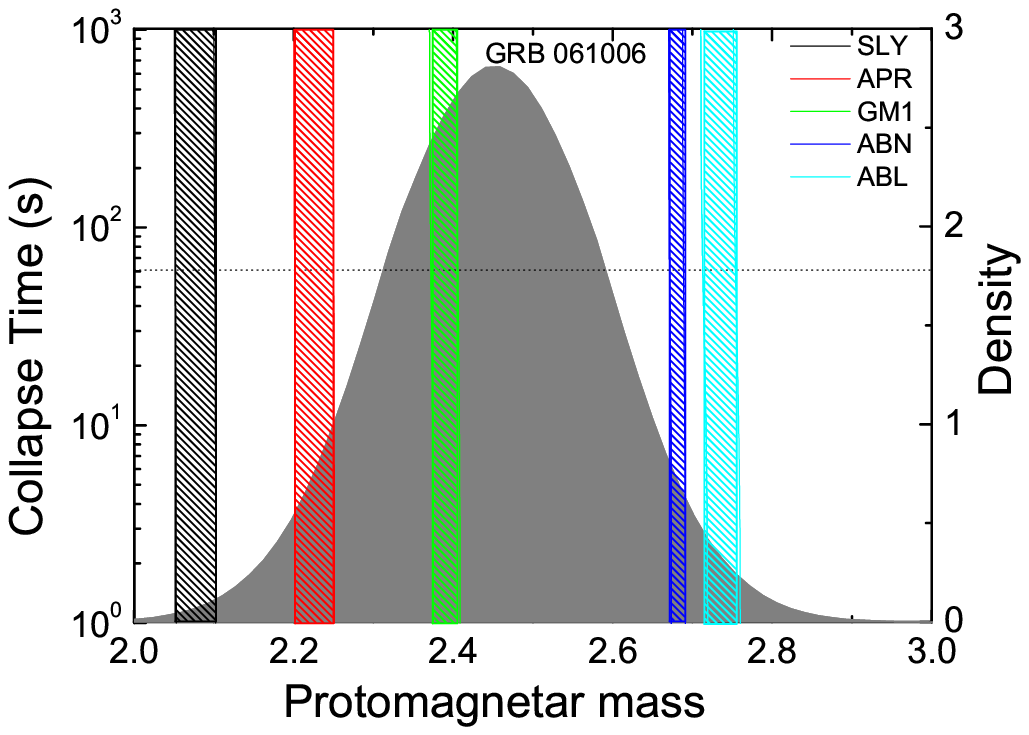}
\includegraphics[angle=0,scale=0.45]{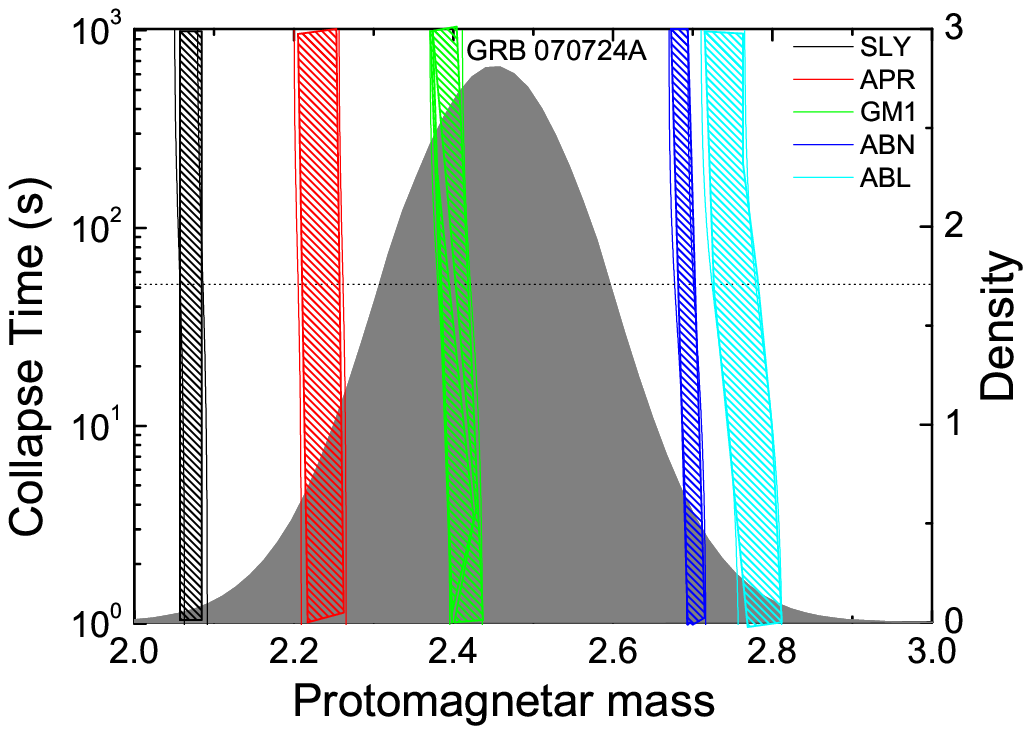}
\includegraphics[angle=0,scale=0.45]{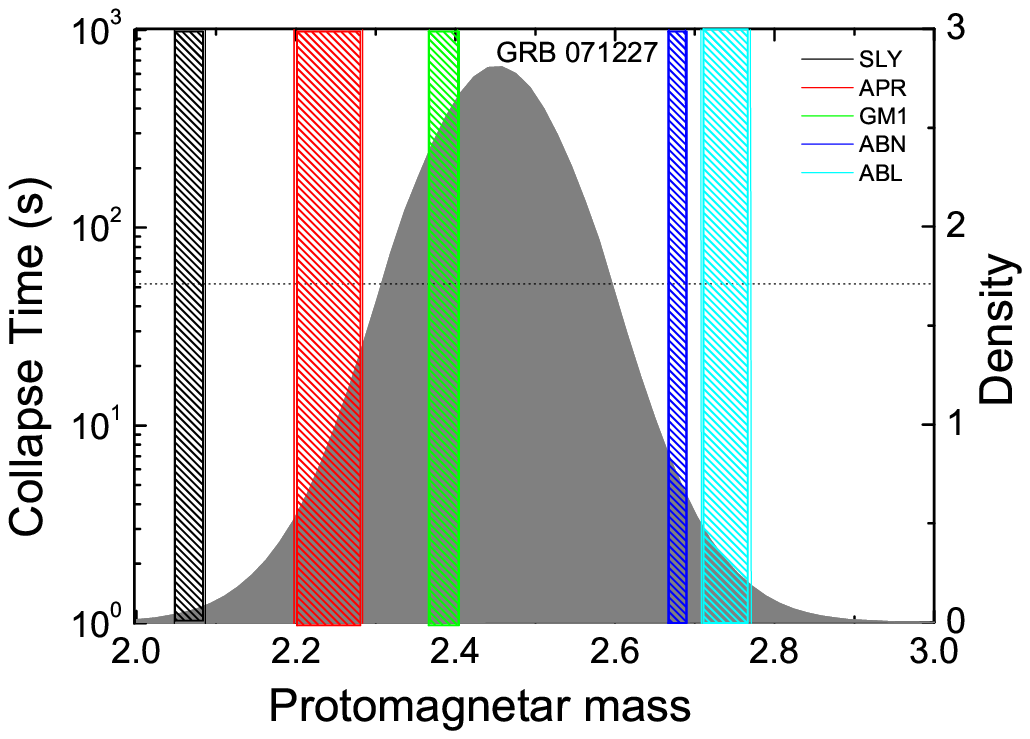}
\includegraphics[angle=0,scale=0.45]{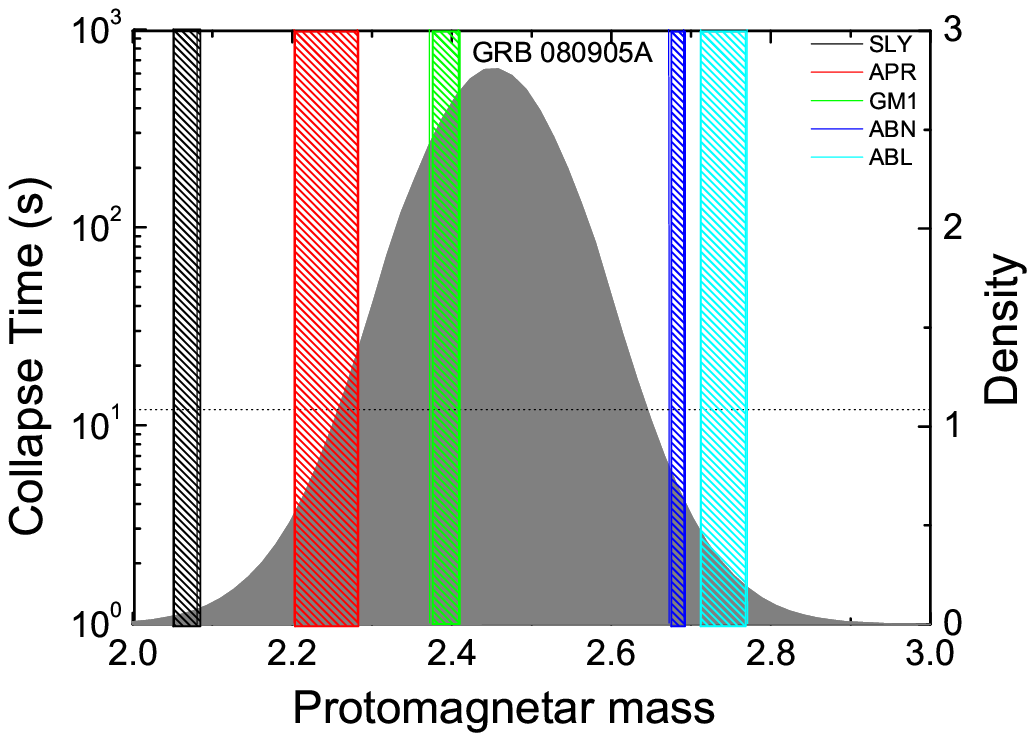}
\includegraphics[angle=0,scale=0.45]{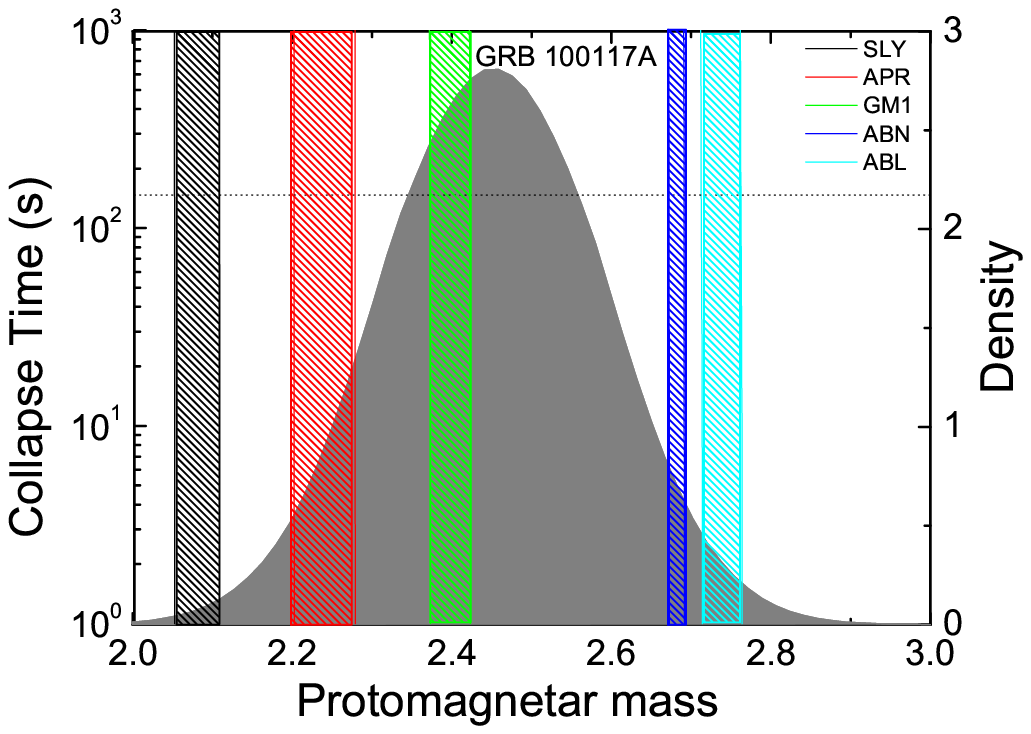}
\includegraphics[angle=0,scale=0.45]{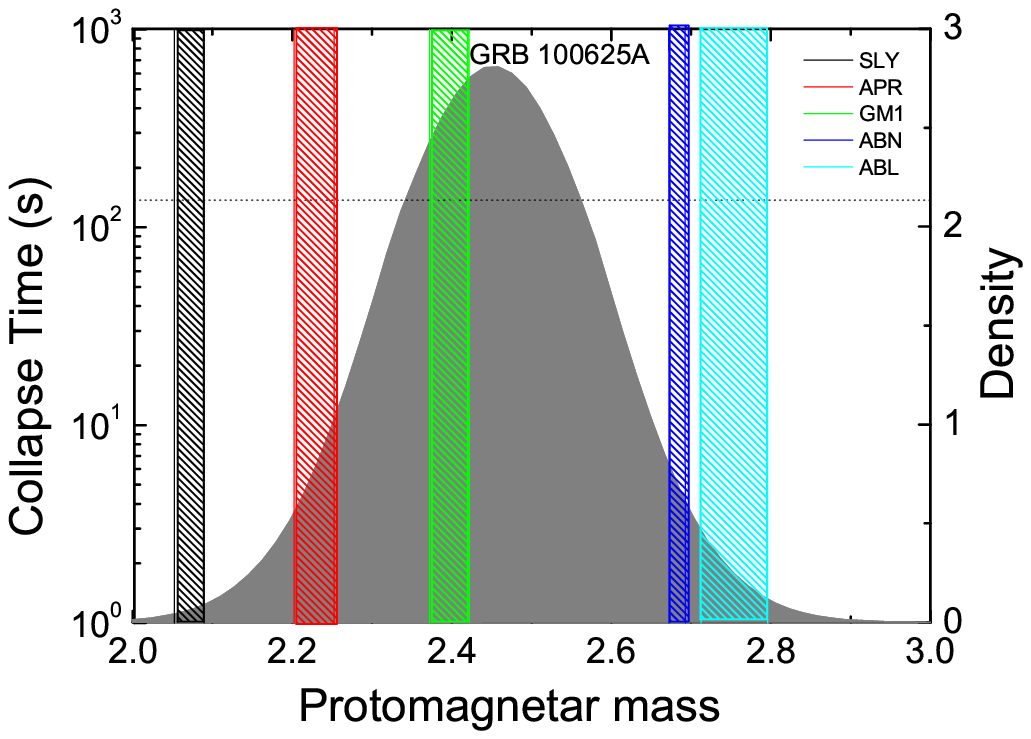}
\hfill
\includegraphics[angle=0,scale=0.45]{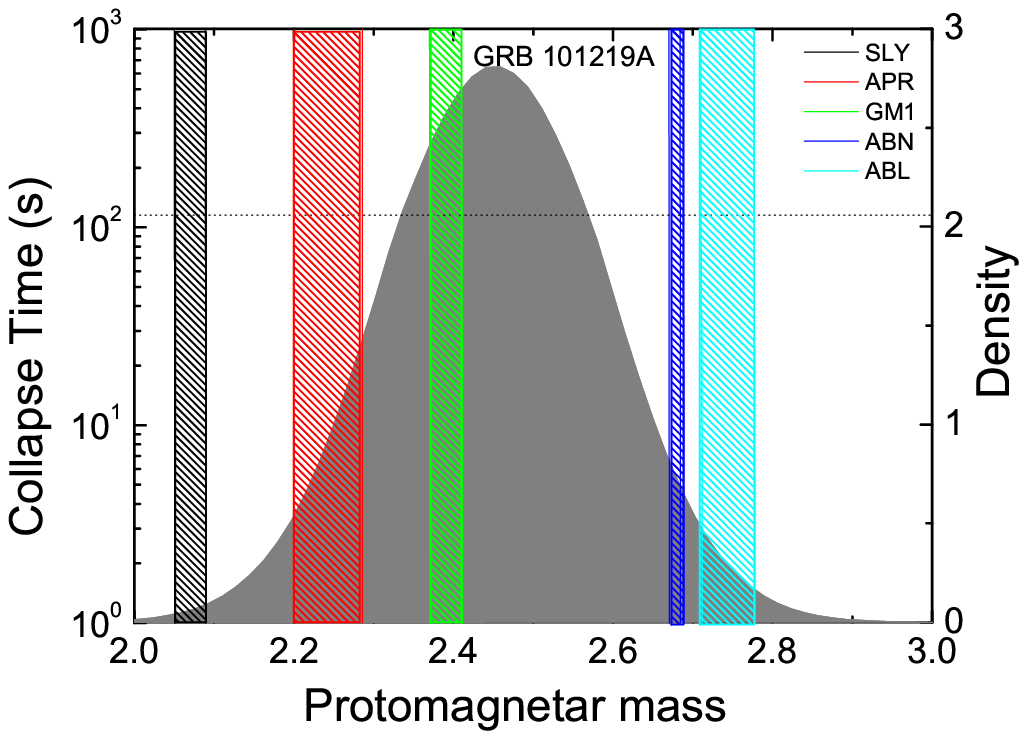}
\caption{Collapse time as a function of the protomagnetar mass.
The shaded region is the protomagnetar mass distribution derived from
the total mass distribution of the Galactic NS$-$NS binary systems. The predicted
results for 5 equations of state are shown in each panel: SLy (black),
APR (red), GM1 (green), AB-N (blue), and AB-L (cyan). The
horizontal dotted line is the observed collapse time for each GRB.}
\end{figure}
%*******************************************************************************************

\end{document}